\pdfoutput=1
\documentclass[english]{article}
\usepackage{geometry}
\usepackage{babel}
\usepackage{xcolor}
\usepackage{graphicx}
\usepackage{amsmath,amsfonts}
\usepackage{xcolor}
\usepackage{graphicx,indentfirst,subfigure,epsfig}
 \usepackage{varioref}
 \usepackage{wrapfig}
 \usepackage{subfigure}
 \usepackage{subfigmat}
 \usepackage{amsthm}
\usepackage{hyperref}
\hypersetup{colorlinks, citecolor=red, linkcolor=blue, urlcolor=red, breaklinks=true}
\usepackage{scrextend}
\usepackage{nomencl}
\makenomenclature
\usepackage{booktabs}
\usepackage{longtable}
\usepackage{soul}
\usepackage{lineno}

\usepackage[italic]{mathastext}
\usepackage{bm}
  \def\clap#1{\hbox to 0pt{\hss#1\hss}}

\providecommand{\mat}[1]{\bm{#1}}%
\renewcommand{\vec}[1]{\mathbf{#1}}

\providecommand{\mD}{\ensuremath{\mat{D}}}

\providecommand{\mF}{\ensuremath{\mat{F}}}
\providecommand{\mG}{\ensuremath{\mat{G}}}

\providecommand{\mK}{\ensuremath{\mat{K}}}

\providecommand{\mS}{\ensuremath{\mat{S}}}

\providecommand{\mX}{\ensuremath{\mat{X}}}

\providecommand{\vf}{\ensuremath{\vec{f}}}

\providecommand{\vh}{\ensuremath{\vec{h}}}

\providecommand{\vr}{\ensuremath{\vec{r}}}
\providecommand{\vs}{\ensuremath{\vec{s}}}
\providecommand{\vt}{\ensuremath{\vec{t}}}

\providecommand{\vx}{\ensuremath{\vec{x}}}

\providecommand{\vz}{\ensuremath{\vec{z}}}

\begin{document}

\title{Bayesian Assessments of Aeroengine Performance \\with Transfer Learning}

\author{Pranay Seshadri$^{\dagger}$\thanks{Address all correspondence to p.seshadri@imperial.ac.uk. Copyright (c) 2021 by Rolls-Royce  plc.}, Andrew B. Duncan$^\dagger$, George Thorne$^{\ddagger}$, \\ Geoffrey Parks$^{\mathsection}$, Ra\'{u}l V\'{a}zquez D\'{i}az$^{\ddagger}$, Mark Girolami$^{\mathsection}$}

\date{\small{$^\dagger$Imperial College London, London, U. K. \\ $^\ddagger$Rolls-Royce plc., Derby, U. K., \\ $^{\mathsection}$University of Cambridge, Cambridge, U. K.}}

\maketitle 

\abstract{Aeroengine performance is determined by temperature and pressure profiles along various axial stations within an engine. Given limited sensor measurements both along and between axial stations, we require a statistically principled approach to inferring these profiles. In this paper we detail a Bayesian methodology for interpolating the spatial temperature or pressure profile at axial stations within an aeroengine. The profile at any given axial station is represented as a spatial Gaussian random field on an annulus, with circumferential variations modelled using a Fourier basis and radial variations modelled with a squared exponential kernel. This Gaussian random field is extended to ingest data from multiple axial measurement planes, with the aim of transferring information across the planes. To facilitate this type of \emph{transfer learning}, a novel planar covariance kernel is proposed, with hyperparameters that characterise the correlation between any two measurement planes. In the scenario where precise frequencies comprising the temperature field are unknown, we utilise a sparsity-promoting prior on the frequencies to encourage sparse representations. This easily extends to cases with multiple engine planes whilst accommodating frequency variations between the planes. The main quantity of interest, the spatial area average is readily obtained in closed form. We term this the Bayesian area average and demonstrate how this metric offers far more precise averages than a sector area average---a widely used area averaging approach. Furthermore, the Bayesian area average naturally decomposes the posterior uncertainty into terms characterising insufficient sampling and sensor measurement error respectively. This too provides a significant improvement over prior standard deviation based uncertainty breakdowns.}

\section{Introduction}
\label{sec:intro}
\begin{figure}
\begin{center}
\includegraphics[scale=0.7]{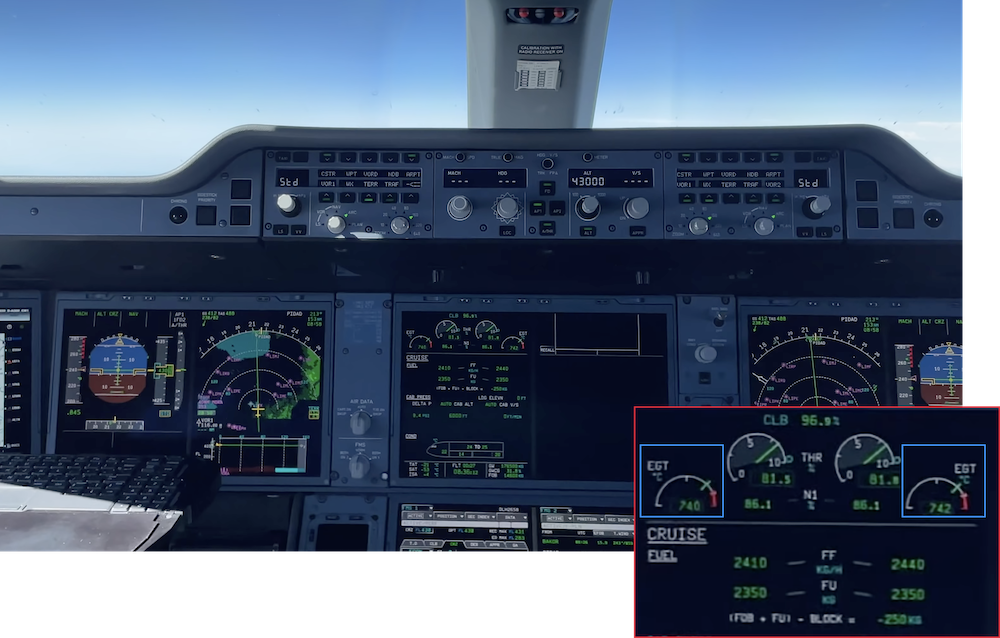}
\caption{Cockpit display of a twin-engine aircraft with a close-up (inset) of the engine performance parameters. The EGT for both engines is shown within the blue boxes.}
\label{fig:cockpit}
\end{center}
\end{figure}

Temperature and pressure measurements are vital in both the prognostics of existing in-flight engines, and the understanding of new engine architectures and component designs. There are two reasons for this. First, over many running cycles an engine will undergo a certain level of degradation. This typically manifests as an increase in blade tip and seal clearances \cite{seshadri2014robust}; an accumulation of dirt and other contaminants within the gas path, and blade surface damage owing to oxidation, sulfidation and the impact of foreign objects \cite{aust2019taxonomy}. These factors increase the amount of work the compressor has to do to achieve a certain pressure rise and the amount of work the turbine has to do to deliver the power required. The consequence of this increased workload is higher temperatures in both the compressor and turbine sections, measured via temperature probes; the pressure rise is measured via pressure probes. One of these measurements, the turbine gas temperature (TGT)\footnote{Also known as the engine gas temperature (EGT)} forms an important metric for forecasting the remaining useful life of an engine \cite{marinai2004gas, bonnet2007avoiding}. It can be found on the engine performance panel in aircraft cockpit displays, as shown in Figure~\ref{fig:cockpit}. Note that this 1D value, among others, is a reported average across a 2D non-uniform spatial field.

\begin{figure}
\begin{center}
\includegraphics[scale=0.8]{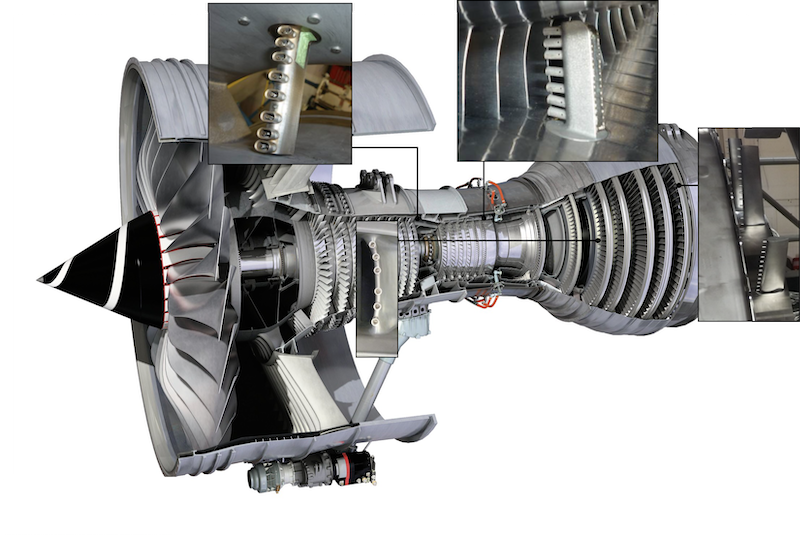}
\caption{Characteristic temperature and pressure rakes at a few locations in an aeroengine.}
\label{fig:intro2}
\end{center}
\end{figure}

\begin{figure}
\begin{center}
\includegraphics[scale=0.4]{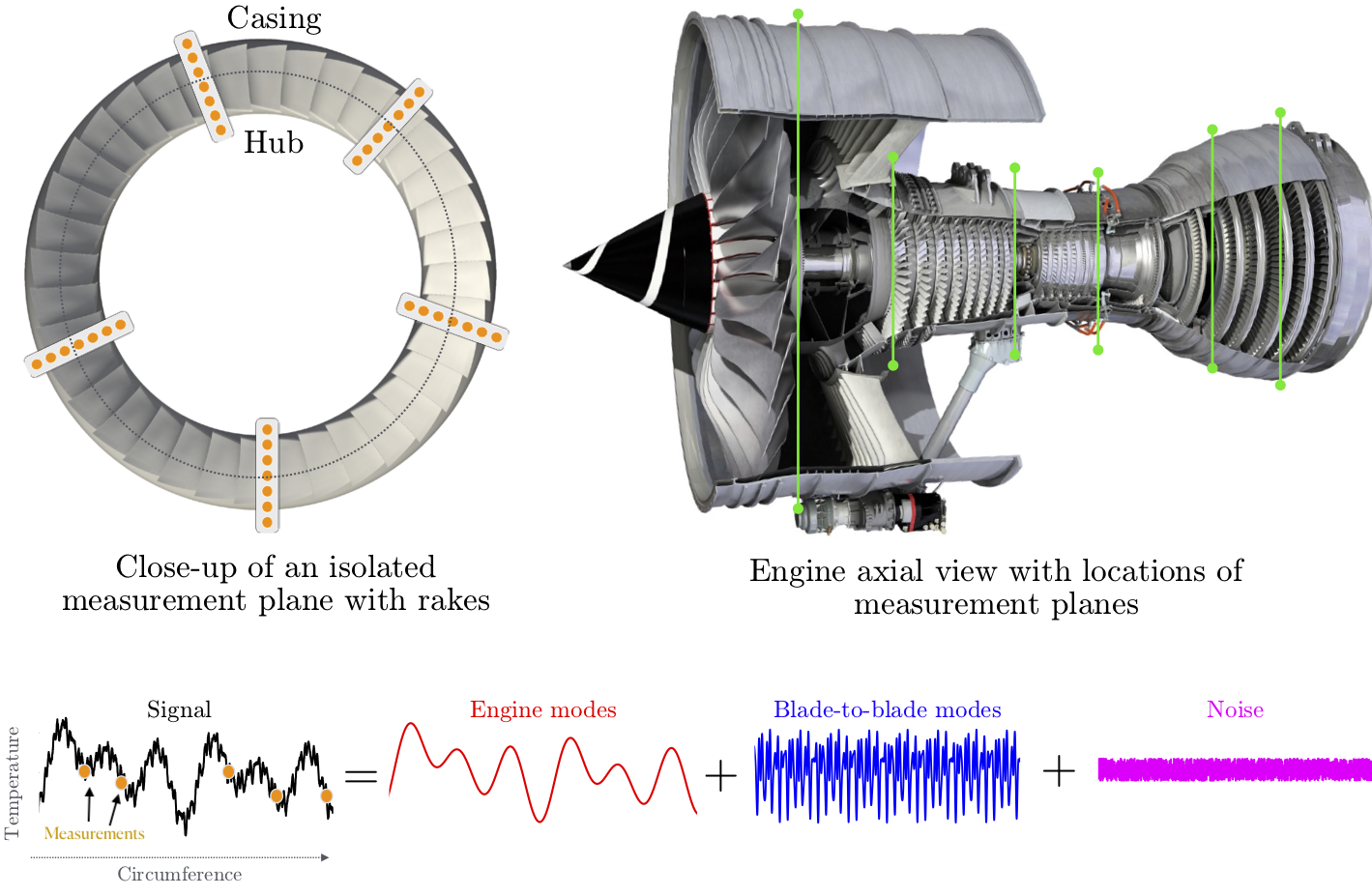}
\caption{Close-up of an axial measurement plane in an engine. Each plane is fitted with circumferentially scattered rakes with radially placed probes. The circumferential variation in temperature (or pressure) can be broken down into various modes, as shown.}
\label{fig:intro}
\end{center}
\end{figure}

The second reason why temperature and pressure measurements are so critical is because they are used to compute sub-system (e.g., low-, intermediate-, and high-pressure compressor and turbine components) efficiencies. This is done by defining a control volume around the sub-system of interest and ascertaining what the average stagnation flow properties are at the inlet and exit, whilst accounting for work being done both into and out of the system. At each measurement plane, circumferentially positioned rakes---with radially varied probes on each rake---are used to measure pressure and temperature values (see Figure~\ref{fig:intro2} and \ref{fig:intro}). These measurements are aggregated through 1D area- or mass-averages of the circumferentially and radially scattered measurements at a given axial plane. Identifying which subsystem needs to be improved based on its efficiency rating, feeds into research and development programmes for current and new engines. Furthermore, if the uncertainty in a given subsystem's calculated efficiency is deemed too large, then it is likely that a decision on adding more instrumentation or improving the precision of the existing sensors will follow. As both (i) research and development programmes for improving the performance of a given sub-system, and (ii) the enhancement of the engine measurement apparatus, are extremely expensive, it is imperative that the decisions made be based upon accurate and precise temperature and pressure values.

%
%

\subsection{1D performance values}
\begin{figure}
\begin{center}
\includegraphics[scale=0.8]{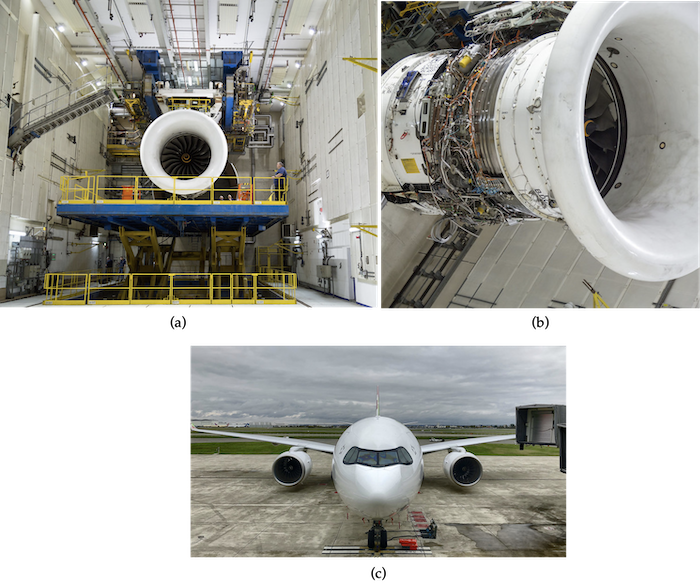}
\caption{Test bed engines in (a) and (b), in-flight engines in (c).}
\label{fig:intro4}
\end{center}
\end{figure}

As in many other engineering disciplines, 1D metrics are often used for performance assessments in turbomachinery. When provided with radial and circumferentially placed temperature or pressure measurements, area-based averages are often the norm for arriving at 1D values. These are typically estimated by assigning each sensor a weight based on the sector area it covers. This weight will depend on the total number of sensors and the radial and circumferential spacing between them \cite{stoll1979effect, francis1989measurement}. This \emph{sector area-average} is computed by taking the weighted sum of each measurement and dividing it by the sum of the weights themselves. In practice, this recipe offers accurate estimates if the spatial distribution of the measured quantity is uniform throughout the measurement plane. For spatially non-uniform flows, the validity of this approach hinges on the circumferential placement of the rakes and the harmonic content of the signal. Should all the rakes be placed so as to capture the trough of the wave forms, then the sector area-average will likely underestimate the \emph{true} area-average. A similar argument holds if the rakes are placed so as to capture only the peaks of the circumferential pattern \cite{seshadri2019a}. It is therefore common to use  empirical corrections to account for the uncertainty in such measurements, however these corrections may introduce biases. It should be noted that in-flight engines may only be fitted with one or two rakes, which may warrant additional corrections. This is different from test-bed (simulated altitude) engines as shown in Figure~\ref{fig:intro4}, which often have more rakes along the same axial plane. Additionally, test-bed engines may also have more axial measurement stations.

\subsection{Limitations with computational fluid dynamics}
A salient point to note here concerns the use and limitations (see \cite{denton2010some}) of a strictly computational approach to estimate the pressures and temperatures. Today, aeroengine computational fluid dynamics (CFD) flow-field approximations via Reynolds averaged Navier Stokes, large eddy simulations \cite{gourdain2014large} and, in some cases, via direct numerical simulations \cite{wheeler2016direct} are being increasingly adopted to gain insight into both component- and sub-system-level design. These CFD solvers with varying fidelities of underpinning equations and corresponding domain discretisations have found success---balancing simulation accuracy with simulation cost---in understanding the flow-physics in the numerous sub-systems of an aeroengine (see Figure 11 in \cite{tyacke2019turbomachinery}). However, in most cases, CFD-experimental validation is carried out using scaled experimental rigs which typically isolate one sub-system or a few stages (rows of rotors and stators) in an engine. Although there has been a tremendous body of work dedicated to incorporating real-engine effects through aleatory \cite{seshadri2015leakage, seshadri2014robust, montomoli2015uncertainty} and epistemic uncertainty quantification \cite{emory2016uncertainty} studies, as a community, we are still far from being able to replicate the aerothermal environment in engines: it is incredibly complex. For instance, the hub and casing are never perfectly circular owing to variability in thermal and fatigue loads; engine structural components introduce asymmetries into the flow that can propagate far downstream into the machine, leading to flow-field distortions; and the pressure and temperature variations induced by bleeds and leakage flows are not circumferentially uniform. The presence of these \emph{engine modes} (also termed engine wave numbers) it challenging to use CFD in isolation to calculate aeroengine performance.

\subsection{Coverage vs accuracy} 
Before we delve into the main ideas that underpin this paper, it will be helpful to understand the experimental coverage versus accuracy trade-off. Sensor placement in an engine is tedious: there are stringent space constraints on the number of sensors, the dimensions of each sensor and its ancillary equipment, along with its axial, radial and tangential location in the engine. However, engine measurements offer the most accurate representation of engine flow physics. Scaled rigs, on the other hand, offer far greater flexibility in sensor number, type and placement, and consequently yield greater \emph{measurement coverage}. While they are unable to capture the engine modes--and thus are limited in their accuracy---they offer an incredibly rich repository of information on the \emph{blade-to-blade} modes. These modes include those associated with periodic viscous mixing (such as from blade tip vortices), overturning boundary layers between two adjacent blades, and the periodic inviscid wakes \cite{sanders2002multi, mailach2008periodical}. Although present in the engine environment too, engines have insufficient measurement coverage to capture these blade-to-blade modes. One can think of the spatial distribution of pressure or temperature as being a superposition of such blade-to-blade modes (visible in rig experiments), engine modes (visible in engine tests) and noise (see Figure~\ref{fig:intro}). Succinctly stated, our best window on flow into an aeroengine---and, in consequence, its composite temperatures and pressures---stems from real engine measurements themselves. The challenge is that they are few and far between.

\subsection{State of the art}
While publicly-available work in the areas of measurement metrology \cite{saravanmuttoo1990h}, \cite{sae2017}, averaging \cite{cumpsty2006averaging, greitzer_tan_graf_2004}, and spatial field approximation \cite{seshadri2019a}, \cite{seshadri2019b} are prevalent, there is no unifying framework for these related concerns. In other words, there is no established workflow that stems from measurements to spatial field approximation to averaging, whilst rigorously accounting for all the sources of uncertainties. There are isolated estimates of uncertainties tailored for specific cases. For instance, \cite{bonham2017combination} state that, for compressors, at least seven measurements are required in the radial direction, and at least five measurements in the circumferential direction to resolve the flow. This is a heuristic, based on the negligible change in isentropic efficiency if more measurements are taken. It should be noted that this assessment is not based on a spatial model, but rather on experimental observations for a compressor with an inlet stagnation temperature of 300 Kelvin and a polytropic efficiency of 85$\%$ at three different pressure ratios. In other words, it is difficult to generalise this across all compressors.

In \cite{seshadri2019b} and \cite{seshadri2019a}, the authors present a regularised linear least squares strategy for estimating the spatial flow-field from a grid of measurements formed by radial and circumferentially placed probes. Their data-driven model represents the spatial flow-field in the circumferential direction via a Fourier series expansion, while capturing flow in the radial direction using a high-degree polynomial. Although an improvement in the state of the art \cite{lou2021reconstructing},  their model does have limitations. For instance, the placement of probes may lead to Runge's phenomenon (see Chapter 13 in \cite{trefethen2013approximation}) in the radial direction, while the harmonic content is set by the Nyquist condition (see Chapter 4 in \cite{strang2012computational}) in the circumferential direction. Another hindrance, one not systemic to their work, but one mentioned in several texts (see 8.1.4.4.3 in \cite{saravanmuttoo1990h} and in \cite{pianko1983propulsion}), is the definition of the uncertainty associated with insufficient spatial sampling and that associated with the imprecision of each sensor. This decomposition of the overall uncertainty is important as it informs aeroengine manufacturers whether they need more measurement sensors or whether they need to improve the precision of existing sensors. At present there are no rigorously derived metrics for this.

\subsection{Paper outline}
In this paper, we argue that an assessment of the area average and a decomposition of the overall uncertainty is only possible with a priori knowledge of the spatial flow field. Thus, we frame part our scope as follows. Given an array of engine sensor measurements at a single or multiple axial stations, our goal is to formulate computationally feasible and statistically rigorous techniques to:
\begin{itemize}
\item construct a spatial model to approximate the flow-field at an axial station given the inherent uncertainty in the measurements and certain physical assumptions (see section~\ref{sec:gp_kernels});
\item compute the area-average of the stagnation pressure and temperature based on this model (section~\ref{sec:area_average});
\item distinguish between uncertainty in the spatial model (and its averages) induced by sensor imprecision, and insufficient spatial sampling (section~\ref{sec:decomp});
\item quantify the dominant circumferential harmonics leveraging some notion of sparsity (section~\ref{sec:spike_and_slab}); 
\item develop methodologies that can transfer information from relatively more heavily instrumented test-bed engines to very sparsely instrumented flight engines at the same plane, and
\item foster the transfer of information between adjacent planes in an engine with the intention of reducing uncertainty (see section~\ref{sec:gp_kernels}). 
\end{itemize}
This latter two aims will be addressed using \emph{transfer learning}---an emerging sub-discipline of machine learning that seeks to transfer information between tasks, intelligently \cite{skolidis2012transfer} especially when one task is afforded more information than the other. In this paper, we explore the topics of spatial field estimation, area averaging, instrumentation sampling vs. precision uncertainty estimation, and transfer learning with Gaussian processes \cite{rasmussen2006gaussian}. 

\section{Gaussian process aeroengine model}
Gaussian processes (GPs) provide a powerful framework for nonparametric regression, where the regression function is  modelled as a random process, such that the distribution of the function evaluated at any finite set of points is jointly Gaussian. GPs are characterised by a mean function and a two-point covariance function. GPs have been widely used to model spatial and temporal varying data since their first application in modeling ore reserves in mining \cite{krige1951statistical}, leading to a method for spatial interpolation known as kriging in the geostatistics community \cite{cressie2015statistics, stein2012interpolation}. The seminal work of Kennedy and O'Hagan provides a mature Bayesian formulation which forms the underpinnings of the approach adopted in this paper. Emulation methods based on GPs are now widespread and find uses in numerous applications ranging from computer code calibration \cite{higdon2004combining} and uncertainty analysis \cite{oakley2002bayesian} to sensitivity analysis \cite{oakley2004probabilistic}. Since then GP regression has enjoyed a rich modern history within uncertainty quantification \cite{kennedy2001bayesian}, with increasingly sophisticated extensions beyond the classical formulation, including latent space models \cite{chen2015uncertainty}, coregional models \cite{alvarez2012kernels} convolutional processes \cite{higdon2002space, alvarez2011computationally}, multi-task processes \cite{bonilla2008multi}, and GPs with incorporated dimension reduction \cite{liu2017dimension, seshadri2019dimension}.

In a multi-task Gaussian process one is given similar but distinct multiple input-output data sets---each referred to as a task. Rather than train a single model for each task (single-task), the idea is to train a single model for all the tasks simultaneously. The advantage is that by constructing the latter, information can be readily shared across tasks in a meaningful manner, thereby aiding in improved inference. This implies, either implicitly or explicitly, that there are features of the model that are either hierarchical or define common structure across the different tasks. As \cite{skolidis2012transfer} remarks, multi-task Gaussian processes can be advantageous when compared to single-task Gaussian processes when there is insufficient data to infer all the model's parameters. It is expected that multi-task Gaussian processes would exploit the common structure prevalent across all tasks for improved parameter inference. Practically, one approach is to express the covariance function as the Kronkecer product of a task-based covariance function and a data-based covariance function (see 54 in \cite{skolidis2012transfer}). While a Kronecker product-based definition of the multi-task kernel does have computational advantages, it restricts one to using the same set of radial and circumferential measurements at each measurement plane. 

We end this brief literature survey with a remark on subtlety between multi-task models and models with transfer learning. All transfer learning models all inherently multi-task, however, not all multi-task models are transfer learning models. The key distinction lies in whether any information is actually transferred across the tasks, and whether that transfer leads to a more well-defined model.

\subsection{Preliminaries and data}
In this subsection, we present a GP aeroengine spatial model---designed to emulate the steady-state temperature and pressure distributions at multiple axial plane. Given the complexity of the flow, our aim is to capture the \emph{primary aerothermal features} rather than resolve the flow-field to minute detail. One can think of the primary aerothermal features as being the engine modes in the circumferential direction. In what follows we detail our GP regression model; our notation closely follows the GP exposition of Rogers and Girolami (see Chapter 8 in \cite{rogers2016first}).

Let us assume that we have sensor measurement location and sensor reading pairs $\left(\vx_i, f_i \right)$ for $i=1, \ldots, N$ and $M$ locations at which we would like to make reading predictions
\begin{equation}
\mX=\left[\begin{array}{c}
\vx_{1}\\
\vdots\\
\vx_{N}
\end{array}\right] \; \; \; \; \vf=\left[\begin{array}{c}
f_{1}\\
\vdots\\
f_{N}
\end{array}\right] \; \; \; \; \text{and} \; \; \; \; \mX^{\ast}=\left[\begin{array}{c}
\vx_{1}^{\ast}\\
\vdots\\
\vx_{M}^{\ast}
\end{array}\right] \; \; \; \; \vf^{\ast}=\left[\begin{array}{c}
f_{1}^{\ast}\\
\vdots\\
f_{M}^{\ast}
\end{array}\right]
\end{equation}
where the superscript $\left( \ast \right)$ denotes the latter. Here $\vx_{i} \in \mathbb{R}^{3}$, thus $\mX \in \mathbb{R}^{N \times 3}$. Without loss in generality, we assume that $\sum_i^{N} f_i = 0$, so that the components correspond to deviations around the mean; physically, being either temperature or pressure measurements taken at the locations in $\mX$. Let the values in $\vf$ be characterised by a symmetric \emph{measurement covariance} matrix $\boldsymbol{\Sigma} \in \mathbb{R}^{N \times N}$ with diagonal measurement variance terms $\sigma_{i}^2$ for $i=1, \ldots, N$. In practice, $\boldsymbol{\Sigma}$, or at least an upper bound on $\boldsymbol{\Sigma}$, can be determined from the instrumentation device used and the correlations between measurement uncertainties, which will be set by an array of factors such as the instrumentation wiring, batch calibration procedure, data acquisition system and filtering methodologies. Thus the true measurements $\vt \in \mathbb{R}^{N}$ are corrupted by a zero-mean Gaussian noise, $\vf = \vt + \boldsymbol{\Sigma}$ yielding the observed sensor values. This noise model, or likelihood, can be expressed as $\mathbb{P} \left( \vf |  \vt , \mX \right) = \mathcal{N} \left( \vf , \boldsymbol{\Sigma} \right)$. 

In the absence of measurements, we assume that $\vf$ is a Gaussian random field with a mean of $\boldsymbol{0}$ and has a two-point covariance function $k\left(\cdot, \cdot \right)$. The joint distribution of $\left(\vf, \vf^{\ast} \right)$ satisfies
\begin{equation}
\left[\begin{array}{c}
\vf\\
\vf^{\ast}
\end{array}\right]\sim\mathcal{N}\left( \mathbf{0},\left[\begin{array}{cc}
\mK_{\circ \circ} +\boldsymbol{\Sigma} & \mK_{\circ \ast}  \\
\mK_{\circ \ast}^{T}  & \mK_{\ast \ast}
\end{array}\right]\right),
\label{equ:joint_dist}
\end{equation}
where the Gram matrices are given by
\begin{align}
[\mK_{\circ \circ}]_{\left(i, j \right)} = k ( \vx_{i}, \vx_{j} ), \; \; \; \; &  [\mK_{\circ \ast }]_{ \left(i, l \right)} = k ( \vx_{i}, \vx^{\ast}_{l} ), \; \; \; \; \textrm{and} \; \; \; \; [\mK_{\ast \ast }]_{\left(l, m \right)} = k ( \vx^{\ast}_{l}, \vx^{\ast}_{m} ),
\end{align}
for $i,j=1, \ldots, N$ and $l,m=1, \ldots, M$. From \eqref{equ:joint_dist}, we can write the predictive posterior distribution of $\vf^{\ast}$ given $\vf$ as
\begin{equation}
\mathbb{P} \left( \vf^{\ast} | \vf , \mX^{\ast}, \mX \right) = \mathcal{N} \left( \boldsymbol{\mu}^{\ast}, \boldsymbol{\Psi}^{\ast} \right),
\end{equation}
where the conditional mean is given by
\begin{align}
\begin{split}
\boldsymbol{\mu}^{\ast} & =   \mK^{T}_{\circ\ast} \left( \mK_{\circ \circ} + \boldsymbol{\Sigma} \right)^{-1} \vf \\
& = \mK^{T}_{\circ \ast}  \mG^{-1} \vf
\end{split}
\label{equ:cond_mean}
\end{align}
with $\mG= \left(\mK_{\circ \circ} + \boldsymbol{\Sigma} \right)$; the conditional covariance is
\begin{align}
\boldsymbol{\Psi}^{\ast} & = \mK_{\ast \ast} - \mK^{T}_{\circ \ast} \mG^{-1}  \mK_{\circ \ast}.
\label{equ:cond_var}
\end{align}
\subsection{Defining the covariance kernels}
\label{sec:gp_kernels}
As our interest lies in applying Gaussian process regression over $P$ engine planes, our inputs $\vx_{i} \in\left\{ \left(r_{i}, \theta_{i}, \rho_{i} \right): r_{i} \in\left[0,1\right], \theta_{i} \in\left[0,2\pi\right), \rho_{i} \in \left[1, P \right] \right\}$ can be parameterized as
\begin{equation}
\mX= \left[\begin{array}{ccc}
r_{1} & \theta_{1} & \rho_{1}\\
\vdots & \vdots & \vdots\\
r_{N} & \theta_{N} & \rho_{N}
\end{array}\right] = \left[\begin{array}{ccc}
\vr & \boldsymbol{\theta} & \boldsymbol{\rho} \\
\end{array}\right], \; \; \; \; \text{and} \; \; \; \; \mX^{\ast}= \left[\begin{array}{ccc}
r_{1} & \theta_{1} & \rho_{1}\\
\vdots & \vdots & \vdots \\
r_{M} & \theta_{M} & \rho_{M}
\end{array}\right] = \left[\begin{array}{ccc}
\vr^{\ast} & \boldsymbol{\theta}^{\ast} & \boldsymbol{\rho}^{\ast}\\
\end{array}\right].
\label{equ:inputs}
\end{equation}
In most situations under consideration, we expect that
\begin{equation}
\mX = \left\{ \left(r_{i}, \theta_{j}, \rho_{l} \right), r_{i} \in \vr, \theta_{j} \in \boldsymbol{\theta} , \rho_{l} \in \boldsymbol{\rho} \right\},
\end{equation}
where $\vr$ is a set of $L$ radial locations, $\boldsymbol{\theta}$ is a set of $O$ circumferential locations and $P$ is the number of measurement planes, such that $N=L \times O \times P$, assuming the measurements across the $P$ planes are taken at the same locations. We define the spatial kernel to be a product of a Fourier kernel $k_c$ in the circumferential direction, a squared exponential kernel $k_r$ in the radial direction, and a planar kernel $k_p$ along the discrete $P$ different planes
\begin{equation}
\begin{split}
k \left( \vx, \vx' \right) & = k\left( \left( \vr, \boldsymbol{\theta} , \boldsymbol{\rho} \right),   \left( \vr', \boldsymbol{\theta}' \boldsymbol{\rho}' \right) \right) \\
& = k_{r} \left( \vr, \vr'  \right) \odot k_{c} \left( \boldsymbol{\theta}, \boldsymbol{\theta}'  \right) \odot k_{p} \left( \boldsymbol{\rho}, \boldsymbol{\rho}'  \right).
\end{split}
\label{equ:joint_kernel}
\end{equation}
where the symbol $\odot$ indicates a Hadamard (element-wise) product\footnote{For computational efficiency, the Kronecker product can also be used in cases where there are no missing entries, i.e., sensor values can be obtained from a grid of measurements.}.

Along the radial direction, the kernel has the form
\begin{equation}
k_s( \vr, \vr' ) = \sigma_f^2 \; \text{exp} \; \left(  - \frac{1}{2l^2} \left(\vr - \vr' \right)^{T}  \left(\vr - \vr' \right)  \right)
\end{equation}
where $\sigma_f$ is the signal variance and $l$ is the length-scale---two hyperparameters that need to be computationally ascertained.

For the planar kernel, define $\vs \in \mathbb{Z}_{+}^{P}$ to be a \emph{similarity} vector of length $P$ comprised of strictly positive integers. Repetitions in $s$ are permitted and are used to indicate which planes are similar. For instance, if we set $\vs=\left(1, 1, 2\right)$, this indicates that the first two planes are similar. We will use the notation $\vs\left(\rho\right)$ to select the similarity value corresponding to a specific plane $\rho$. The number of unique integers in $\vs$ may be thought of as the number of independent planes; let this be given by $Q$, implying $Q \leq P$.

Seeing as there are $Q$ independent planes, we require a metric that serves to \emph{correlate} the different independent plane combinations. To this end, consider a symmetric matrix $\mS \in \mathbb{R}^{Q \times Q}$ with a diagonal formed of $\eta$ values. The $\eta$ values denote the correlation between planes that are similar, and by construction it is a tunable hyperparameter. In practice, unless the planes are identical, their correlation will be less than unity, i.e., $\eta < 1$. Next, set $W=Q(Q-1)/2$, corresponding to the number of upper (or lower) triangular off-diagonal elements in a $Q \times Q$ matrix. As each off-diagonal entry represents a pairwise correlation between two independent planes, it needs to be represented via another appropriate hyperparameter. Let $\boldsymbol{\xi}=\left(\xi_1, \ldots, \xi_{W}\right)$ be this hyperparameter, yielding 
\begin{equation}
\mS=\left[\begin{array}{cccc}
\eta & \xi_{1} & \ldots & \xi_{Q-1}\\
\xi_{1} & \ddots & \ldots & \vdots\\
\vdots & \ldots & \ddots & \xi_{W}\\
\xi_{Q-1} & \ldots & \xi_{W} & \eta
\end{array}\right].
\end{equation}

Then, the planar kernel us given by
\begin{equation}
k_{\rho}\left(\boldsymbol{\rho}_{i},\boldsymbol{\rho}_{j}'\right)=\begin{cases}
\begin{array}{cc}
1 & \;\; \; \text{if}\; \; \boldsymbol{\rho}_{i}=\boldsymbol{\rho}_{j}'\\
\left[\mS\right]_{\vs\left(\boldsymbol{\rho}_{i}\right), \vs\left(\boldsymbol{\rho}_{j}'\right)} & \text{otherwise.}
\end{array}\end{cases}
\label{equ:kernel_planar}
\end{equation}
In summary, the planar kernel establishes the correlation between all the $P$ measurement planes. It is invariant to the radial and circumferential values and is only dependent upon the planes chosen.

Prior to defining the kernel along the circumferential direction, a few additional definitions are necessary. Let $\boldsymbol{\omega} = \left( \omega_1, \ldots, \omega_{K}\right)$ indicate the $K$ wave numbers present along the circumferential direction for a given plane. These can be a specific set, i.e., $\boldsymbol{\omega} = \left(1,4, 6, 10 \right)$, or can be all wave numbers up to a particular cut-off, i.e., $\boldsymbol{\omega} = \left(1, 2, \ldots, 25 \right)$. We define a \emph{Fourier design matrix} $\mF \in \mathbb{R}^{\left(2K + 1 \right) \times N}$, the entries of which are given by
\begin{equation}
\mF_{ij} \left( \boldsymbol{\theta} \right) =\begin{cases}
\begin{array}{cc}
1 & \text{if}\; \; i=1,\\
\sin\left(\boldsymbol{\omega}_{\frac{i}{2}}\pi\boldsymbol{\theta}_{j}/180^{\circ}\right) & \text{if} \; \; i>1\; \; \text{when} \; i \; \text{is even},\\
\cos\left(\boldsymbol{\omega}_{\frac{i-1}{2}}\pi\boldsymbol{\theta}_{j}/180^{\circ}\right) & \text{if} \; \; i>1\; \; \text{when} \; i \; \text{is odd},
\end{array}\end{cases}.
\end{equation}
Note that the number of columns in $\mF$ depends on the size of the inputs $\boldsymbol{\theta}$. To \emph{partially} control the amplitude and phase of the Fourier modes and the value of the mean term, we introduce a set of diagonal matrices $\mathcal{D} = \left(\mD_1, \ldots, \mD_{Q} \right)$. Each matrix has dimension $\mathbb{R}^{(2K+1) \times (2K+1)}$, with entries $\mD_i = \text{diag}\left(\delta^2_{i,1}, \ldots, \delta^2_{i,2K+1} \right)$ for $i=1, \ldots, Q$. Note, we use the word partially, as these parameters are not indicative of the amplitude or phase directly, and depend on the measured data. Furthermore, note that the matrices in $\mathcal{D}$, and thus number of tunable hyperparameters, scale as a function of the number of independent planes $Q$ and not by the total number of planes $P$. The kernel in the circumferential direction may then be written as
\begin{equation}
k_{c}\left(\left(\boldsymbol{\theta} , \boldsymbol{\rho}_{i} \right),  \left(\boldsymbol{\theta}' , \boldsymbol{\rho}_{j}' \right)\right) = \mF\left(\boldsymbol{\theta} \right)^{T} \sqrt{\mD_{\vs\left(\boldsymbol{\rho}_{j}\right)}} \sqrt{\mD_{\vs\left(\boldsymbol{\rho}_{j}'\right)}} \mF\left(\boldsymbol{\theta}' \right),
\label{equ:fourier_kernel}
\end{equation}
where the notation $\mD_{\vs\left(\boldsymbol{\rho}_{i} \right)}$ corresponds to the diagonal matrix index by $\vs\left(\boldsymbol{\rho}_{i} \right)$. We remark here that as written in \eqref{equ:fourier_kernel} the Fourier modes across all the $P$ planes are fixed, though the amplitudes and phases can vary.

Having established the definition of the radial, planar and circumferential kernels, it is worthwhile to make a note of our aim. Using the GP regression framework implies that our model prior is Gaussian $\mathbb{P} \left( \vt | \mX \right) = \mathcal{N} \left( \boldsymbol{0}, \mK_{\circ \circ} \right)$. We have already established that our likelihood function is also Gaussian. The central objective of our effort is to determine the posterior $\mathbb{P}\left( \vf | \mX , \sigma_{m}^2, \sigma_{f}^2, l^2, \delta_{i,1}^2, \ldots, \delta_{i,2K+1}^2, \eta, \xi_1, \ldots, \xi_W \right)$. In the following section we will prescribe priors on the hyperparameters to reflect a priori assumptions on the profiles.

\section{Priors}
In this section, we impose priors on the hyperparameters in \eqref{equ:joint_kernel}. Priors for the measurement noise and the squared exponential kernel are given by
\begin{equation}
\begin{split}
\sigma_{m} & \sim \mathcal{U} \left[ 0, \epsilon \right], \\
\alpha & \sim \mathcal{N}^{+} \left(0, 1 \right), \\
l & \sim  \mathcal{N}^{+} \left(0, 1 \right),
\end{split}
\label{equ:hyperparameter_priors}
\end{equation}
where $\epsilon$ is an estimate of the standard deviation of the instrumentation measurement uncertainty, $\mathcal{N}^{+}$ represents a half-Gaussian distribution and $\mathcal{U}$ represents a uniform distribution. For the planar kernel, in this paper we set $\eta=0.96$ and assign
\begin{equation}
\xi_{i} \sim \mathcal{U}\left[-1, 1 \right]
\end{equation}
for $i=1, \ldots, W$. Priors for the Fourier kernel are detailed below.
\subsection{Simple prior}
There are likely to be instances where the precise harmonics $\boldsymbol{\omega}$ are known, although this is typically the exception and not the norm. In such cases, the Fourier priors for a given plane index $i$ may be given by $\delta{i,j}^2 \sim \mathcal{N}^{+} \left(0, 1 \right)$, for $j=1, \ldots, 2K+1$. 

\subsection{Sparsity promoting prior}
\label{sec:spike_and_slab}
In the absence of further physical knowledge, we constrain the posterior by invoking an assumption of sparsity, i.e., the spatial measurements can be adequately explained by a small subset of the possible harmonics. This is motivated by the expectation of a sparse number of Fourier modes as contributing to the total variation. In adopting this assumption, we expect to reduce the variance at the cost of a possible misfit. Here, we engage the use of sparsity promoting priors, which mimic the shrinkage behaviour of the least absolute shrinkage and selection operator (LASSO) \cite{tibshirani1996regression, buhlmann2011statistics} in the fully Bayesian context.

A well-known shrinkage prior for regression models is the spike-and-slab prior \cite{ishwaran2005spike}, which involves discrete binary variables indicating whether or not a particular frequency is employed in the regression. While this choice of prior would result in a truly sparse regression model, where Fourier modes are selected or de-selected discretely, sampling methods for such models tend to demonstrate extremely poor mixing. This motivates the use of continuous shrinkage priors, such as the horseshoe \cite{carvalho2009handling} and regularized horseshoe \cite{piironen2017sparsity} prior. In both of these a global scale parameter $\tau$ is introduced for promoting sparsity; large values of $\tau$ will lead to diffuse priors and permit a small amount of shrinkage, while small values of $\tau$ will shrink all of the hyperparameters towards zero. The regularized horseshoe is given by
\begin{equation}
\begin{split}
c & \sim \mathcal{IG} \left( \frac{\gamma}{2}, \frac{\gamma s^2}{2} \right), \\
\tilde{\lambda_{i}} & \sim \mathcal{C}^{+} \left(0, 1 \right), \\
\delta_{i,j}^{2} & = \frac{c \tilde{\delta}_{i,j}^2 }{c + \tau^2 \tilde{\delta}^2_{i,j}}, \; \; \; \; \text{for}\; \; j = 1, \ldots, 2K+1,\; \; \; \text{and} \; \; \; i=1, \ldots, Q.
\end{split}
\label{equ:sparse_hyperparameter_priors}
\end{equation}
where $\mathcal{C}^{+}$ denotes a half-Cauchy distribution; $\mathcal{IG}$ denotes an inverse gamma distribution, and where the constants
\begin{equation}
\tau = \frac{ \beta \sigma_{m} }{\left( 1 - \beta \right) \sqrt{N}}, \; \; \;  \gamma = 30 \; \; \;  \text{and} \; \; \; s=1.0.
\label{equ:tau}
\end{equation}
The scale parameter $c$ is set to have an inverse gamma distribution---characterized by a light left tail and a heavy right tail---designed to prevent probability mass from aggregating close to zero \cite{piironen2017sparsity}. This parameter is used when a priori information on the scale of the hyperparameters is not known; it addresses a known limitation in the horseshoe prior where hyperparameters whose values exceed $\tau$ would not be regularized. Through its relationship with $\tilde{\delta}_{i,j}$, it offers a numerical way to avoid shrinking the standard deviation of the Fourier modes that are far from zero. Constants $\gamma$ and $s$ are used to adjust the mean and the variance of the inverse gamma scale parameter $c$, while constant $\beta$ controls the extent of sparsity; large values of $\beta$ imply that more harmonics will participate in the Fourier expansion, while smaller values of $\beta$ would offer a more parsimonious representation. 

There are two additional remarks regarding the hierarchical nature of the priors above. First, we assume that all planes have the same harmonics and that they adhere to the same sparsity structure. Second, if two measurement planes are similar as classified by $\vs$, then they have the same set of Fourier hyperparameters. 

\section{Posterior inference}
We generate approximate samples from the posterior distribution jointly on $\vf^{\ast}$ and the hyperparameters using Hamiltonian Monte Carlo (HMC) \cite{duane1987hybrid, horowitz1991generalized}. In this work, we specifically use the No-U-Turn (NUTS) sampler of Hoffman and Gelman \cite{hoffman2014no}, which is a widely adopted extension of HMC. The main advantage of this approach is that it mitigates the sensitivity of sampler performance on the HMC step size and the number of leapfrog steps.

\subsection{Predictive posterior inference for the area average}
\label{sec:area_average}
The analytical area-weighted average of a spatially varying temperature or pressure function $y \left( \vx \right)$ at an isolated measurement plane indexed by $l \subset \left[1, P\right]$, where $r \in\left[0, 1\right]$ and $\theta\in\left[0,2\pi\right)$, is given by
\begin{equation}
\mu_{\textrm{area}, l}  = \nu_l \int_{0}^{1} \int_{0}^{2\pi} T \left( r, \theta \right) h \left( r \right) d r d \theta
\label{equ:area_avg}
\end{equation}
where $T$ represents the spatially varying temperature or pressure at a given axial measurement plane, and
\begin{equation}
\nu_l = \frac{r_{\textrm{outer}, l} - r_{\textrm{inner}, l}}{ \pi \left(r_{\textrm{outer}, l}^2 - r_{\textrm{inner}, l}^2 \right) } \; \; \; \text{and} \; \; \; h_{l} \left( r \right) = r \left(r_{\textrm{outer},l } - r_{\textrm{inner},l} \right) + r_{\textrm{inner},l},
\label{equ:nu_h}
\end{equation}
where $r_{\textrm{inner},l}$ is the inner radius and $r_{\textrm{outer},l}$ the outer radius for plane $l$. For the joint distribution \eqref{equ:joint_dist} constructed across $P$ axial planes, one can express the area average as
\begin{align}
\left[\begin{array}{c}
\vf \left( \mX \right) \\
 \int f \left( \vz \right) \vh \left( \vz \right) d \vz \cdot \boldsymbol{\nu} \
\end{array}\right]  \sim\mathcal{N}\left( \mathbf{0},\left[\begin{array}{cc}
\mK_{\circ \circ} +\boldsymbol{\Sigma} &  \int  \mK \left( \mX, \vz \right) \odot \vh \left( \vz \right) d \vz \cdot \boldsymbol{\nu} \\[2.5mm]
\int \mK \left( \vz, \mX \right) \odot \vh \left( \vz \right) d \vz \cdot \boldsymbol{\nu} & \left( \boldsymbol{\nu}^{T} \int \int \mK \left( \vz, \vz \right) \odot \vh^2 \left( \vz \right) d \vz d \vz \cdot \boldsymbol{\nu} \right)
\end{array}\right]\right).
\label{equ:area_avg3}
\end{align}
where $\boldsymbol{\nu} = \left( \nu_1, \ldots, \nu_{P} \right)$, $\vh =\left(h_1, \ldots, h_{P} \right)$ and $\vz\in\left\{ \left(r,\theta\right):r\in\left[0,1\right],\theta\in\left[0,2\pi\right), \rho \in \left[1, P\right] \right\} $. Through this construction, we can define the area-average spatial quantity as multivariate Gaussian distribution with mean
\begin{equation}
\boldsymbol{\mu}_{\textrm{area}} \left[ f \right] = \left( \boldsymbol{\nu} \int \mK \left( \vz, \mX \right) \odot \vh \left( \vz \right) d \vz \right) \mG^{-1} \vf,
\end{equation}
where $\boldsymbol{\mu}_{\textrm{area}} \in \mathbb{R}^{P}$. It should be clear that the posterior is obtained by averaging over the various hyperparameters. The covariance is given by
\begin{equation}
\begin{split}
\boldsymbol{\Sigma}^2_{\textrm{area}} \left[ f \right] = \left( \boldsymbol{\nu}^{T} \int \int \mK \left( \vz, \vz \right) \odot \vh^2 \left( \vz \right) d \vz d \vz \cdot \boldsymbol{\nu} \right)- & \left(\int \mK \left( \vz, \mX \right) \odot \vh \left( \vz \right) d \vz \cdot  \boldsymbol{\nu}  \right) \cdot \mG^{-1} \cdot \\
& \left( \int \mK \left( \mX, \vz \right)h \left( \vz \right) d \vz \cdot \boldsymbol{\nu}  \right).
\end{split}
\end{equation}
One point to note here is that although the integral of the harmonic terms is zero, the hyperparameters associated with those terms do not drop out and thus do contribute to the overall variance.

\section{Decomposition of uncertainty}
\label{sec:decomp}
To motivate this section, we consider the following questions:
\begin{enumerate}
\item Can we ascertain whether the addition of instrumentation will alter the area-average of a single measurement plane (and its uncertainty)?
\item How do we determine whether we require more sensors of the present variety, or higher precision sensors at present measurement locations at a given plane?
\item In the case of the former, can we determine where these additional sensors should be placed?
\end{enumerate}
As instrumentation costs in aeroengines is expensive, statistically justified reductions in instrumentation can lead to substantial savings per engine test. Thus, the answers to the questions above are important. At the same time, greater accuracy in both the spatial pattern and its area-average can offer improved aerothermal inference. To aid our mathematical exposition, for the remainder of the methodology section of this paper, we restrict our analysis to a single measurement plane. In other words, $P=1$ and thus the planar kernel does not play a role in the Gaussian random field.
\subsection{Spatial field covariance decomposition}
To offer practical solutions to aid our inquiry, we utilize the \emph{law of total covariance} which breaks down the total covariance into its composite components $\text{cov} \left[ \mathbb{E} \left( \vf^{\ast} | \vf, \mX \right) \right]$ and $\mathbb{E} \left( \text{cov} \left[ \vf^{\ast} | \vf , \mX \right] \right)$. These are given by
\begin{equation}
\text{cov} \left[ \mathbb{E} \left( \vf^{\ast} | \vf, \mX \right) \right] = \mK_{\circ \ast}^{T} \mK_{\circ \circ}^{-1} \boldsymbol{\Psi}_{\vf}  \mK_{\circ \circ} \mK_{\circ \ast}
\label{equ_measurement}
\end{equation}
and
\begin{equation}
\mathbb{E} \left( \text{cov} \left[ \vf^{\ast} | \vf, \mX \right] \right) =  \mK_{\ast \ast }- \mK_{\circ \ast}^{T} \mK_{\circ \circ}^{-1} \mK_{\circ \ast},
\label{equ_sampling}
\end{equation}
where
\begin{equation}
\boldsymbol{\Psi}_{\vf} = \left(\mK_{\circ \circ}^{-1} + \boldsymbol{\Sigma}^{-1} \right)^{-1} \; \; \; \; \text{and} \; \; \; \; \boldsymbol{\mu}_{\vf} = \boldsymbol{\Sigma}^{-1} \boldsymbol{\Psi}_{\vf} \vf,
\end{equation}
where once again we are marginalizing over the hyperparameters. We term the uncertainty in \eqref{equ_measurement} the \emph{impact of measurement imprecision}, i.e., the contribution owing to measurement imprecision. Increasing the precision of each sensor should abate this uncertainty. The remaining component of the covariance is given in \eqref{equ_sampling}, which we define as \emph{spatial sampling uncertainty}, i.e., the contribution owing to limited spatial sensor coverage (see \cite{pianko1983propulsion}). Note that this term does not have any measurement noise associated with it. Adding more sensors, particularly in regions where this uncertainty is high, should diminish the contribution of this uncertainty.

\subsection{Decomposition of area average uncertainty}
Extracting 1D metrics that split the contribution of the total area-average variance into its composite spatial sampling $\sigma^2_{\textrm{area},\textrm{s}}$ and impact of measurement imprecision $\sigma^2_{\textrm{area}, \textrm{m}}$ is a direct corollary of the law of total covariance, i.e.,
\begin{equation}
\begin{split}
\sigma^2_{\textrm{area}, \text{s}} = \left( \nu_{l}^2 \int \int \mK \left( \vz, \vz \right) h_{l}^2 \left( \vz \right) d \vz d \vz \right) - & \left( \nu_{l} \int  \mK \left( \vz , \mX \right)h \left( \vz \right) d \vz \right) \cdot \mK_{\circ \circ}^{-1} \\
& \cdot \left( \nu_{l} \int \mK \left( \mX, \vz \right)h_{l} \left( \vz \right) d \vz \right)
\end{split}
\label{equ:area_avg_sampling}
\end{equation}
and
\begin{equation}
\sigma^2_{\textrm{area}, \textrm{m}}= \left( \nu_{l} \int \mK \left( \vz , \mX \right)h_{l} \left( \vz \right) d \vz \right) \cdot \mK_{\circ \circ}^{-1} \boldsymbol{\Psi}_{\vf} \mK_{\circ \circ}^{-1} \cdot \left( \nu_{l} \int \mK \left( \mX, \vz \right)h_{l} \left( \vz \right) d \vz \right),
\label{equ:area_avg_measurement}
\end{equation}
where $\nu_{l}$ and $h_{l}$ were defined previously in \eqref{equ:nu_h}. We remark here that whole-engine performance analysis tools usually require an estimate of sampling and measurement uncertainty---with the latter often being further decomposed into contributions from static calibration, the data acquisition system and additional factors. Sampling uncertainty has been historically defined by the sample variance (see 8.1.4.4.3 in \cite{saravanmuttoo1990h}). We argue that our metric offers a more principled and practical assessment.

Guidelines on whether engine manufacturers need to (i) add more instrumentation, or (ii) increase the precision of existing measurement infrastructure can then follow, facilitating a much-needed step-change from prior efforts \cite{saravanmuttoo1990h, pianko1983propulsion}.

\section{Isolated plane studies with the simple prior}
\begin{figure}
\begin{center}
\subfigure[]{\includegraphics[scale=0.45]{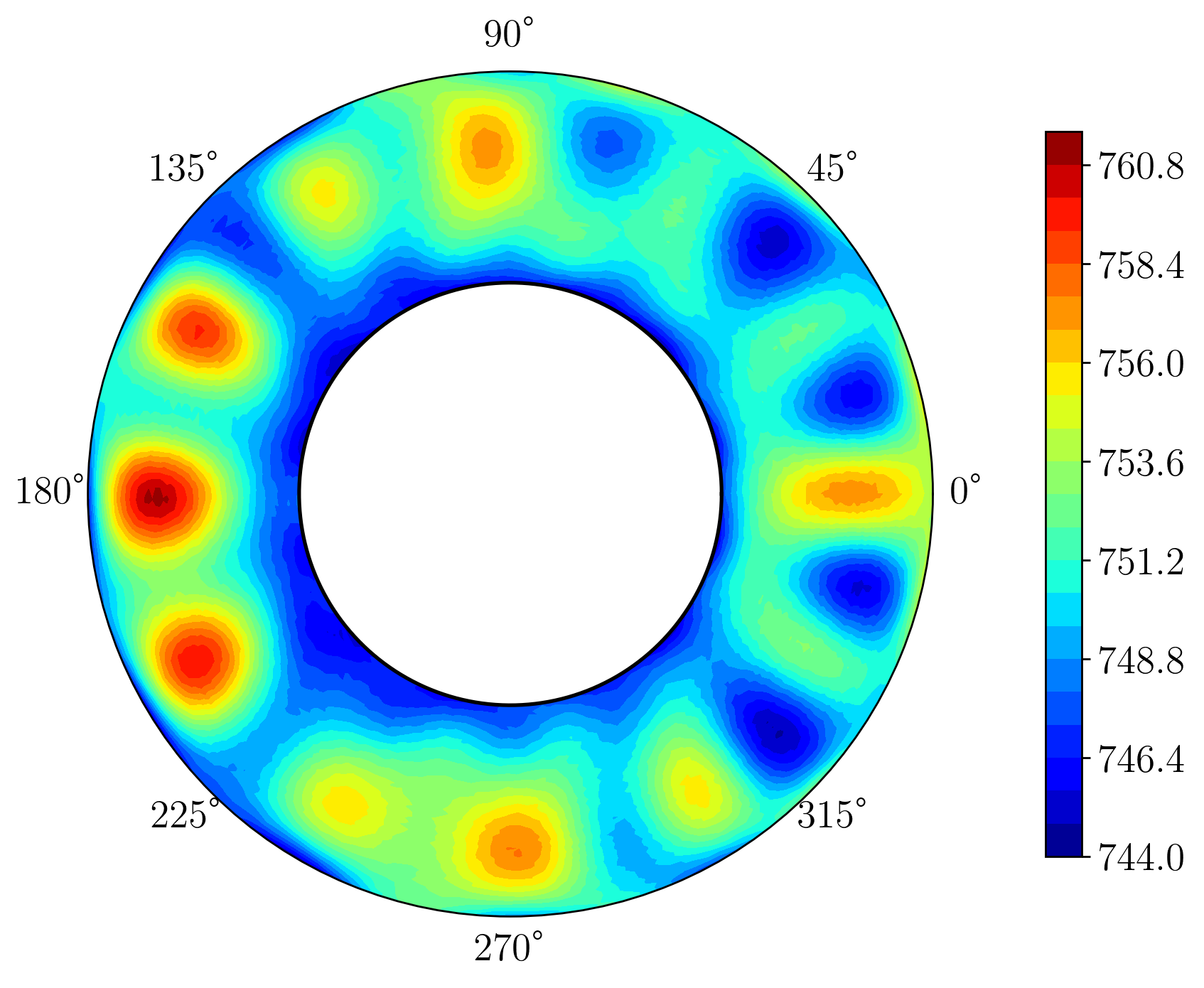}}
\caption{Ground truth spatial distribution of temperature.}
\label{fig:spatial_true}
\end{center}
\end{figure}

To set the stage for an exposition of our formulations and algorithms, we design the spatial temperature distribution shown in Figure~\ref{fig:spatial_true}. This field comprises of five circumferentially varying harmonics $\boldsymbol{\omega} = \left(1, 4, 7, 12, 14\right)$ that have different amplitudes and phases going from the hub to the casing. A small zero-mean Gaussian noise with a standard deviation of 0.1 Kelvin is added to the spatial field. The computed area average mean of the field is 750.94 Kelvin.

\subsection{Spatial field estimation}
Consider a six-rake arrangement given by instrumentation placed as per Table~\ref{table:sampling_locs}, representative of certain planes in an engine. Note that rake arrangements in engines are driven by structural, logistical (access) and flexibility constraints, and thus, it is not uncommon for them to be periodically positioned. As will be demonstrated, the rake arrangements have an impact on the spatial random field and the area average.

\begin{figure}
\begin{center}
\begin{subfigmatrix}{2}
\subfigure[]{\includegraphics[]{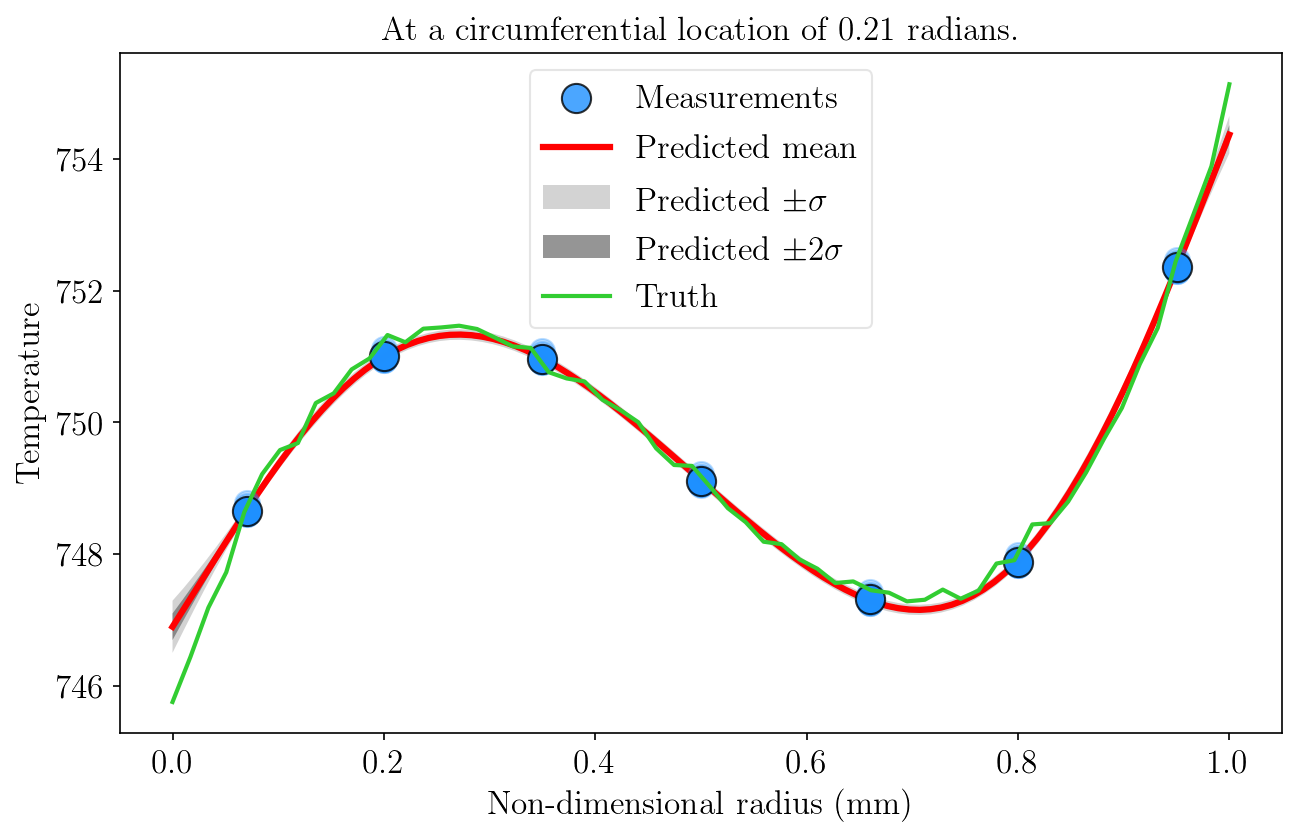}}
\subfigure[]{\includegraphics[]{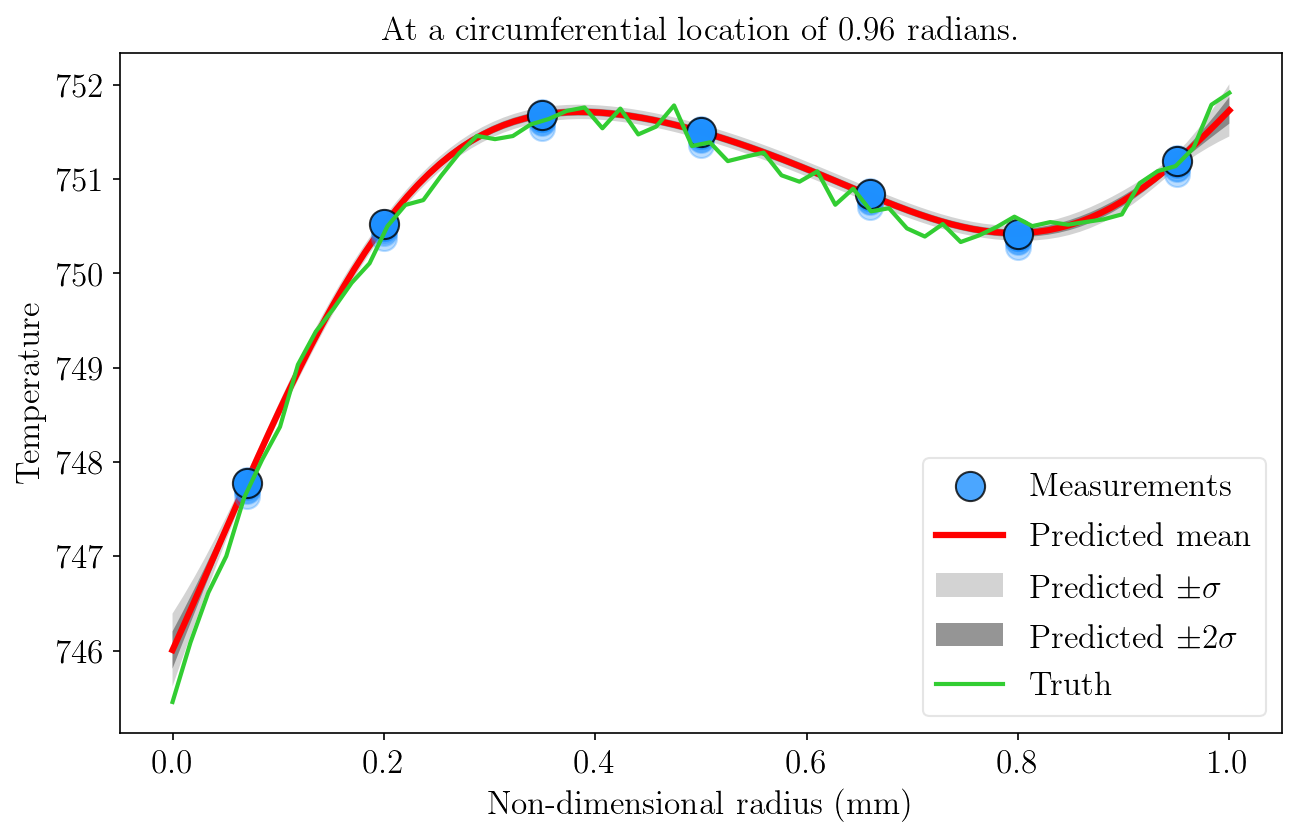}}
\subfigure[]{\includegraphics[]{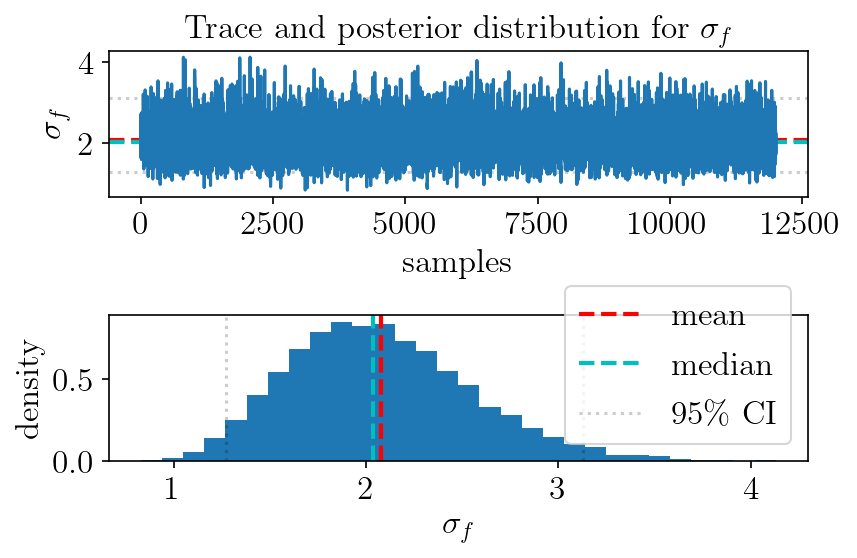}}
\subfigure[]{\includegraphics[]{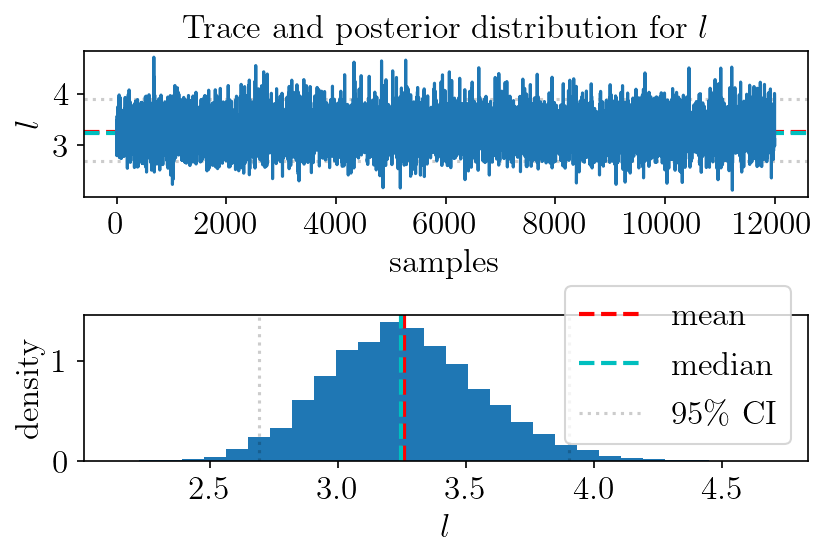}}
\end{subfigmatrix}
\caption{Traceplots for the MCMC chain for some of the  hyperparameters (a) $\lambda_{0}$; (b) $\lambda_1$; (c) $\sigma_f$; (d) $l$.}
\label{fig:example_mcmc}
\end{center}
\end{figure}

\begin{figure}
\begin{center}
\begin{subfigmatrix}{2}
\subfigure[]{\includegraphics[]{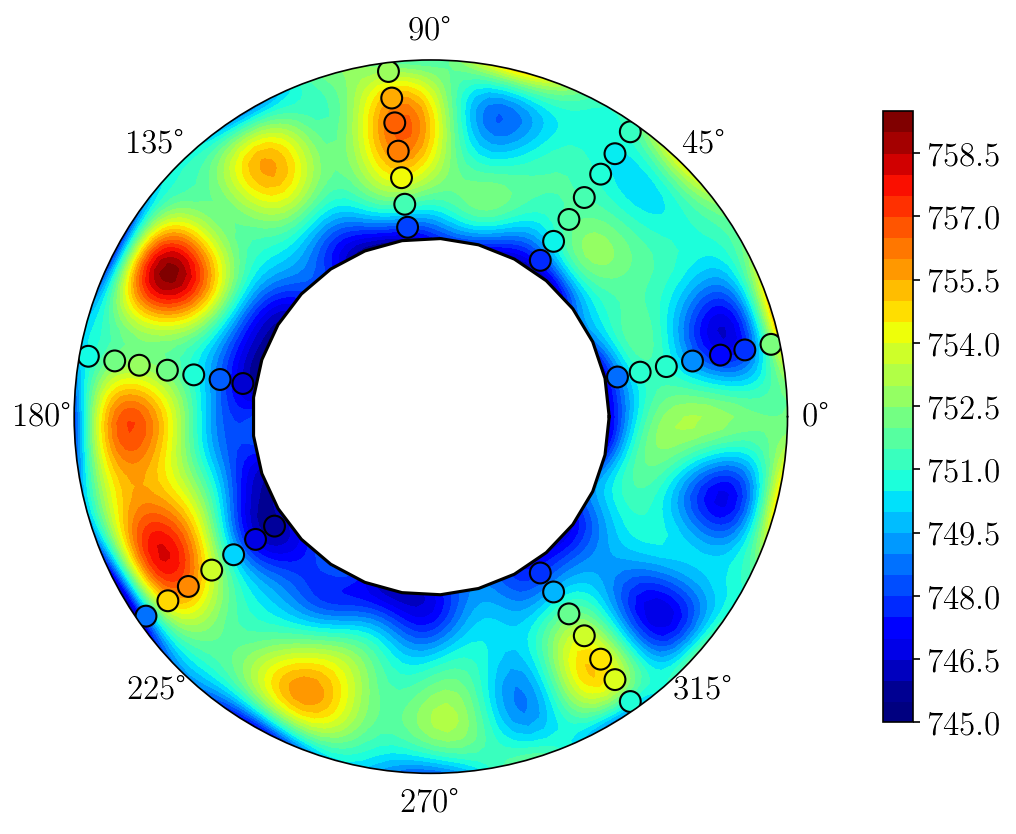}}
\subfigure[]{\includegraphics[]{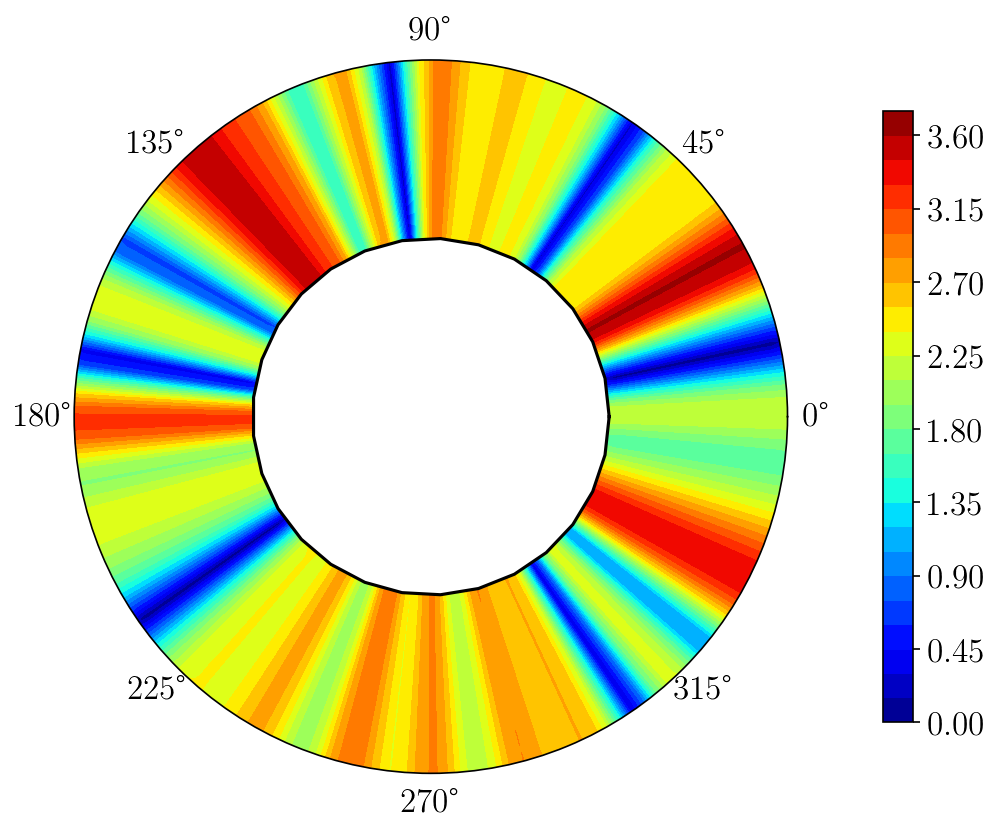}}
\subfigure[]{\includegraphics[]{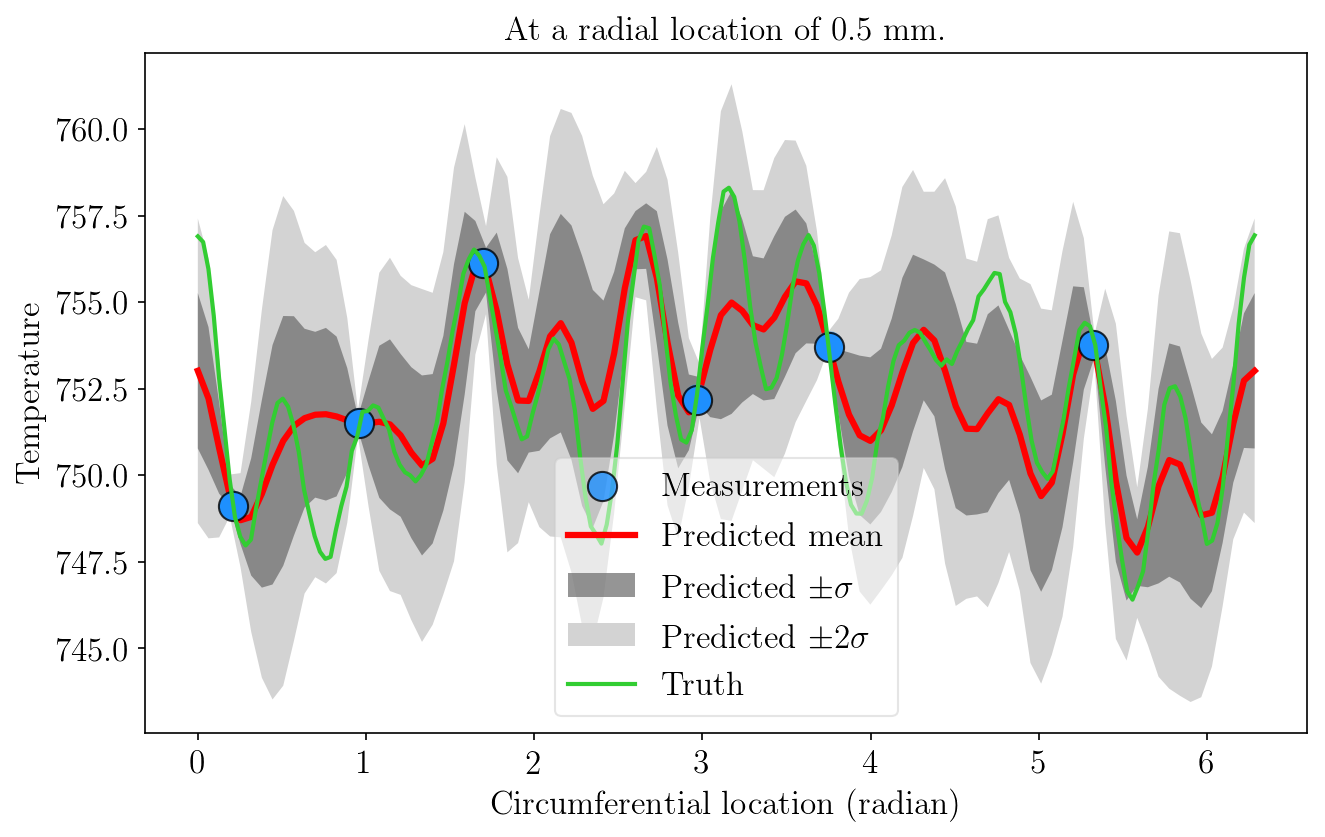}}
\subfigure[]{\includegraphics[]{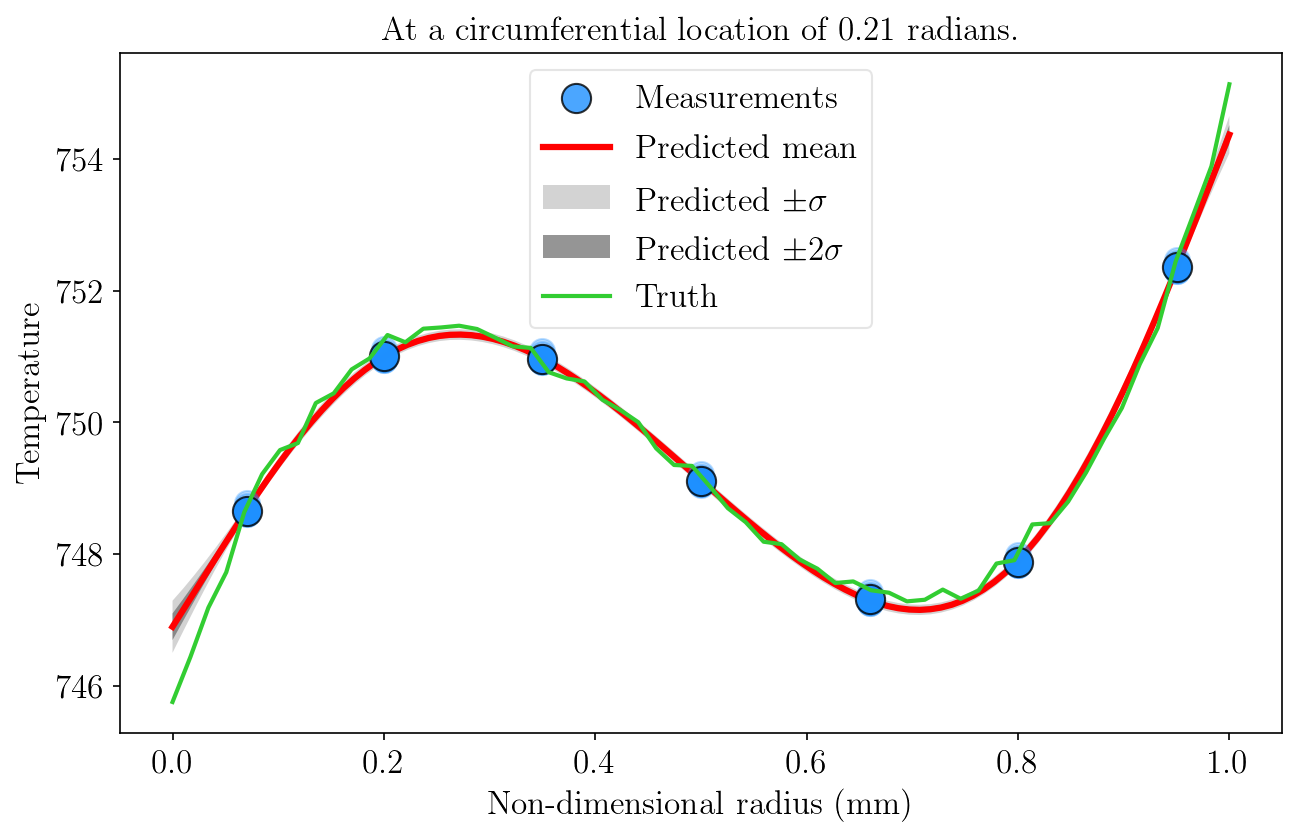}}
\end{subfigmatrix}
\caption{Spatial distributions for (a) the mean and (b) the standard deviation, generated using an ensemble average of the iterates in the MCMC chain, and a circumferential slice at (c) 0.5 mm and a radial slice at (d) 0.21 radians.}
\label{fig:mcmc_resultsA}
\end{center}
\end{figure}

We set our \emph{simple priors} (non-sparsity promoting) as per \eqref{equ:hyperparameter_priors}; harmonics to $\boldsymbol{\omega} = \left(1, 4, 7, 12, 14\right)$, and extract training data from the circumferential and radial locations provided in Table~\ref{table:sampling_locs}.
\begin{table}
\begin{center}
 \begin{tabular}{|l|l|l|}
 \hline
 Property name & Symbol & Value(s) \\ \hline
 Rake arrangement & $\boldsymbol{\theta}$ & $\left( 12^{\circ}, 55^{\circ}, 97^{\circ}, 170^{\circ}, 215^{\circ}, 305^{\circ} \right)$ \\
 Probe locations (non-dimensional) & $\vr$ & $\left( 0.07, 0.2, 0.35, 0.5, 0.66, 0.8, 0.95 \right)$ \\ \hline
 \end{tabular}
\end{center}
\caption{Summary of sampling locations for the default test case.}
\label{table:sampling_locs}
\end{table}
Traceplots for the NUTS sampler for hyperparameters $\lambda_0, \lambda_1, \sigma_f$ and $l$ are shown in Figure~\ref{fig:example_mcmc} for the chosen rake placement. Note that these plots exclude the first few burn-in samples and are the outcome of four parallel chains. The visible stationarity in the these traces, along with their low auto correlation values give us confidence in the convergence of NUTS for this problem.
The Gelman-Rubin statistic for all hyperparameters above was found to be 1.00; the Geweke z-scores were found to be well-within the two standard deviation limit. Figure~\ref{fig:mcmc_resultsA}(a) plots the mean of the resulting spatial distribution (ensemble averaged), while (b) plots its standard deviation. In comparing Figure~\ref{fig:mcmc_resultsA}(a) with Figure~\ref{fig:spatial_true}, we note that in addition to adequately approximating the radial variation (cooler hub and warmer casing), our methods are able to delineate the relatively hotter left half-annulus and its three hot spots at $150^{\circ}, 180^{\circ}$ and $210^{\circ}$. This is especially surprising given the fact that we have 5 spatial harmonics and only 6 and 7 rakes, and not the 11 needed as per the Nyquist bound. A circumferential slice of these plots is shown in Figure~\ref{fig:mcmc_resultsA}(c) at a radial height of 0.5 mm; a radial slice is shown in Figure~\ref{fig:mcmc_resultsA}(d) at a circumferential location of 0.21 radians. Here, we note that the true spatial variation (shown as a green line) lies within the $\pm \sigma$ intervals in the circumferential direction, demonstrating that our approach is able to provide sufficiently accurate uncertainty estimates in this case.

\begin{figure}
\begin{center}
\begin{subfigmatrix}{2}
\subfigure[]{\includegraphics[]{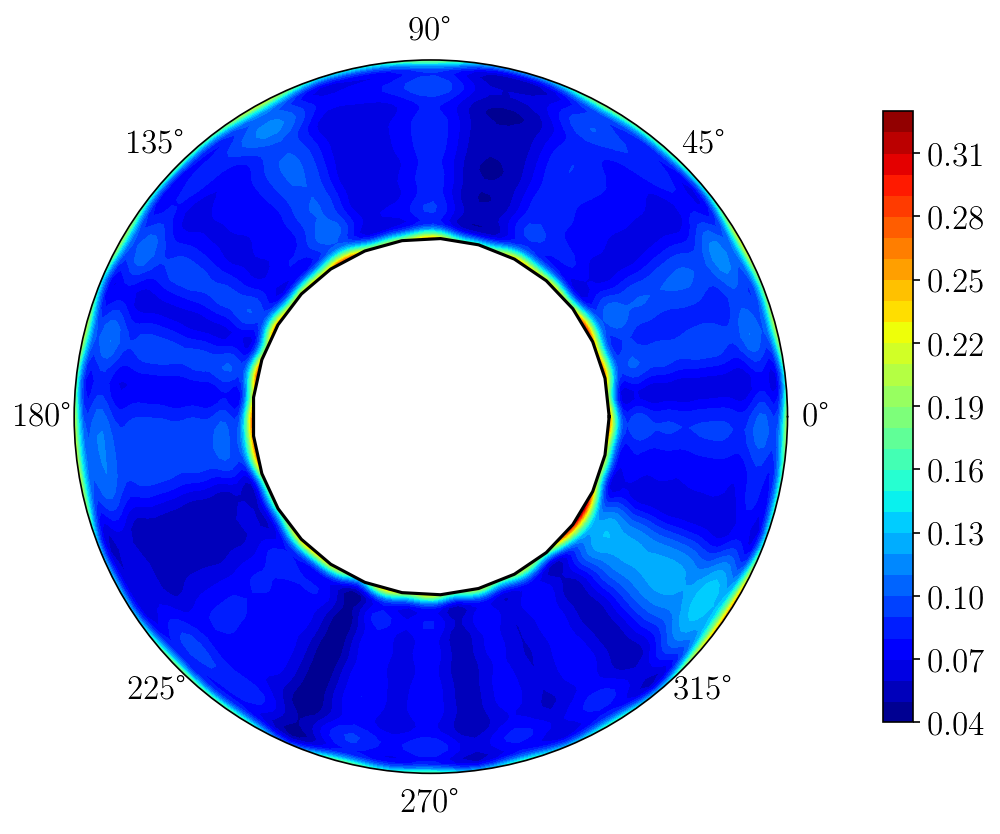}}
\subfigure[]{\includegraphics[]{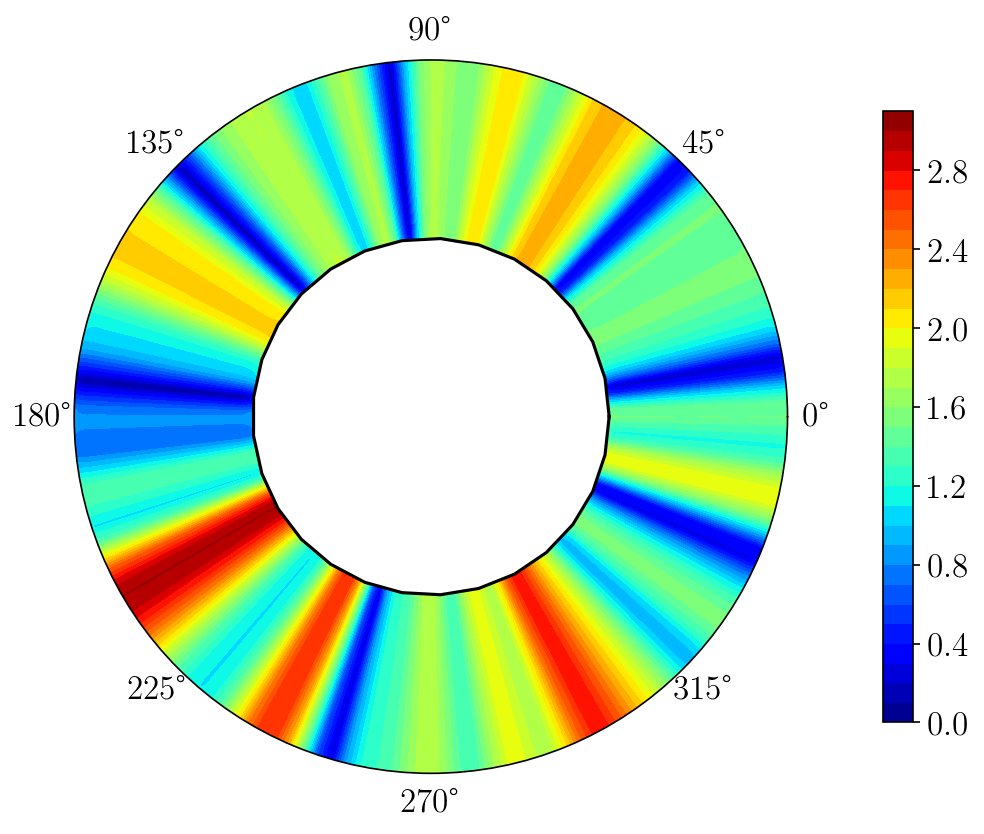}}
\end{subfigmatrix}
\caption{Decomposition of the standard deviations in the temperature: (a) impact of measurement imprecision, and (b) spatial sampling.}
\label{fig:mcmc_results_decomp3}
\end{center}
\end{figure}

For completeness we plot the decomposition of the uncertainty in Figure~\ref{fig:mcmc_results_decomp3}, where the contribution of \emph{impact of measurement imprecision} is, on average, an order of magnitude lower than that of \emph{spatial sampling}. When inspecting these plots one can state that reductions in the overall uncertainty can be obtained by adding additional rakes at $215^{\circ}$ and $300^{\circ}$ (see Figure~\ref{fig:mcmc_results_decomp3}(b)).

\subsection{Spatial field uncertainty variations}
\begin{figure}
\begin{center}
\begin{subfigmatrix}{3}
\subfigure[]{\includegraphics[]{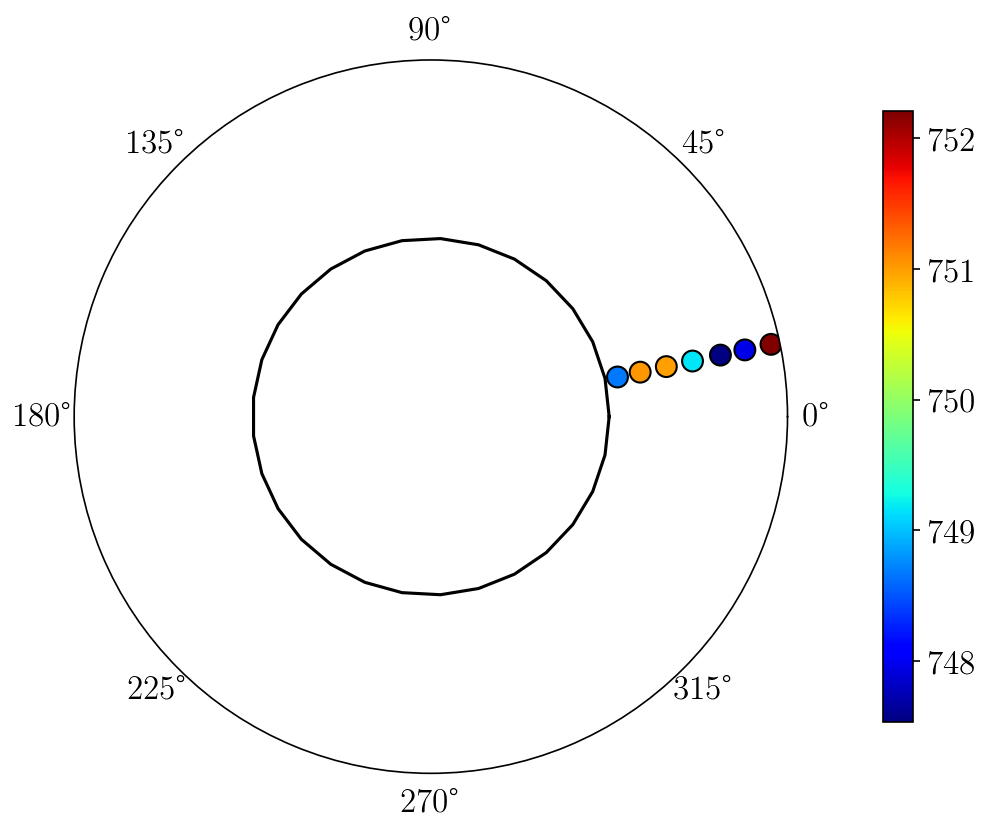}}
\subfigure[]{\includegraphics[]{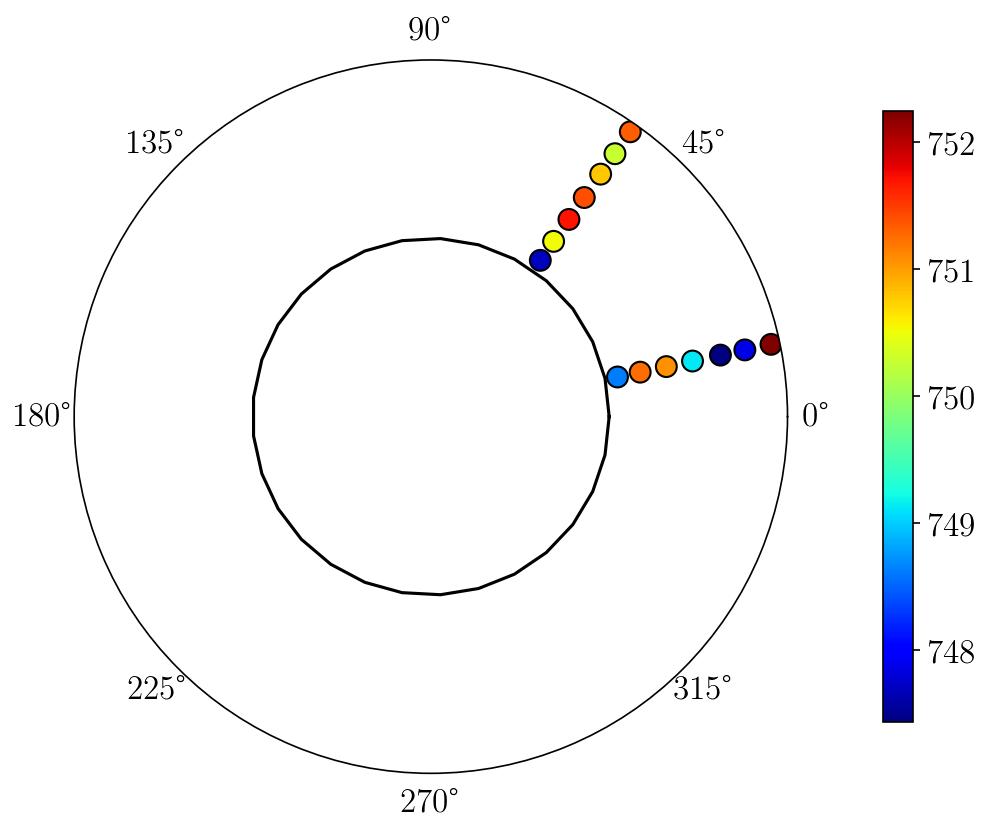}}
\subfigure[]{\includegraphics[]{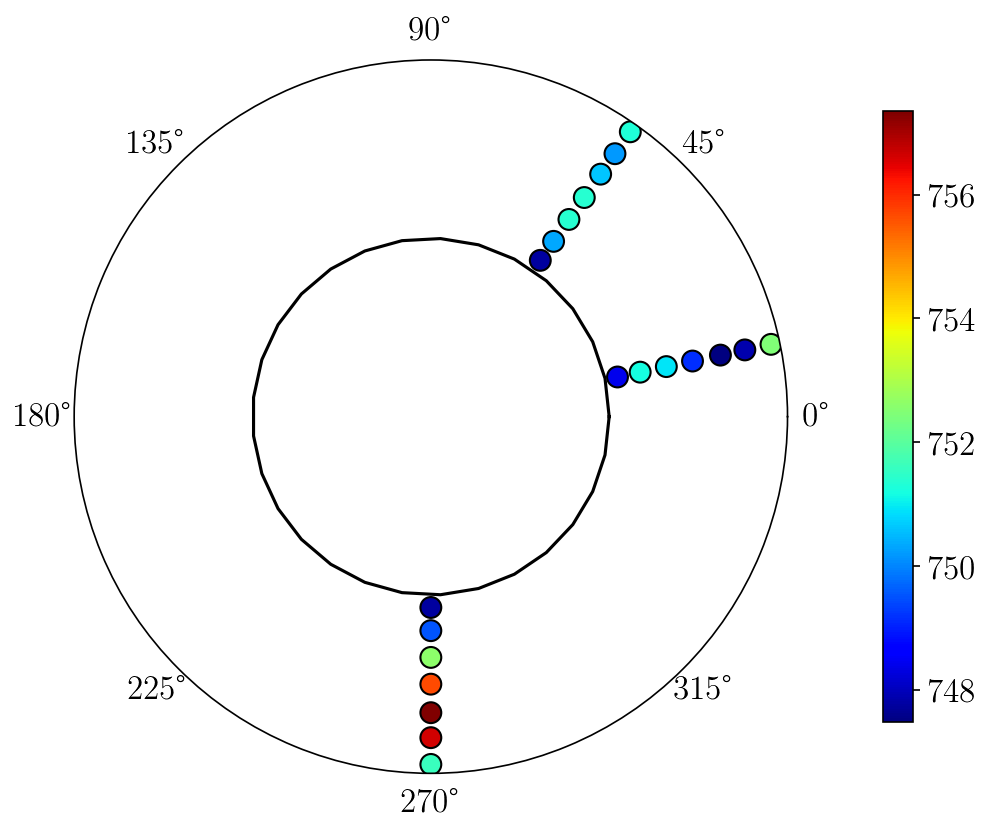}}
\subfigure[]{\includegraphics[]{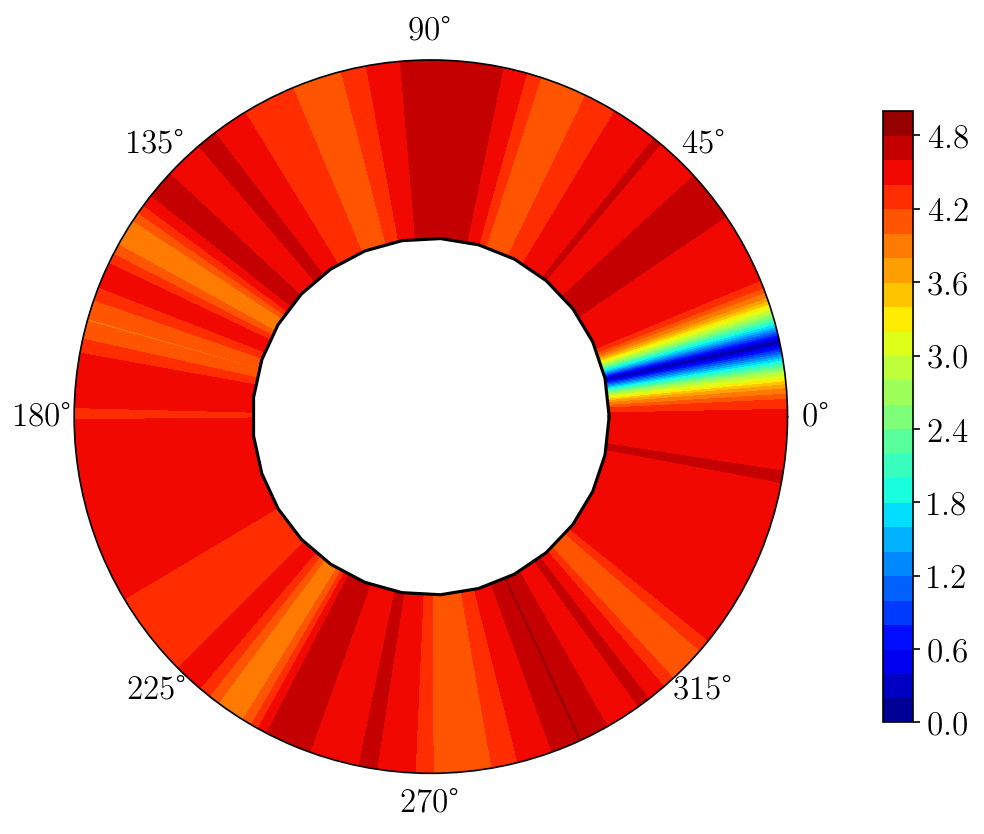}}
\subfigure[]{\includegraphics[]{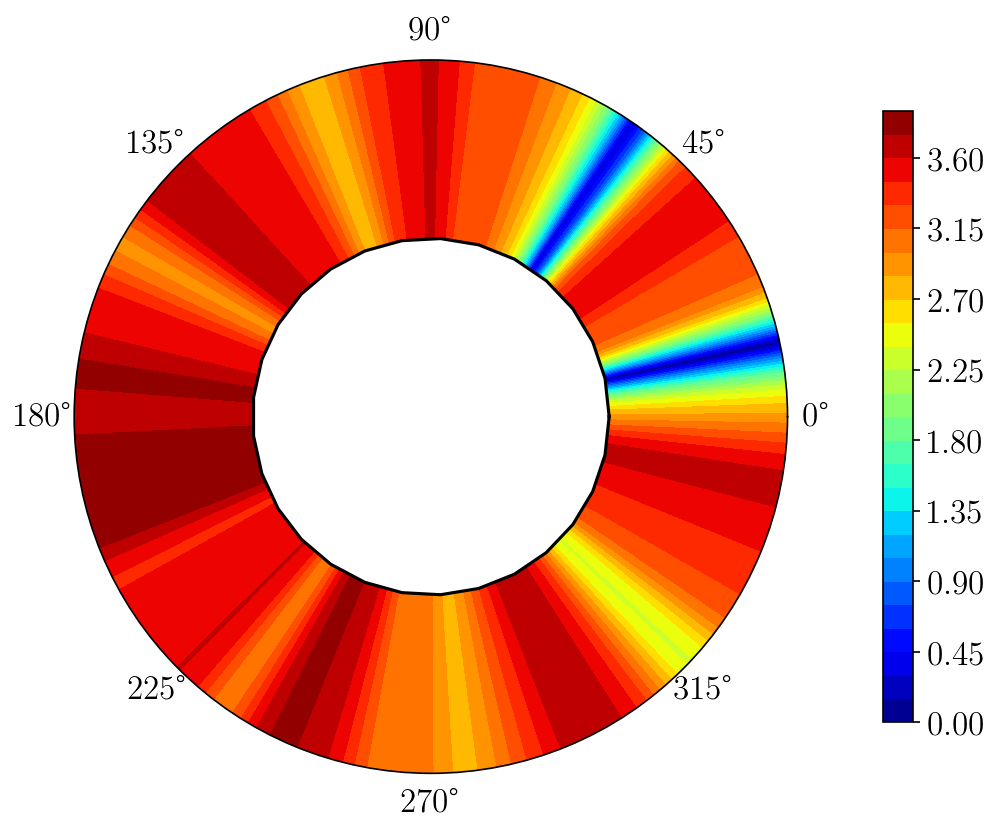}}
\subfigure[]{\includegraphics[]{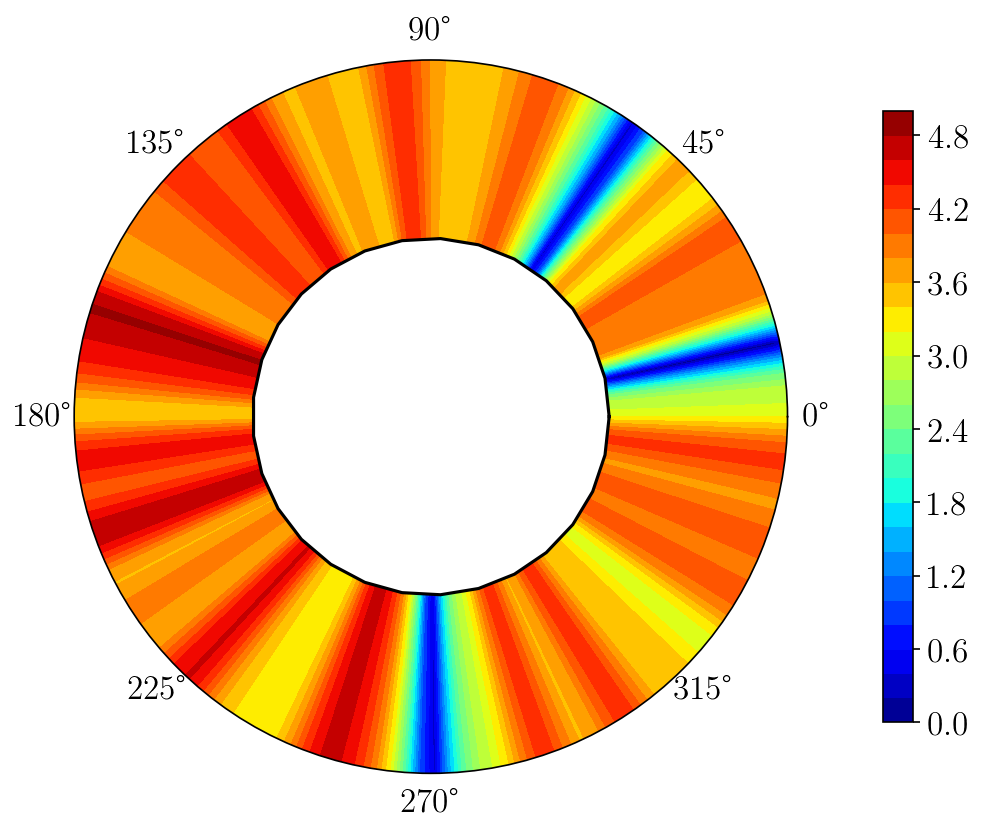}}
\subfigure[]{\includegraphics[]{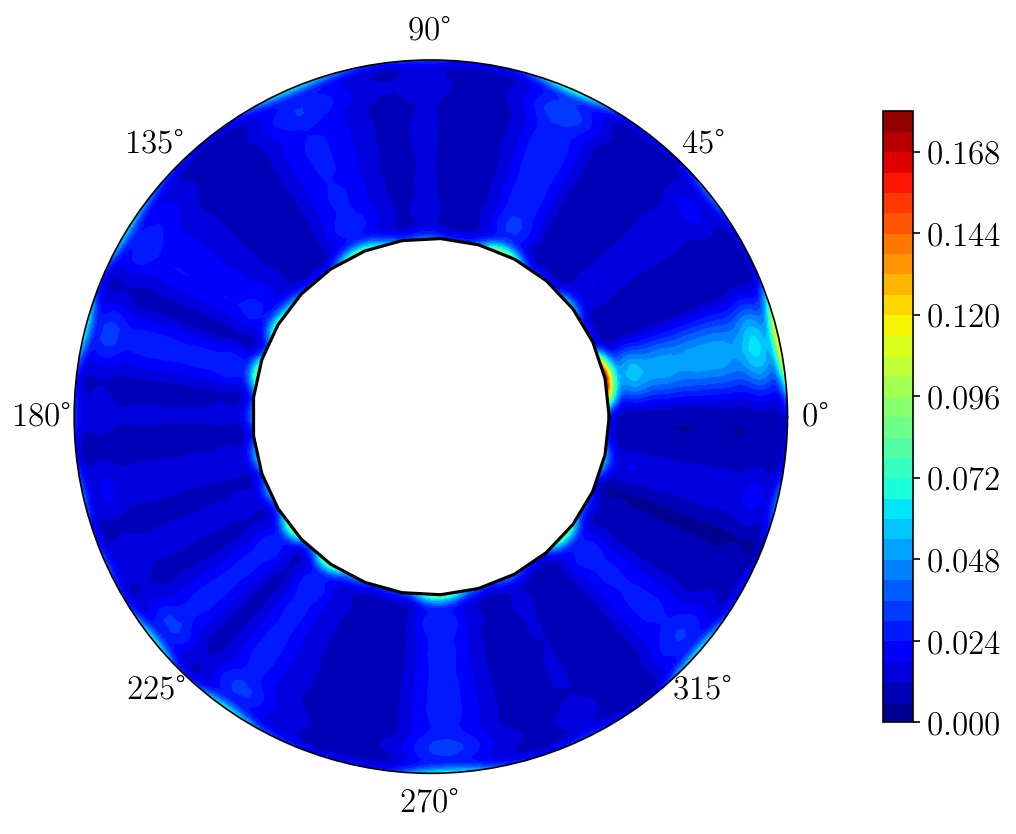}}
\subfigure[]{\includegraphics[]{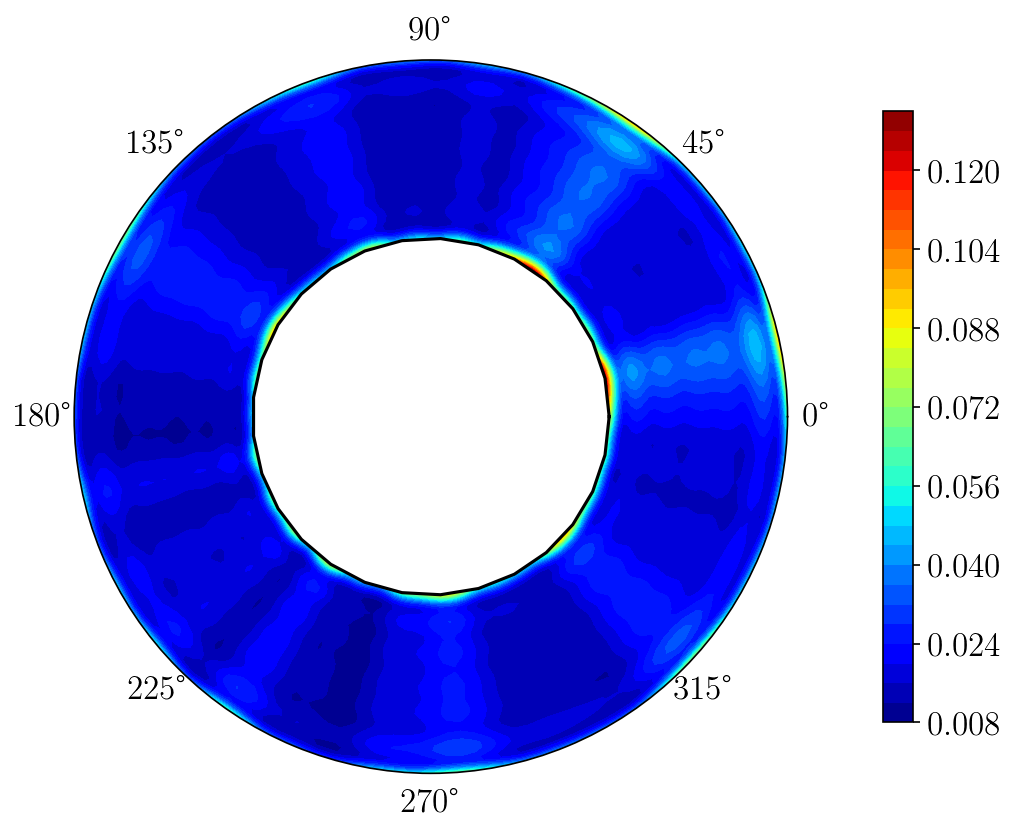}}
\subfigure[]{\includegraphics[]{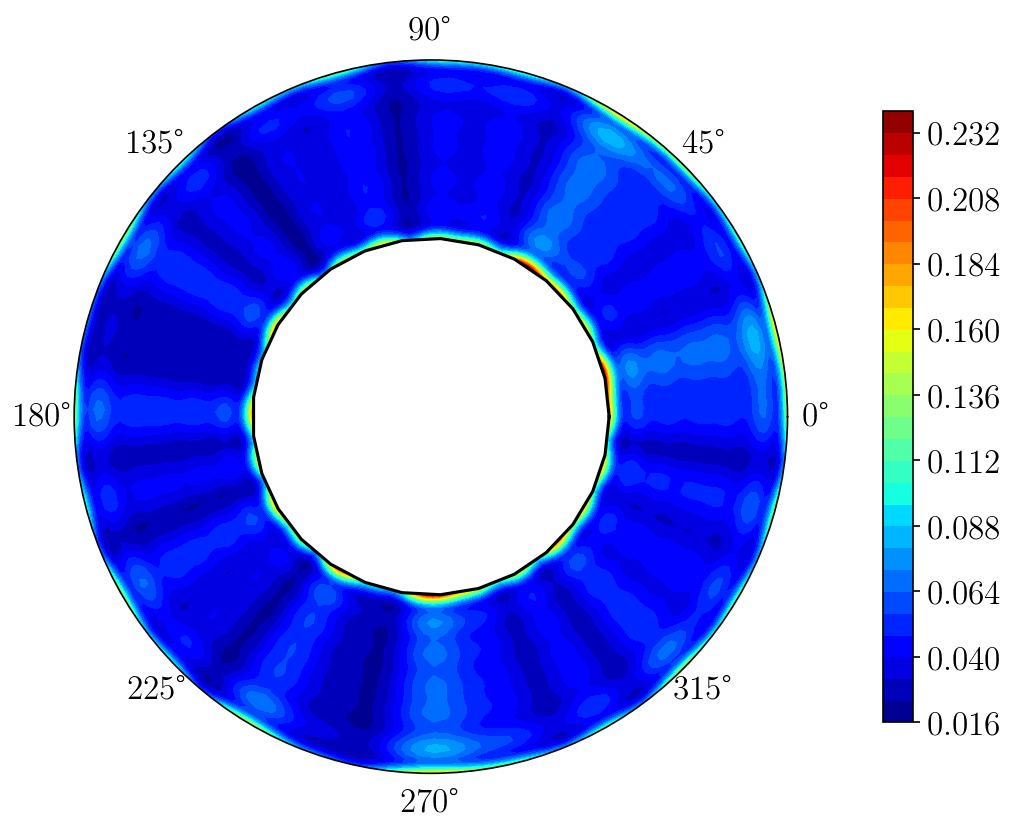}}
\end{subfigmatrix}
\caption{Decomposition of the standard deviations in the temperature for different number of rakes where the top row shows the measurement locations, the middle row illustrates the spatial sampling uncertainty, and the bottom row shows the impact of measurement imprecision. Results are shown for (a,d,g) one rake; (b,e,h) two rakes; (c,f,i) three rakes.}
\label{fig:decomp_study_1}
\end{center}
\end{figure}

\begin{figure}
\begin{center}
\begin{subfigmatrix}{3}
\subfigure[]{\includegraphics[]{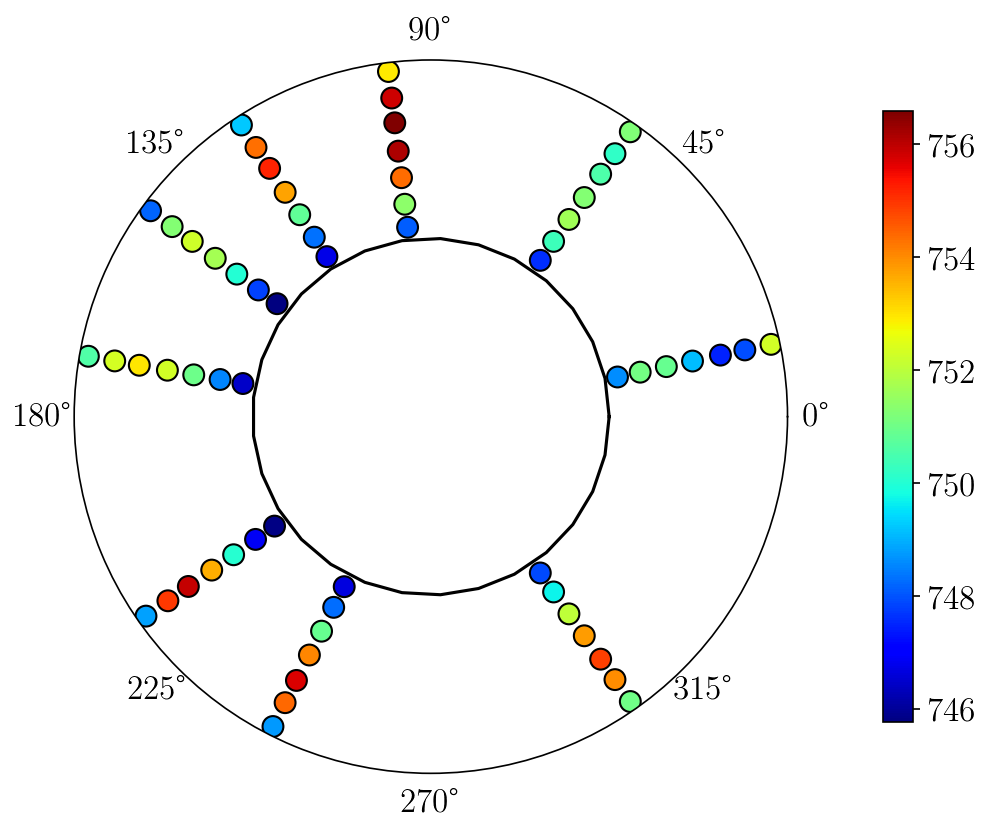}}
\subfigure[]{\includegraphics[]{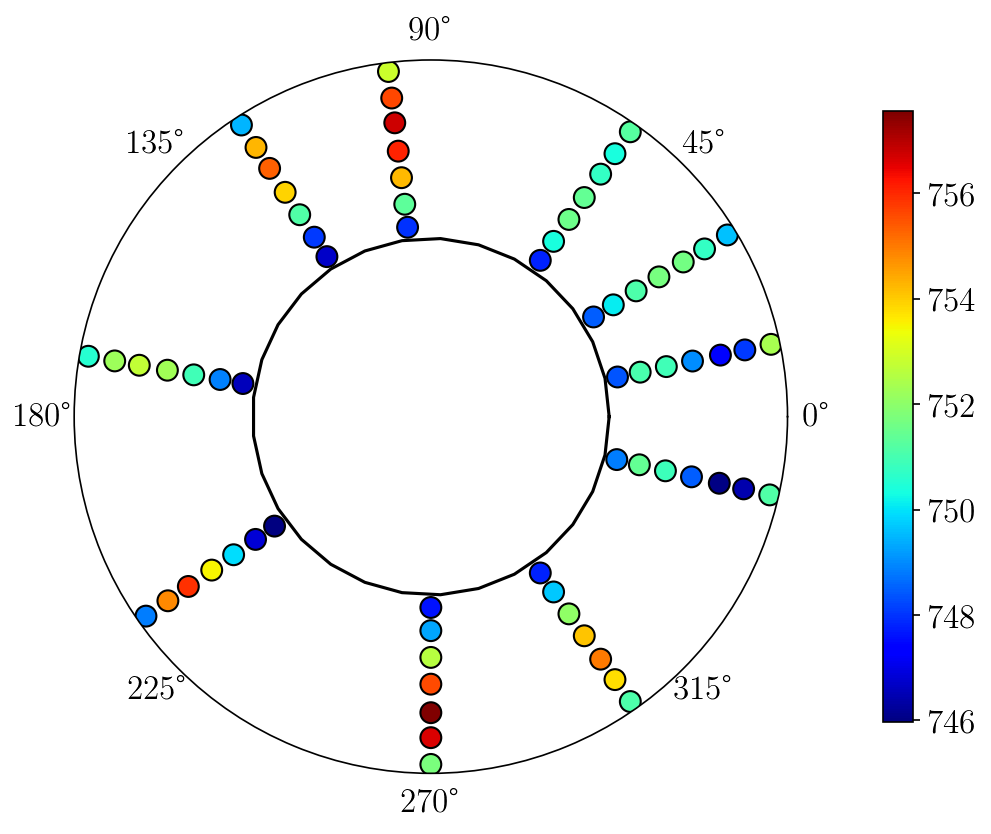}}
\subfigure[]{\includegraphics[]{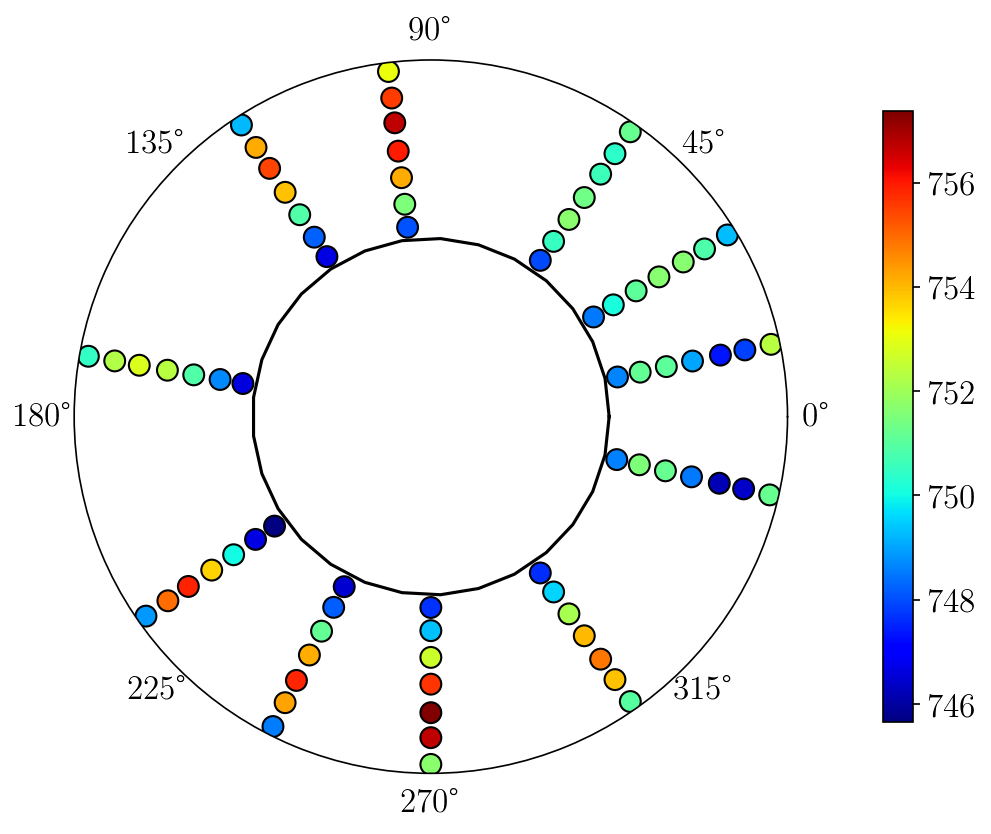}}
\subfigure[]{\includegraphics[]{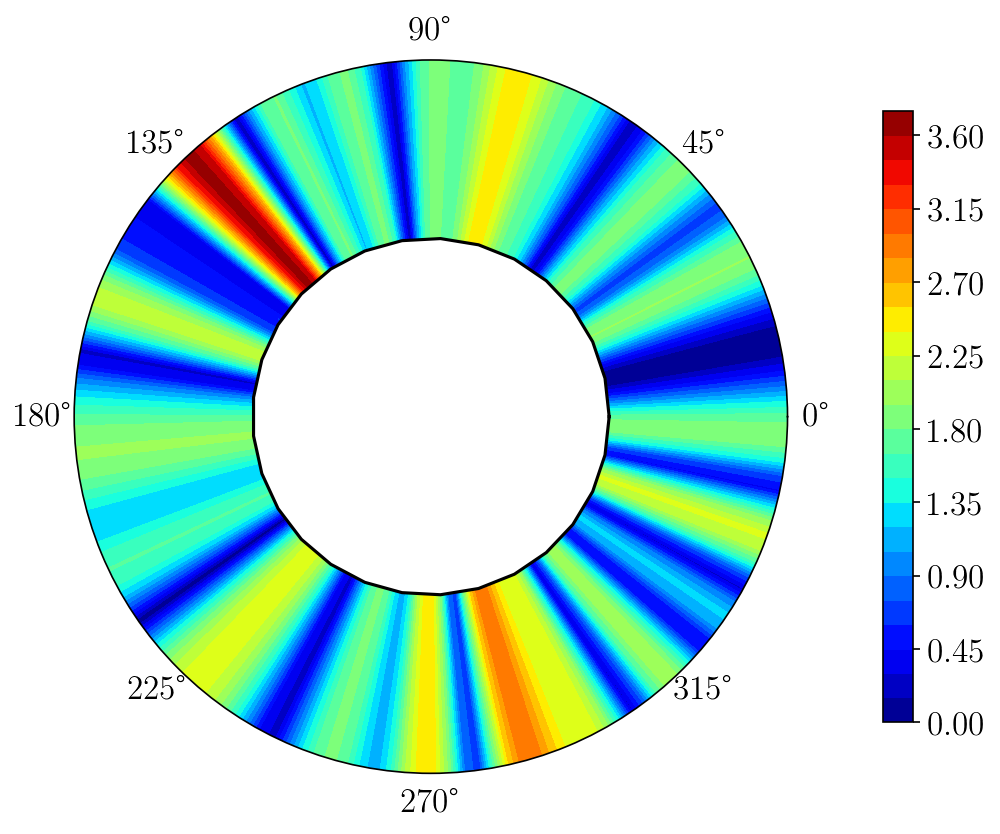}}
\subfigure[]{\includegraphics[]{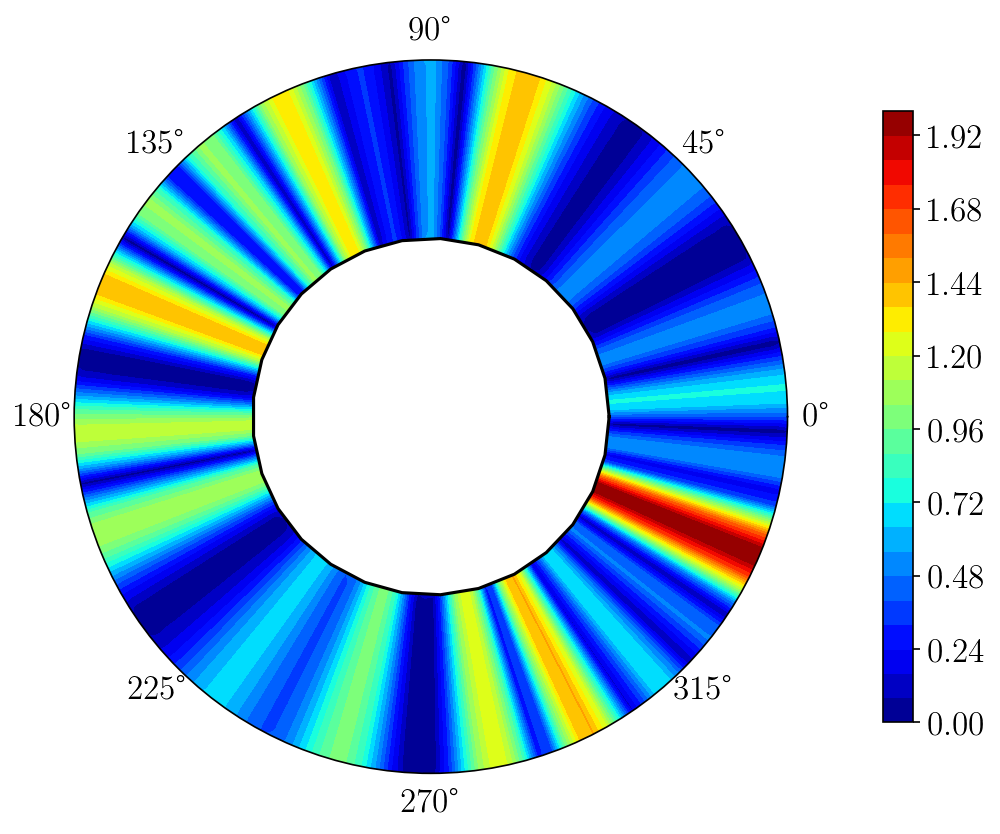}}
\subfigure[]{\includegraphics[]{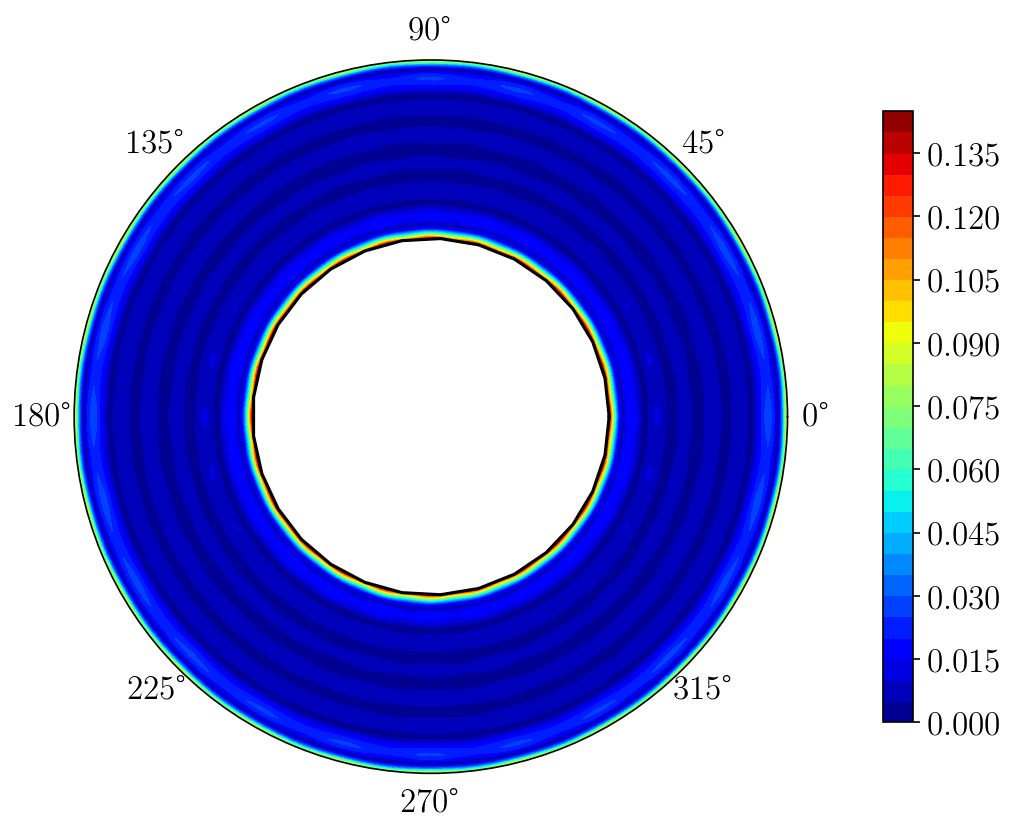}}
\subfigure[]{\includegraphics[]{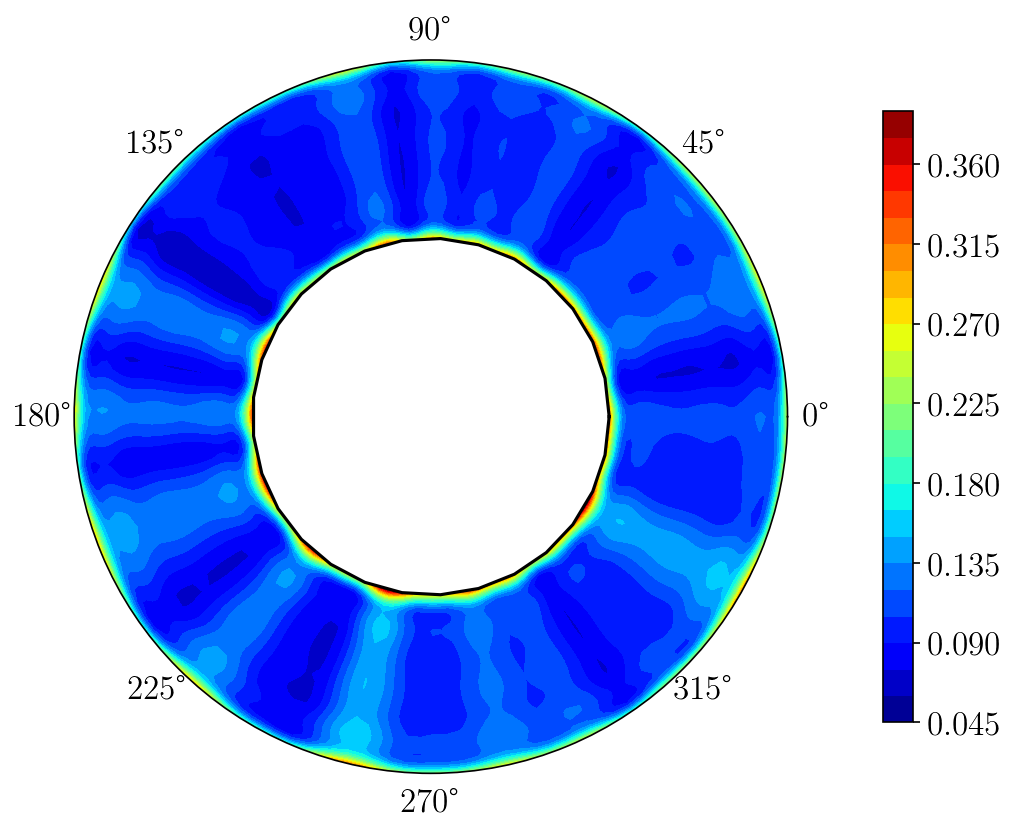}}
\subfigure[]{\includegraphics[]{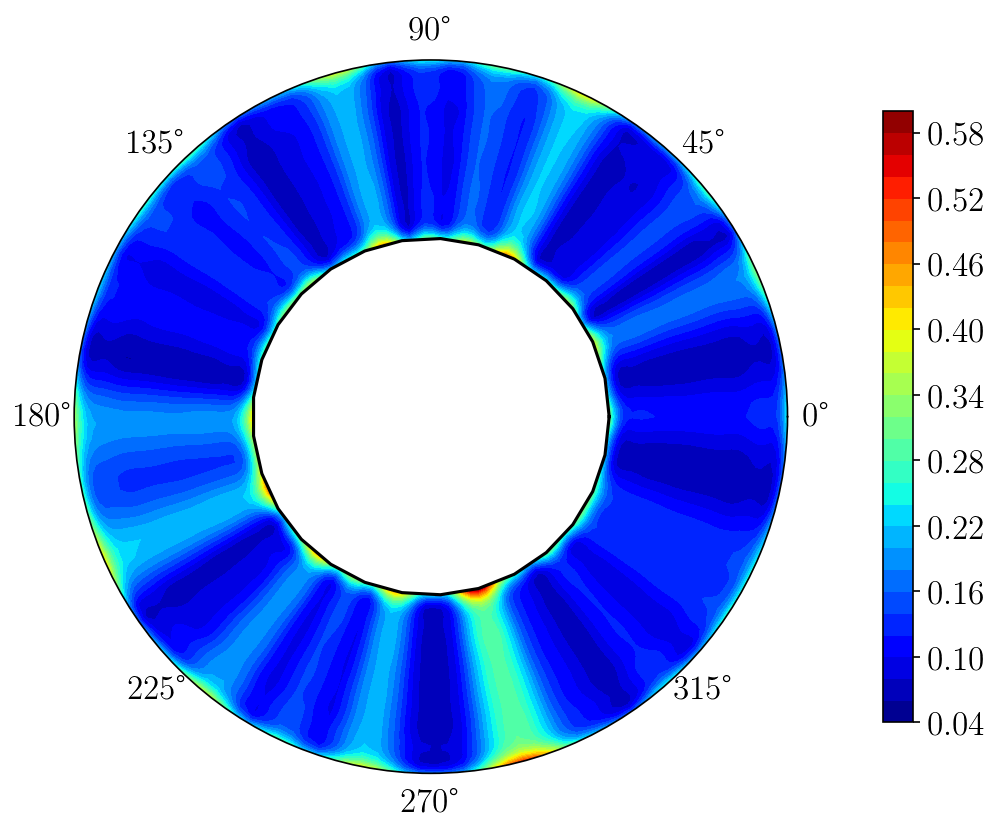}}
\subfigure[]{\includegraphics[]{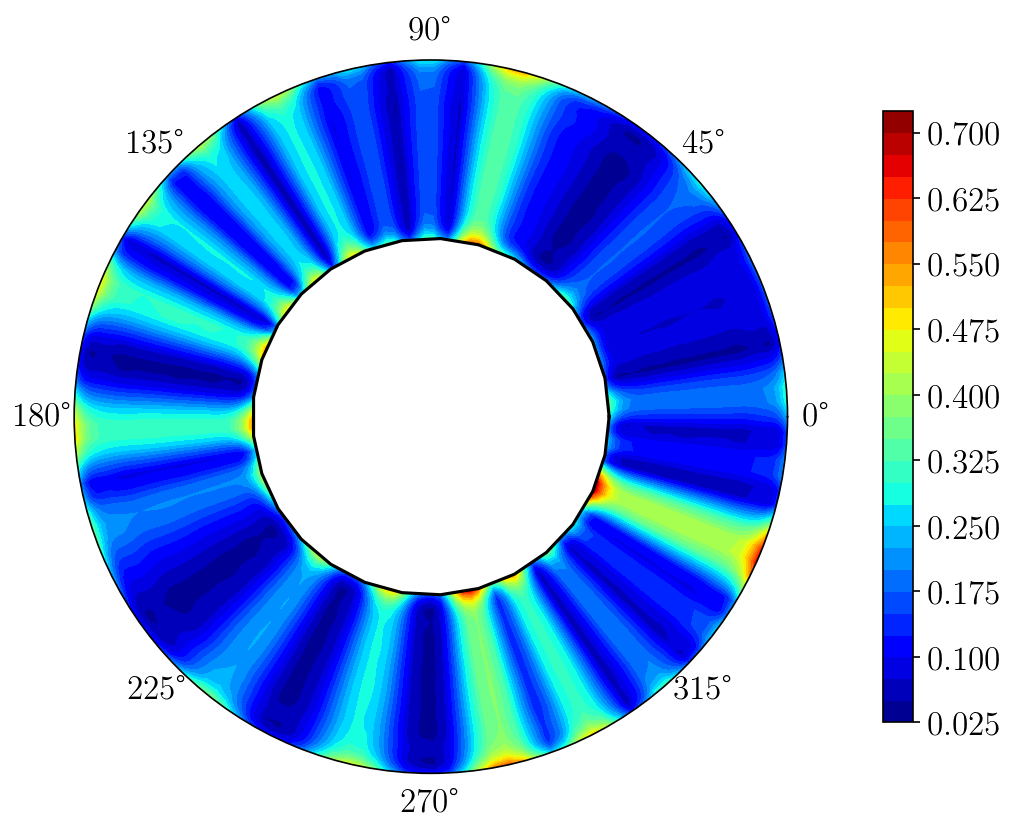}}
\end{subfigmatrix}
\caption{Decomposition of the standard deviations in the temperature for different number of rakes where the top row shows the measurement locations, the middle row illustrates the spatial sampling uncertainty, and the bottom row shows the impact of measurement imprecision. Results are shown for (a,d,g) nine rakes; (b,e,h) ten rakes; (c,f,i) eleven rakes.}
\label{fig:decomp_study_2}
\end{center}
\end{figure}

To assist in our understanding of the spatial uncertainty decompositions above, we carry out a study varying the number of rakes and their spatial locations. Figure~\ref{fig:decomp_study_1} plots the two components of uncertainty for one, two and three rakes, while Figure~\ref{fig:decomp_study_2} plots them for nine, ten and eleven rakes. There are several interesting observations to report.

First, the impact of measurement uncertainty deviates from the location of the sensor with the accumulation of more rakes. For instance, in the case with one rake in Figure~\ref{fig:decomp_study_1}(a), light blue and red contours can be found near each sensor measurement. However, as we add more rakes, there seems to be a phase shift that is introduced to this pattern. This is because the measurement uncertainty will not necessarily lie around the rakes themselves---especially if knowledge about a sensors' measurement can be obtained from other rakes---but rather be in regions that are most sensitive to that particular sensor's value. Furthermore, in the case with an isolated rake, the impact of measurement imprecision locally will be very close to the $\sigma_{m}$ value assigned as the measurement noise. However, with the addition of more instrumentation, the impact of measurement imprecision will increase, as observed in Figure~\ref{fig:decomp_study_1}(g-i), before decreasing again once the spatial pattern is fully known (see Figure~\ref{fig:decomp_study_2}(g-i)).

Second, when the number of rakes is equal to eleven, the spatial sampling uncertainty in the circumferential direction will not vary, and thus the only source of spatial sampling uncertainty will be due to having only seven radial measurements. The former is due to the fact that with five harmonics, we have eleven circumferential unknowns. This is clearly seen in Figure~\ref{fig:decomp_study_2}(f). It should be noted that the position of the rakes can abate the uncertainties observed. This raises a very important point concerning experimental design, and how within a Bayesian framework, sampling uncertainty can be significantly reduced when the rakes are accordingly positioned.

\subsection{Bayesian area average}
Bayesian area average estimates are obtained by integrating the spatial approximation (as per \eqref{equ:area_avg3}) at each iteration of the previously presented MCMC chain, and averaging over sample realisations. The deficiency of the sector area-average compared to the Bayesian area average is apparent when one studies its convergence.

\begin{figure}
\begin{center}
\begin{subfigmatrix}{2}
\subfigure[]{\includegraphics[]{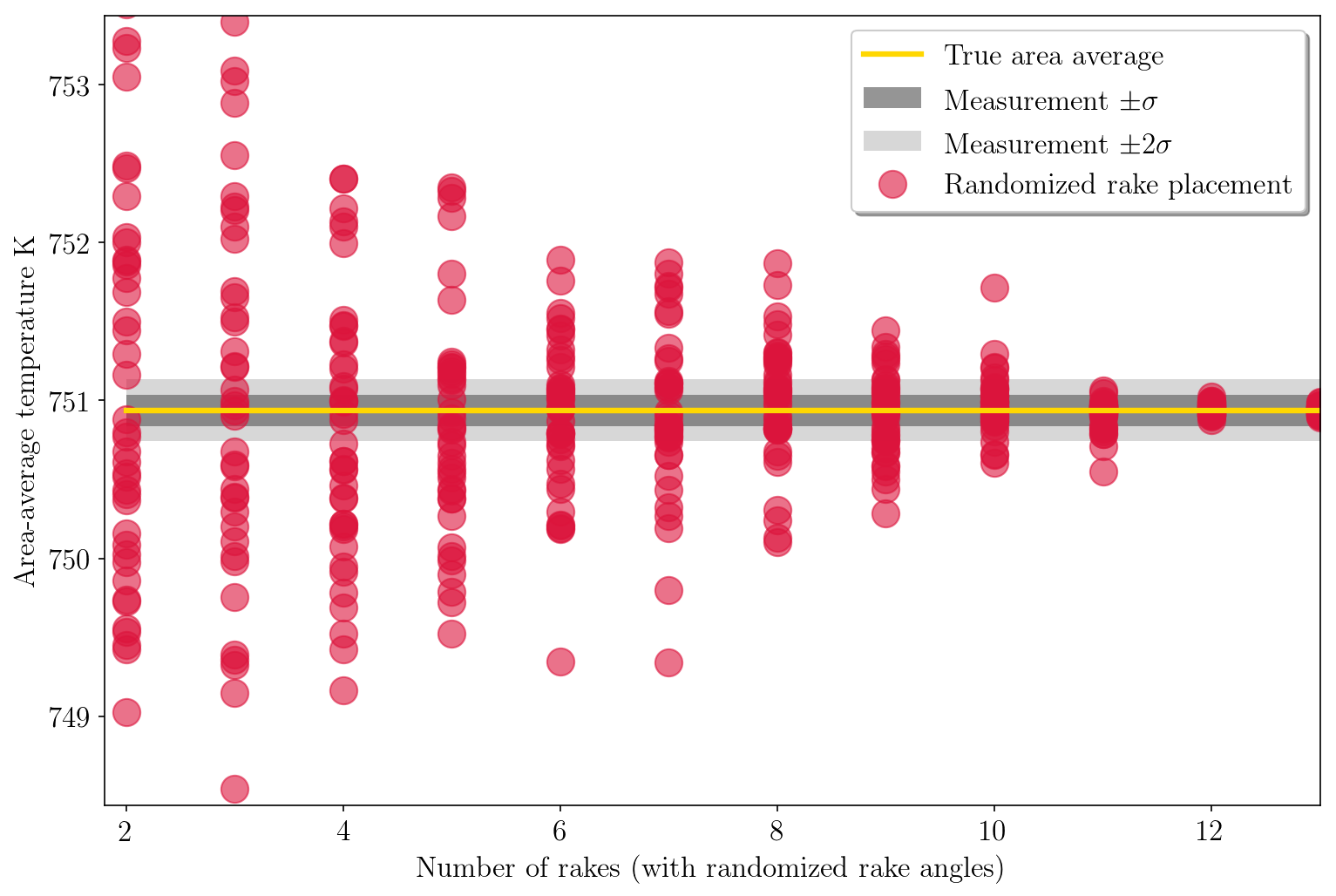}}
\subfigure[]{\includegraphics[]{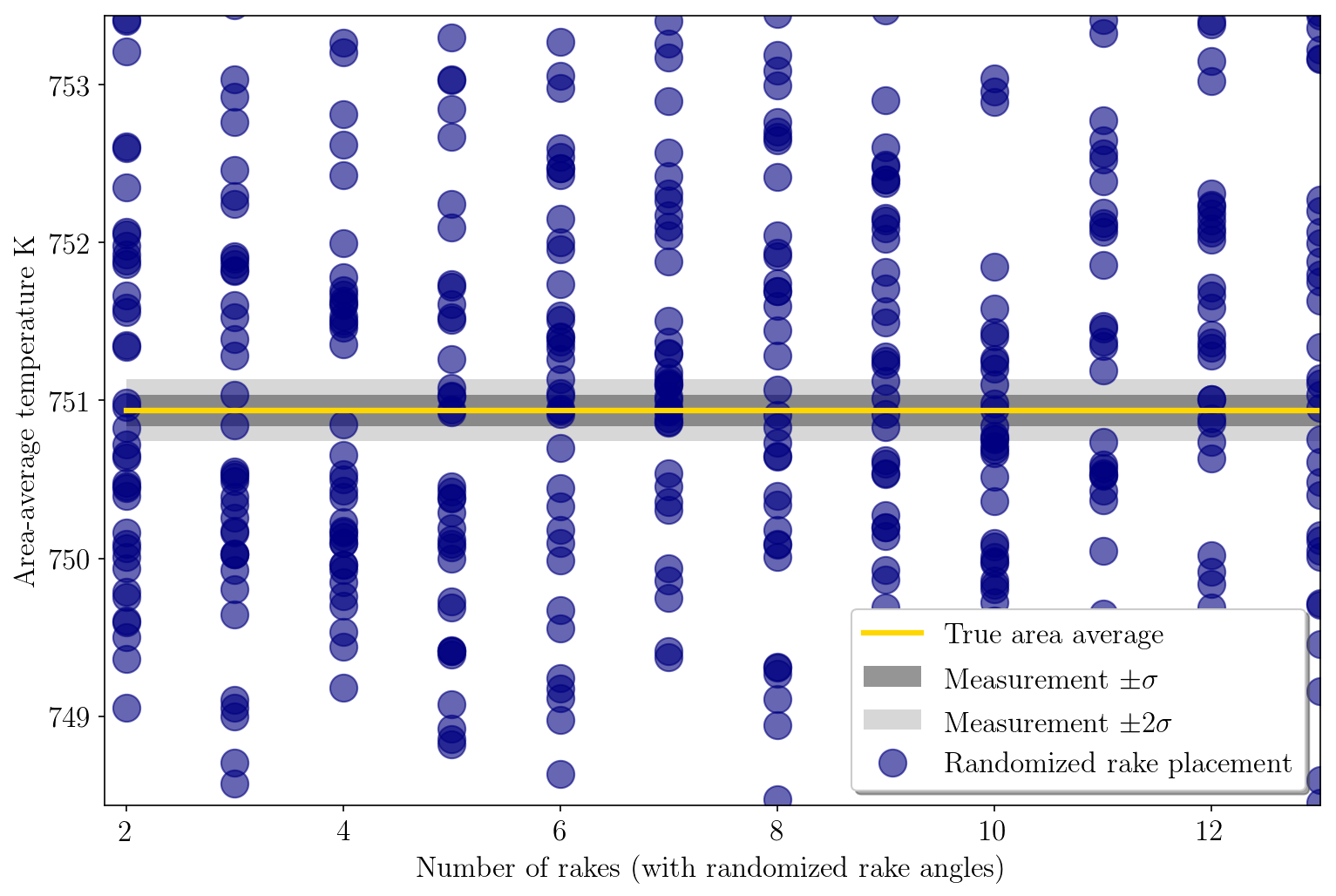}}
\end{subfigmatrix}
\caption{Convergence of (a) the sector weighted area-average and (b) the Bayesian area-average (only mean reported) for forty randomised arrangements of rake positions.}
\label{fig:area_avg_comparison}
\end{center}
\end{figure}

To do this, we sample our true spatial distribution at forty different randomised circumferential locations for different numbers of rakes, while maintaining the number of radial probes and their locations. The circumferential locations are varied by randomly selecting rake positions between $0^{\circ}$ and $355^{\circ}$ inclusive, in increments of $5^{\circ}$. Figure~\ref{fig:area_avg_comparison}(a) plots the resulting sector area-average. The yellow line represents the true area-average and the shaded grey intervals around it reflect the measurement noise. It is clear that the addition of rakes does not necessarily result in any convergence of the area-average temperature. Furthermore, the reported area-average is extremely sensitive to the placement of the rakes; in some cases a $\pm 2$ Kelvin variation is observed. In Figure~\ref{fig:area_avg_comparison}(b) we plot the reported mean for each randomised trial using our Bayesian framework. Not only is the scatter less, but, in fact, after 10 rakes we see that reported area-averages lie not to far from the measurement noise. This makes a compelling case for replacing the practice for computing area-averages in turbomachinery via sector weights with a more rigorous Bayesian treatment.

\begin{figure}
\begin{center}
\includegraphics[scale=0.35]{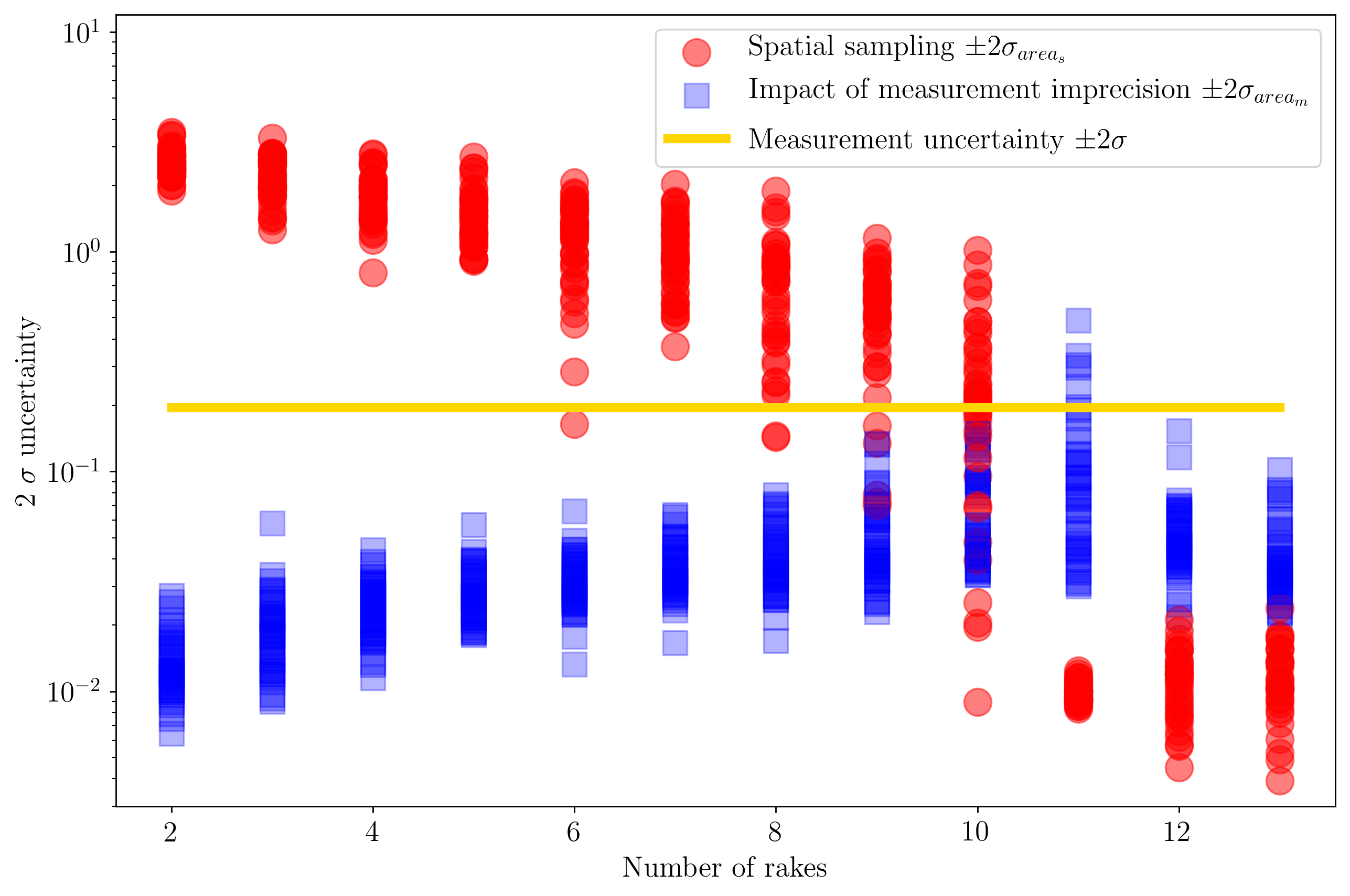}
\caption{Decomposition of area-average spatial sampling and impact of measurement imprecision area-average values for 40 randomised arrangements of rake positions.}
\label{fig:problem_2}
\end{center}
\end{figure}

We study the decomposition of the area-average variance in these randomised experiments and plot their \emph{spatial sampling} and \emph{impact of measurement imprecision} components (see \eqref{equ:area_avg_sampling} and \eqref{equ:area_avg_measurement}) in Figure~\ref{fig:problem_2}. As before, the measurement noise is demarcated as a solid yellow line. There are interesting observations to make regarding these results.

First, the impact of measurement uncertainty increases with more instrumentation, till the model is able to adequately capture all the Fourier harmonics (after eleven rakes); we made an analogous finding when studying the spatial decomposition plots. This intuitively makes sense, as the more instrumentation we add, the greater the impact of measurement uncertainty. It is also worth noting that numerous rake arrangements can be found that curtail this source of uncertainty, many far below the threshold associated with the measurement noise.

Second, across the forty rake configurations tested, spatial sampling uncertainty contributions were found to be very similar when using only two to three rakes. The variability in spatial sampling uncertainty decreases significantly when the number of rakes is sufficient to capture the circumferential harmonics. Thereafter, it is relatively constant, as observed by the collapsing of the red circles in Figure~\ref{fig:problem_2}.

\section{Transfer learning results with the sparsity promoting prior}
In this results section, we present the results of our transfer learning framework with sparsity promoting priors.

\subsection{Transfer learning by splitting instrumentation}
\begin{figure}
\begin{center}
\begin{subfigmatrix}{2}
\subfigure[]{\includegraphics[width=0.37\textwidth]{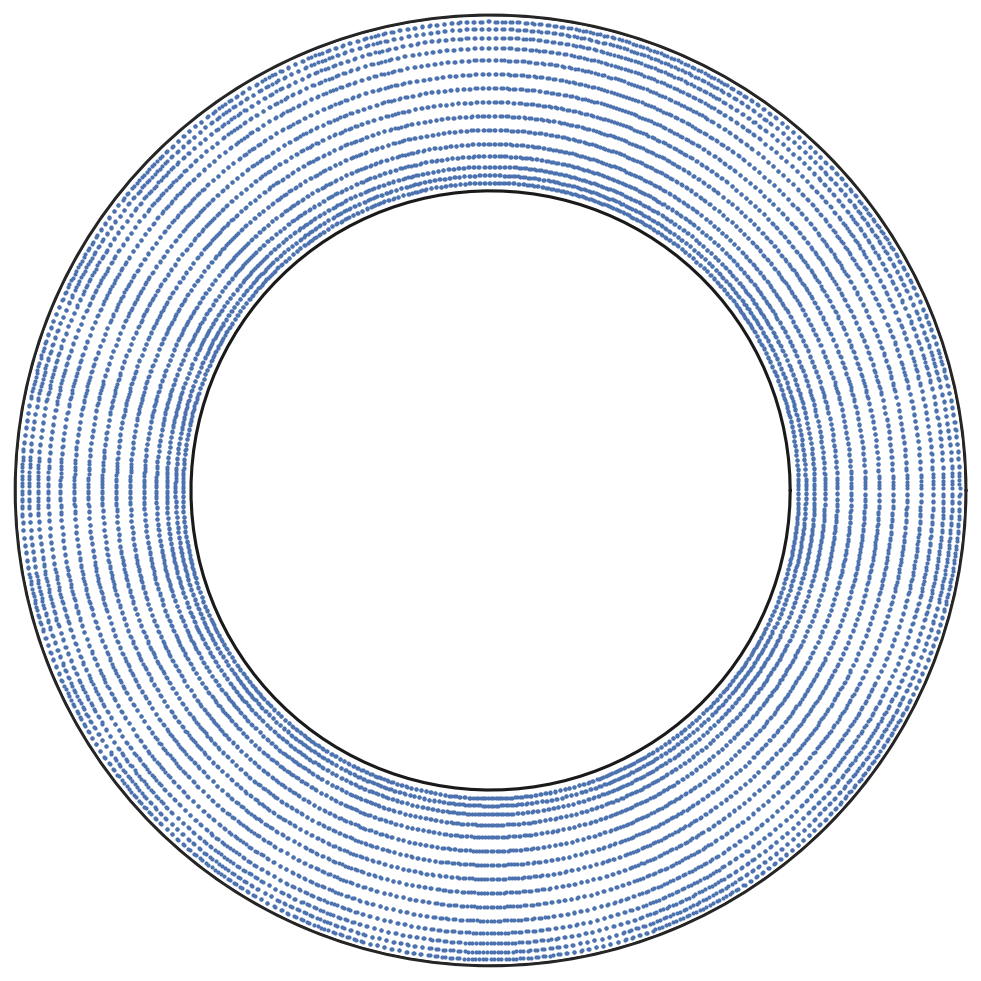}}
\subfigure[]{\includegraphics[]{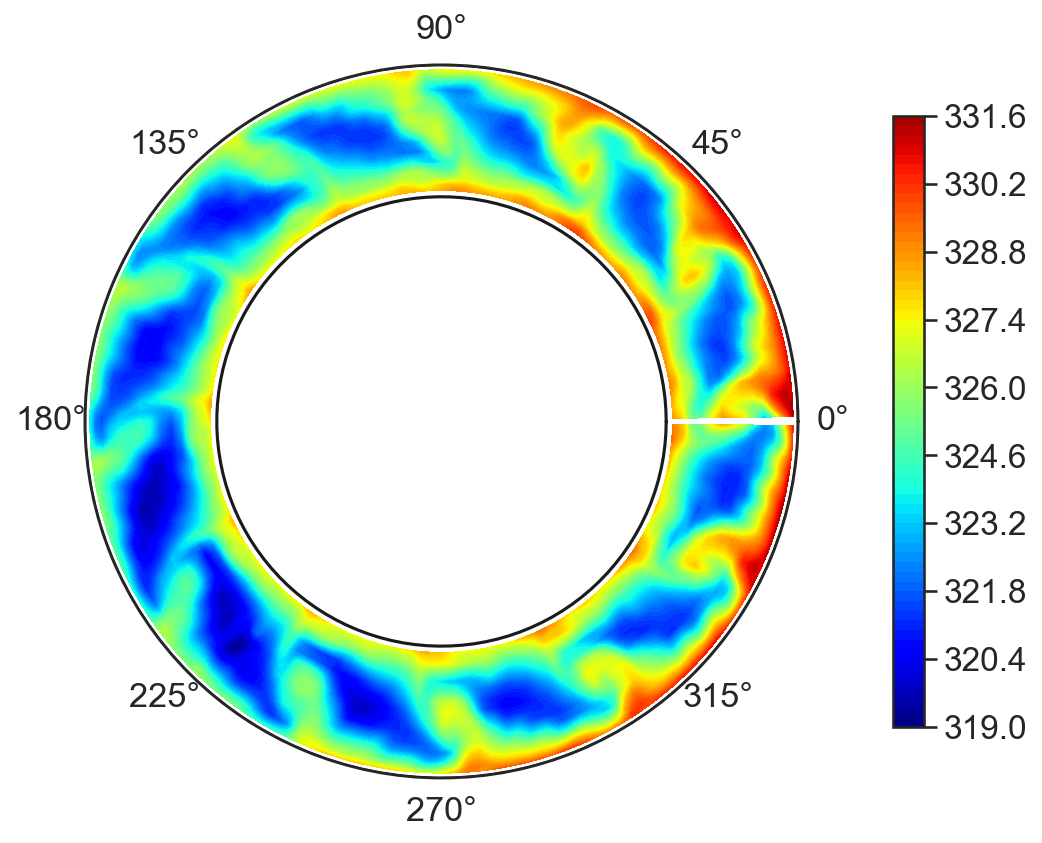}}
\end{subfigmatrix}
\begin{subfigmatrix}{2}
\subfigure[]{\includegraphics[]{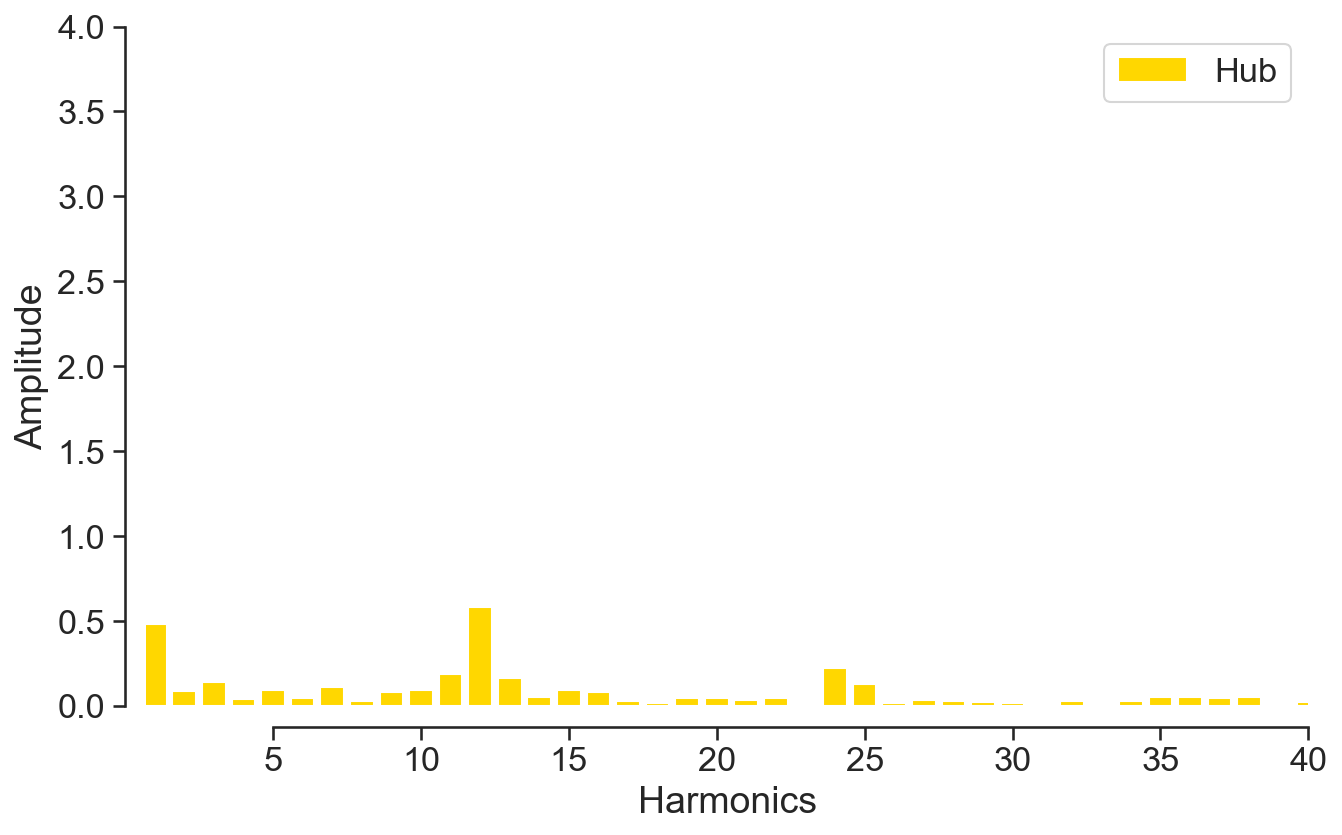}}
\subfigure[]{\includegraphics[]{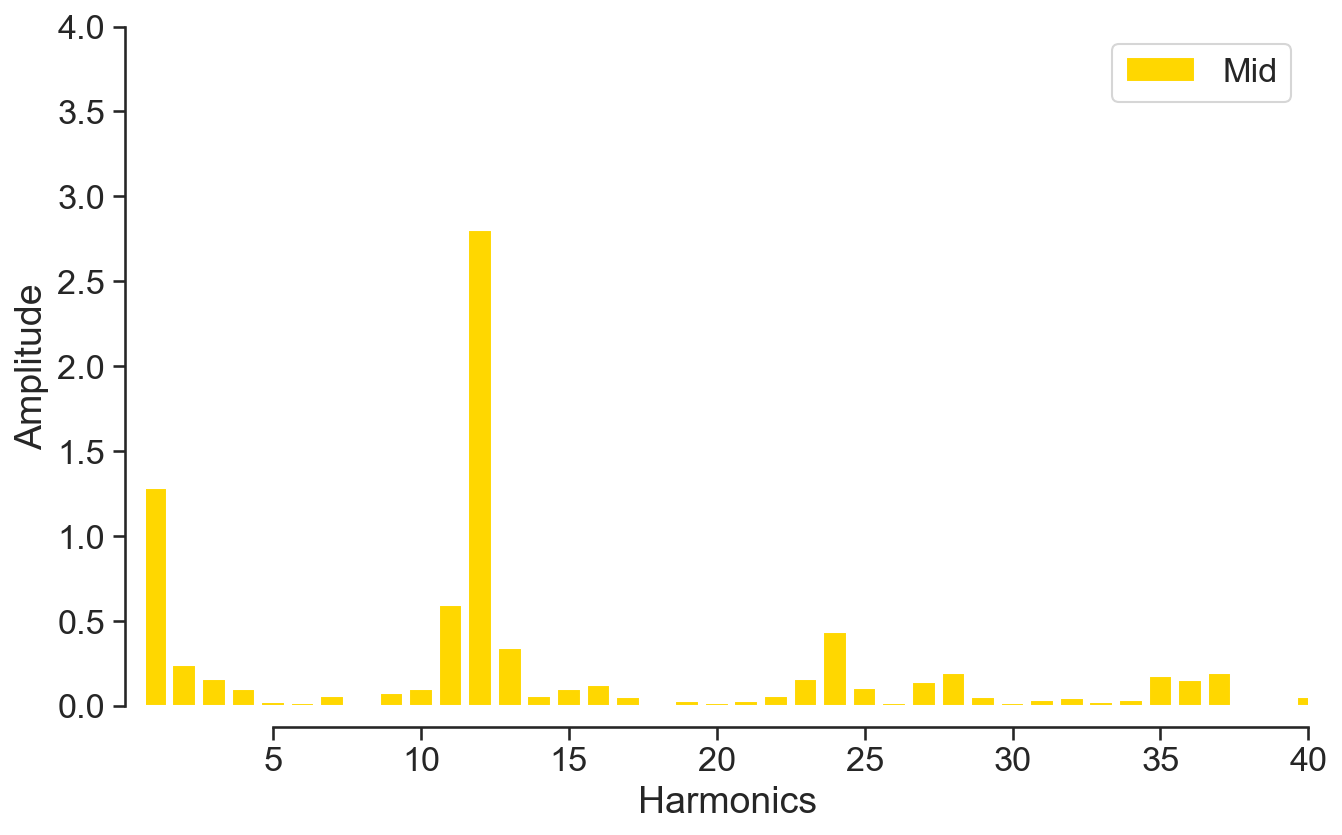}}
\subfigure[]{\includegraphics[]{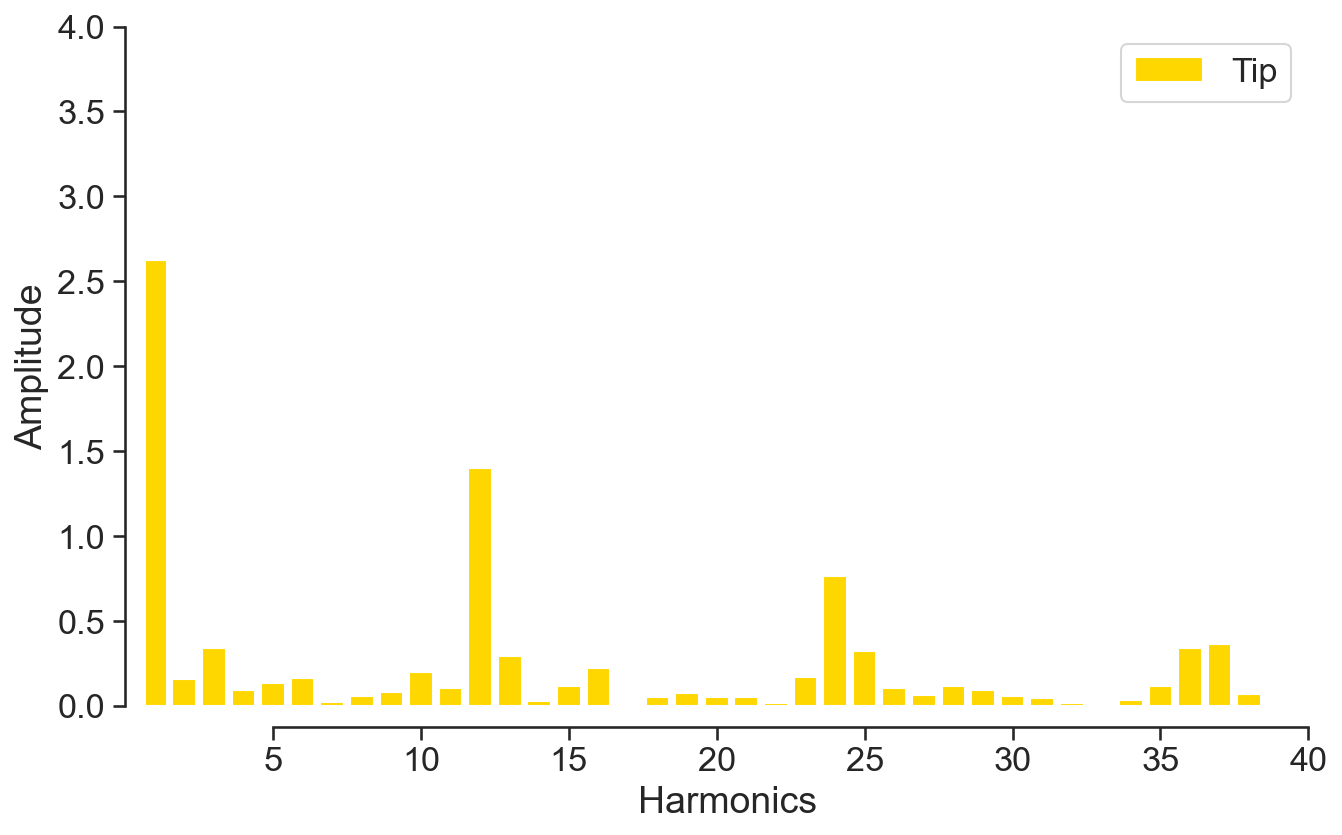}}
\end{subfigmatrix}
\caption{Experimental data from an exit station in a high pressure turbine test rig: (a) traverse locations; (b) true temperature; Fourier amplitudes at the (c) hub, (d) mid-span, and (f) tip.}
\label{fig:experiment}
\end{center}
\end{figure}

In this first case study, we consider traverse temperature measurements taken from a research turbine rig. Figure~\ref{fig:experiment}(a) shows the traverse locations at a temperature station, while \ref{fig:experiment}(b) shows the resulting steady-state temperature field. A fast Fourier transform was carried out on the temperature field at the hub, mid-span and tip along the circumferential direction; the resulting amplitudes are captured in Figures~\ref{fig:experiment}(c, d, e). It is clear that wave numbers 1, 12, and 24 are dominant. Additionally, we note that the signal is generally sparse, and thus utilisation of the aforementioned sparsity promoting priors seems like a sensible decision.

Let us assume that we can sample this spatial field using only 4 circumferential rakes, each fitted with 6 probes. Assume further that we are permitted to do this twice, with different circumferential rake placements. In both cases the rakes are clocked with respect to the upstream components, i.e., they are aligned to be at the same pitchwise location, so as not to capture any upstream wakes. 

\begin{figure}
\begin{center}
\begin{subfigmatrix}{2}
\subfigure[]{\includegraphics[]{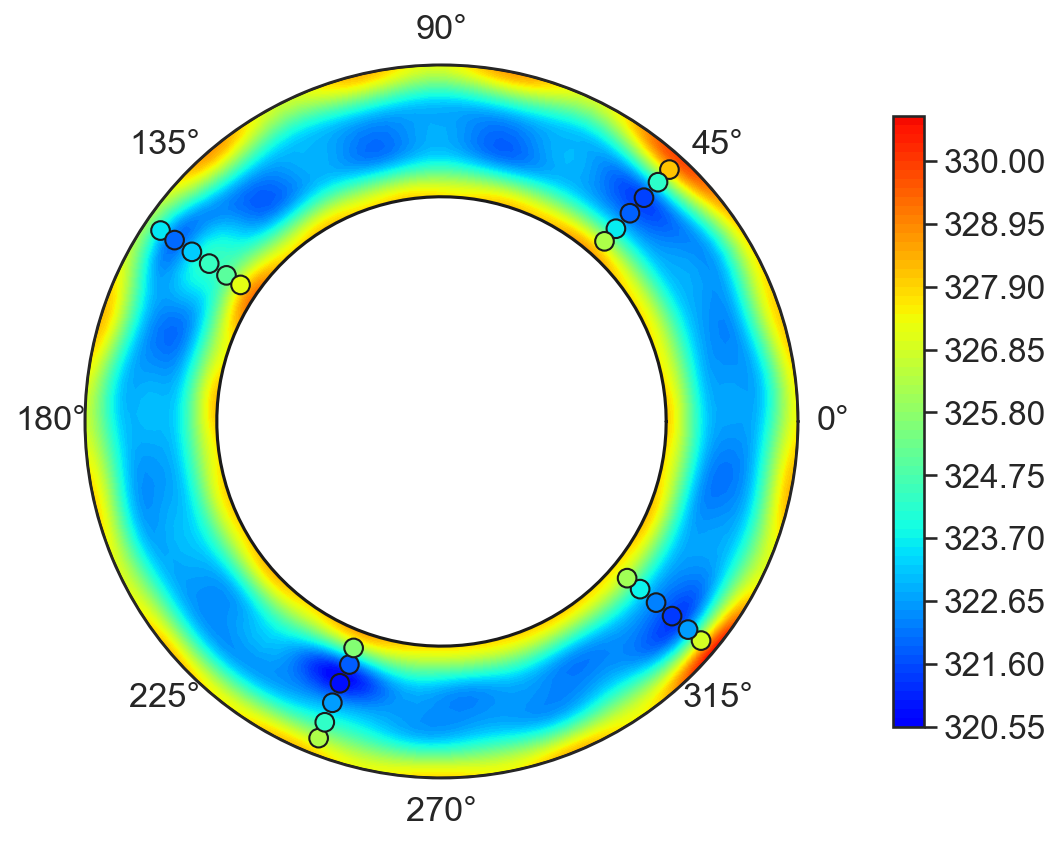}}
\subfigure[]{\includegraphics[]{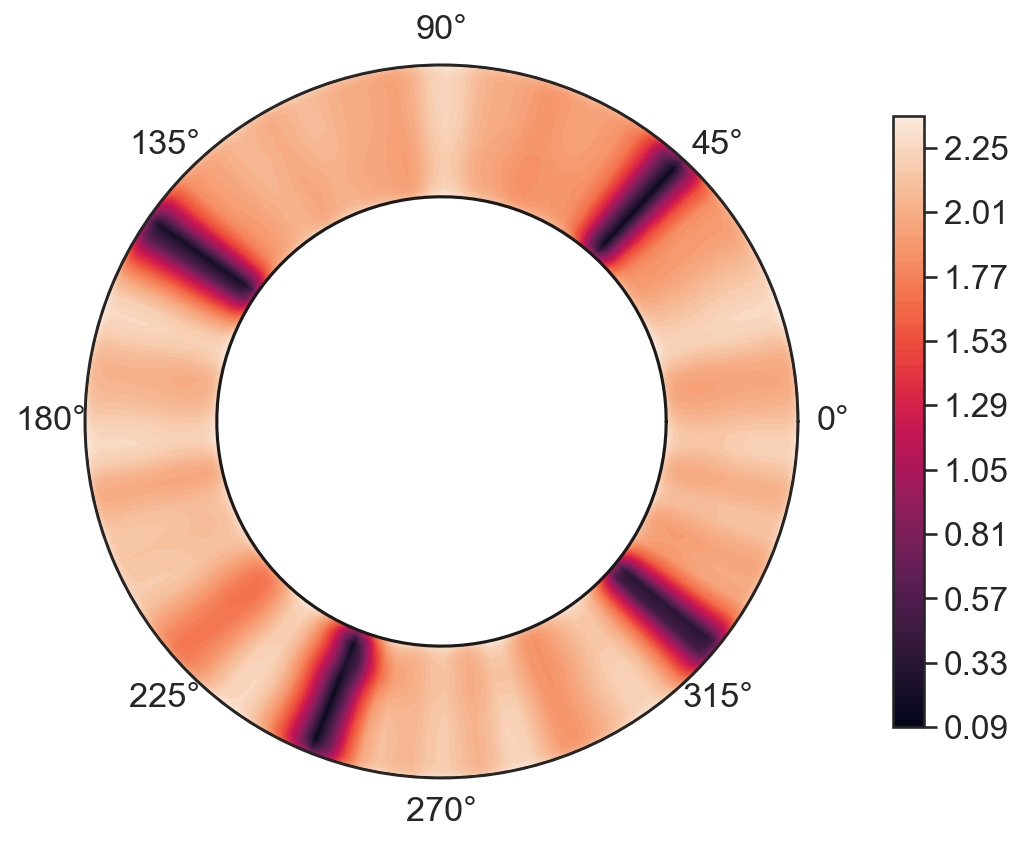}}
\subfigure[]{\includegraphics[]{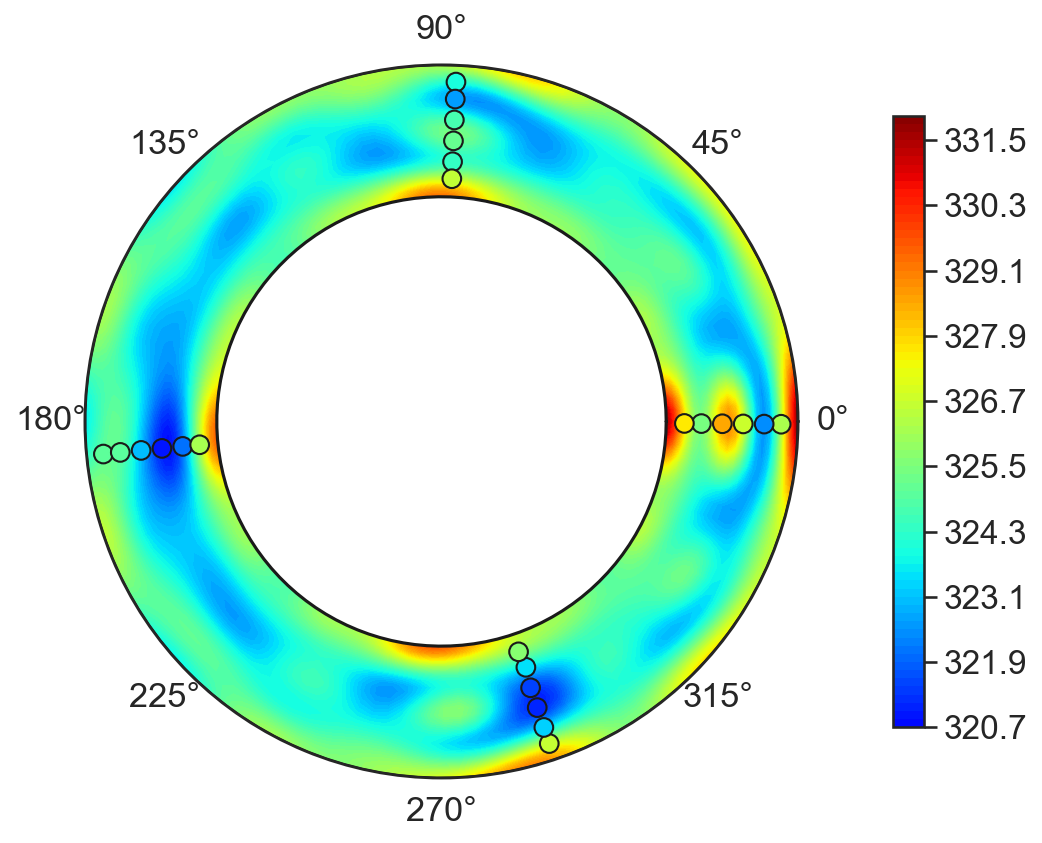}}
\subfigure[]{\includegraphics[]{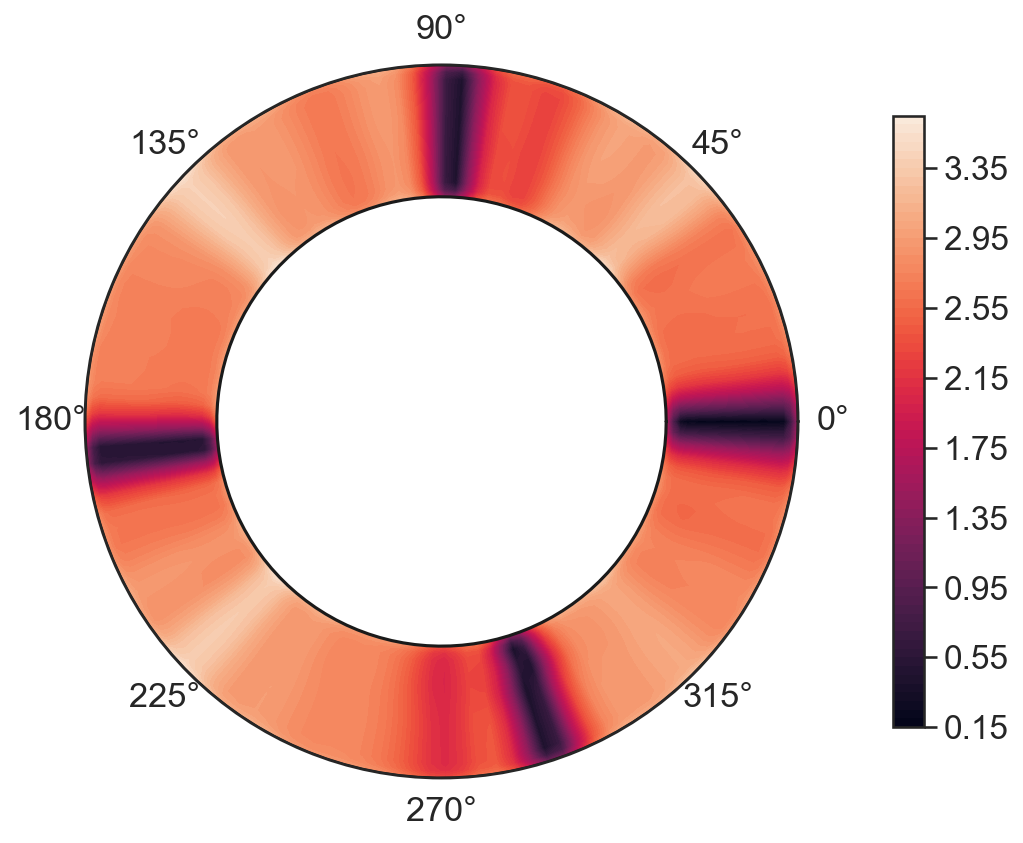}}
\end{subfigmatrix}
\caption{Single plane calculations for the first rake arrangement (top row) and the second rake arrangement (bottom row). Posterior annular mean in (a, c); standard deviation in (b, d).}
\label{fig:single_plane_rig}
\end{center}
\end{figure}

Running the isolated plane model---i.e., with no planar kernel---with sparsity promoting priors with $\nu = \left(1, 2, \ldots, 15 \right)$ for the first of the chosen rake arrangements, we obtain annular mean and standard deviation plots as shown in Figure~\ref{fig:single_plane_rig}(a, b). While the posterior Gaussian random field does interpolate the measurements, by inspection it is readily apparent that the spatial pattern in Figure~\ref{fig:single_plane_rig}(a) does not match the truth in Figure~\ref{fig:experiment}(b). Results run for the second rake arrangement are shown in Figure~\ref{fig:single_plane_rig}(c, d).

\begin{figure}
\begin{center}
\begin{subfigmatrix}{2}
\subfigure[]{\includegraphics[]{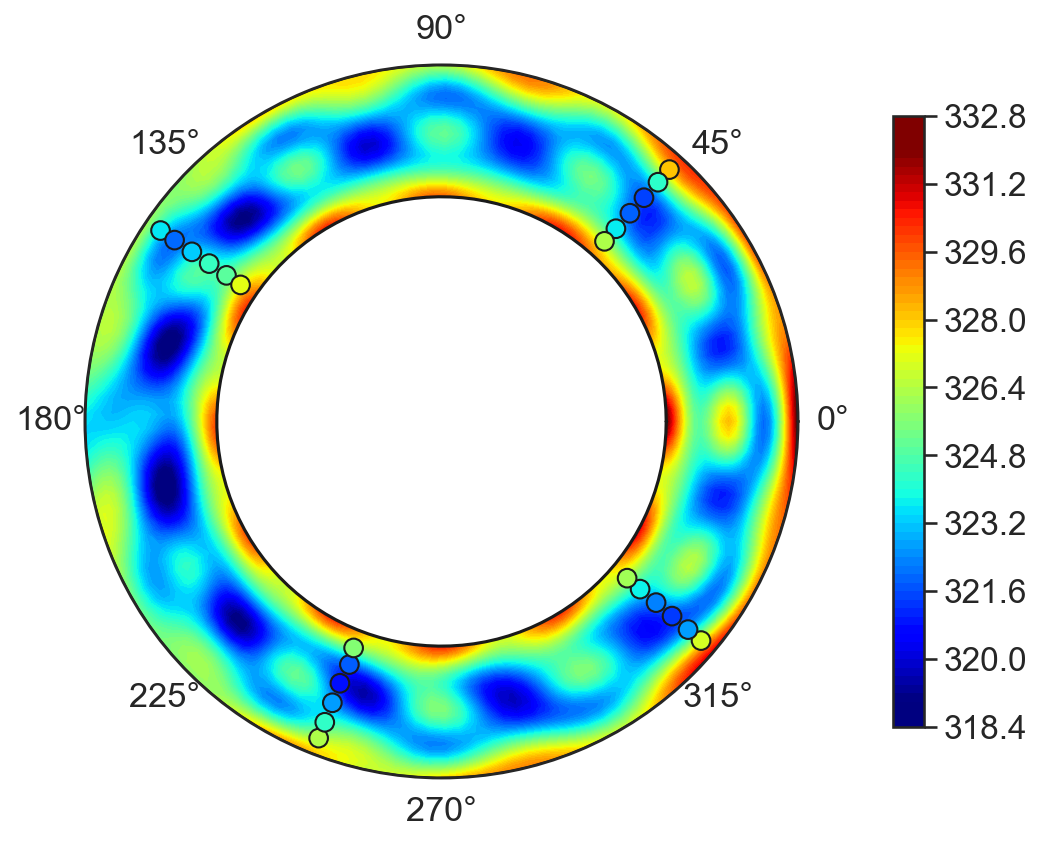}}
\subfigure[]{\includegraphics[]{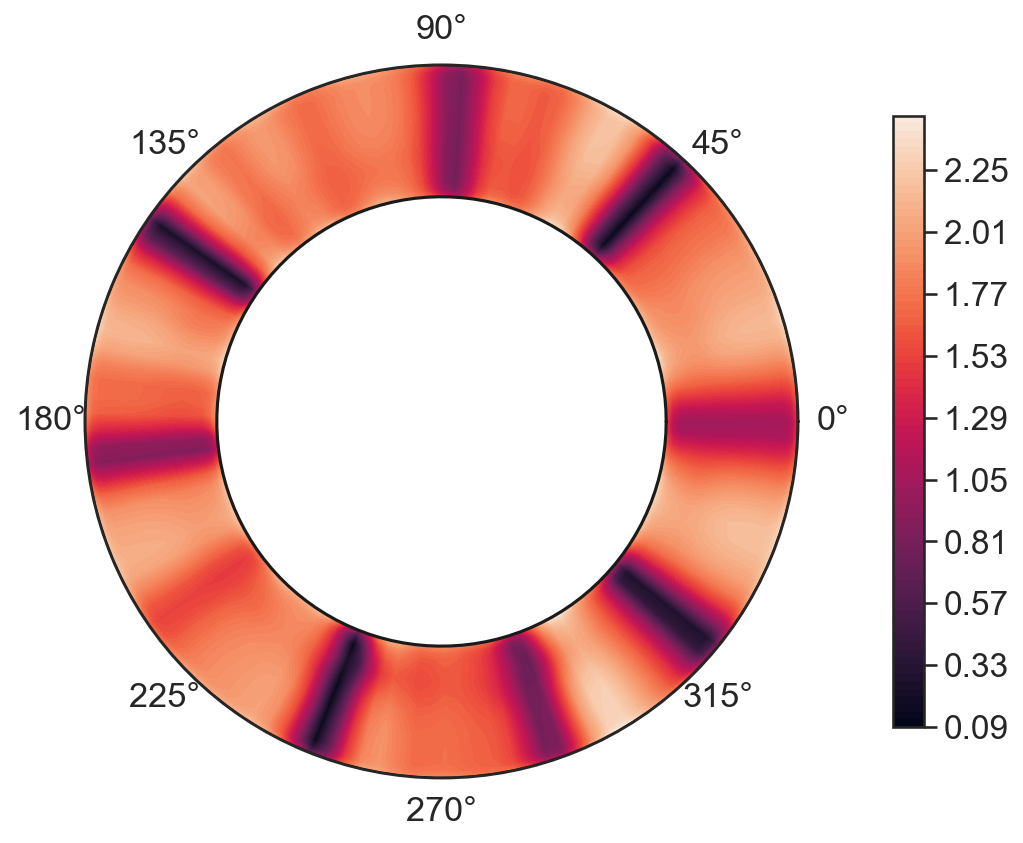}}
\subfigure[]{\includegraphics[]{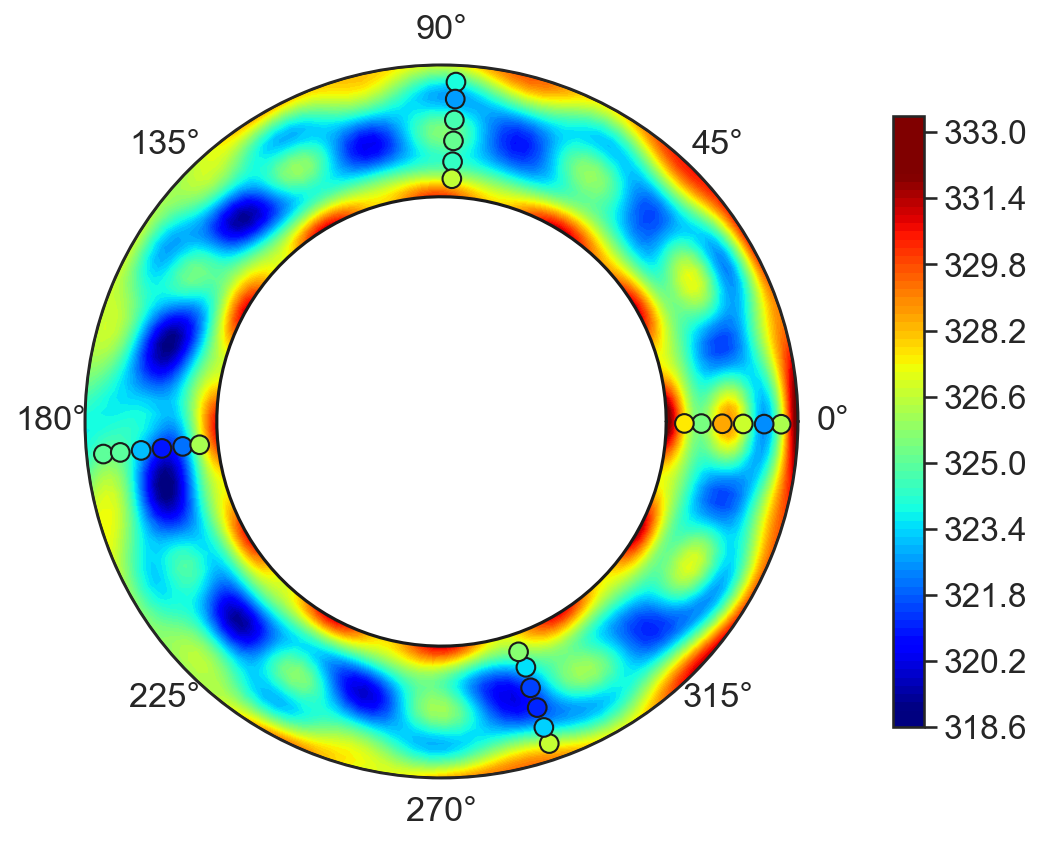}}
\subfigure[]{\includegraphics[]{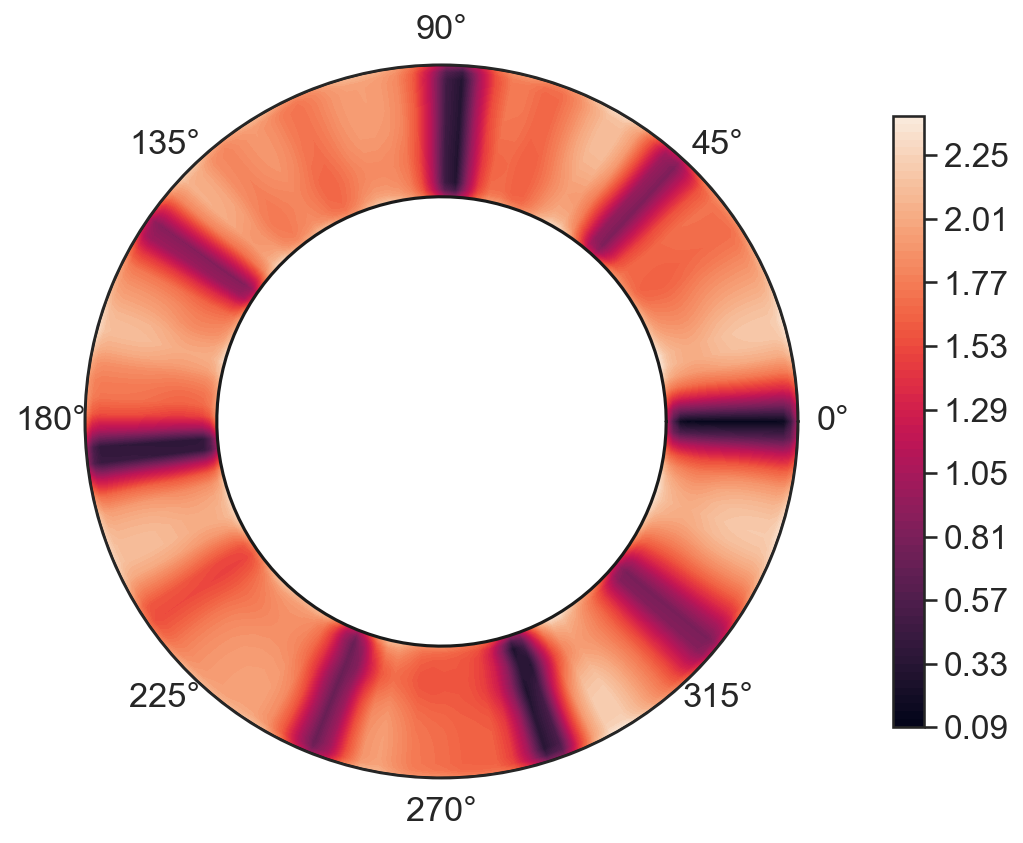}}
\end{subfigmatrix}
\caption{Multi-plane calculations for the first rake arrangement (top row) and the second rake arrangement (bottom row). Posterior annular mean in (a, c); standard deviation in (b, d).}
\label{fig:multi_plane_rig}
\end{center}
\end{figure}

To ascertain if the transfer learning approach works, we pass these two rake arrangements as two separate measurement planes in the multi-plane model. As these measurements are from the same physical measurement station, they are both assigned the same value in the similarity vector, i.e., $\vs=(1, 1)$. Once again, we resort to sparsity promoting priors for inference. The results are shown in Figure~\ref{fig:multi_plane_rig}. Beyond the greater resemblance to the truth in Figure~\ref{fig:experiment}(b), a slight reduction in the spatial uncertainty is also observed. It is clear that the model has successfully transferred information across the two measurement planes to arrive at a more precise estimate of the temperature distribution. At the same time, the model still has sufficient flexibility to offer slightly different temperature distributions for each plane individually; as we will see in the next case study, this is an extremely useful characteristic.

\begin{figure}
\begin{center}
\begin{subfigmatrix}{2}
\subfigure[]{\includegraphics[]{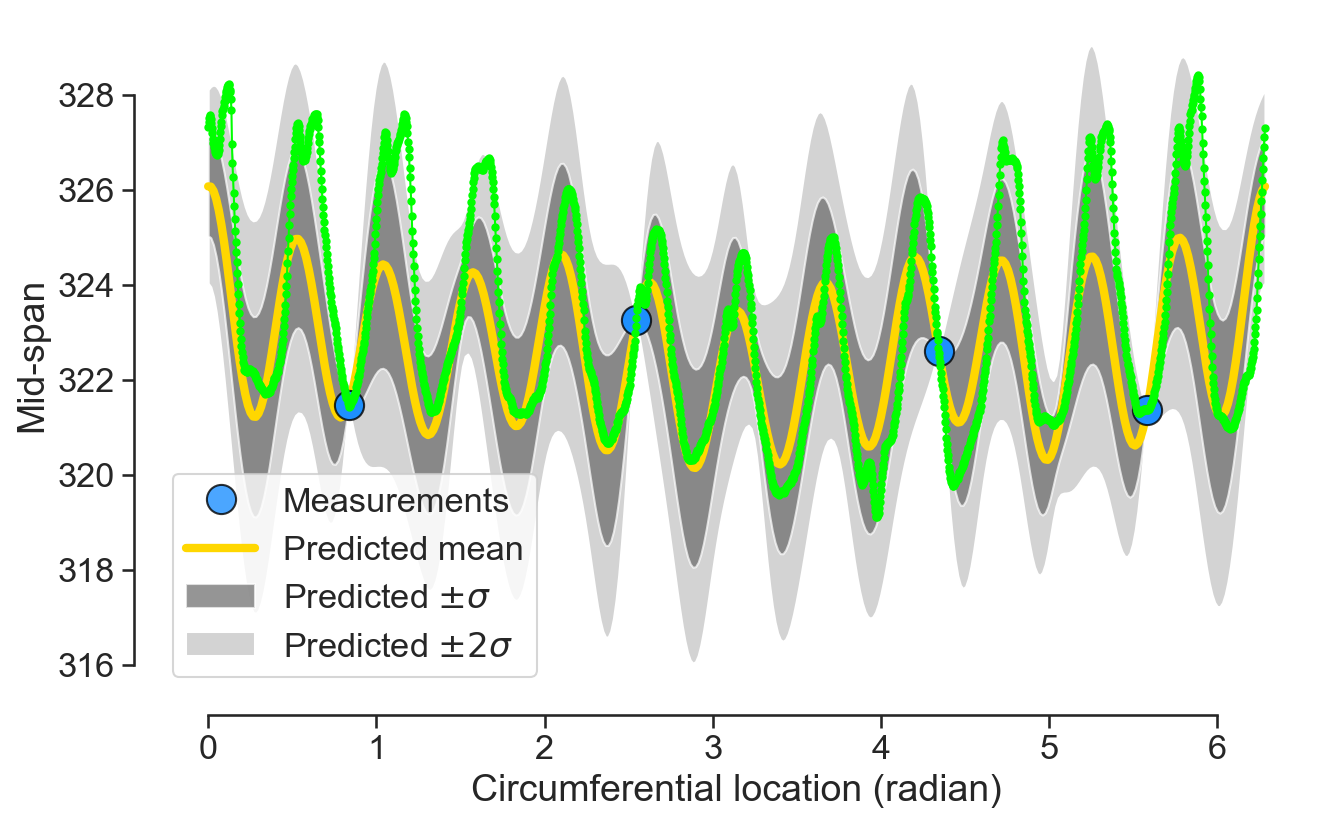}}
\subfigure[]{\includegraphics[]{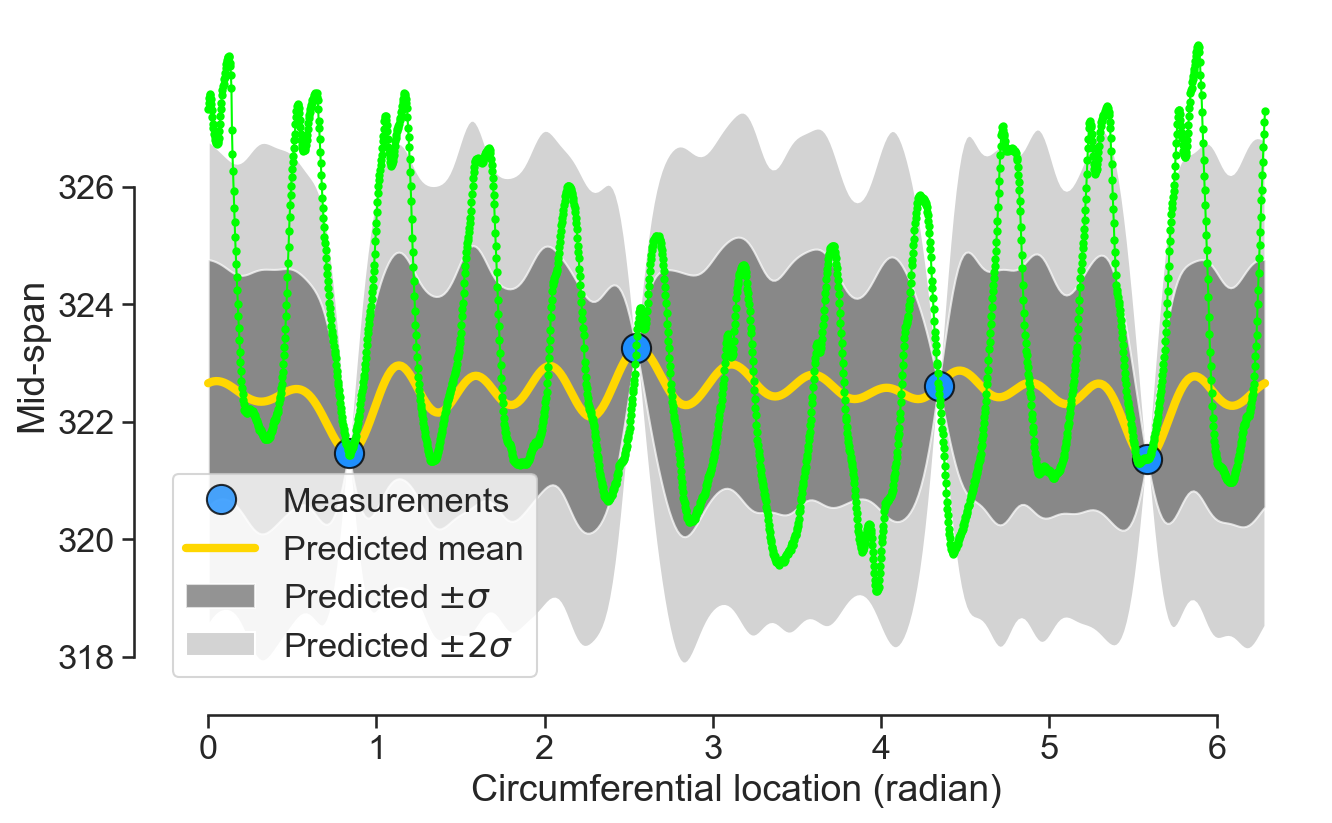}}
\subfigure[]{\includegraphics[]{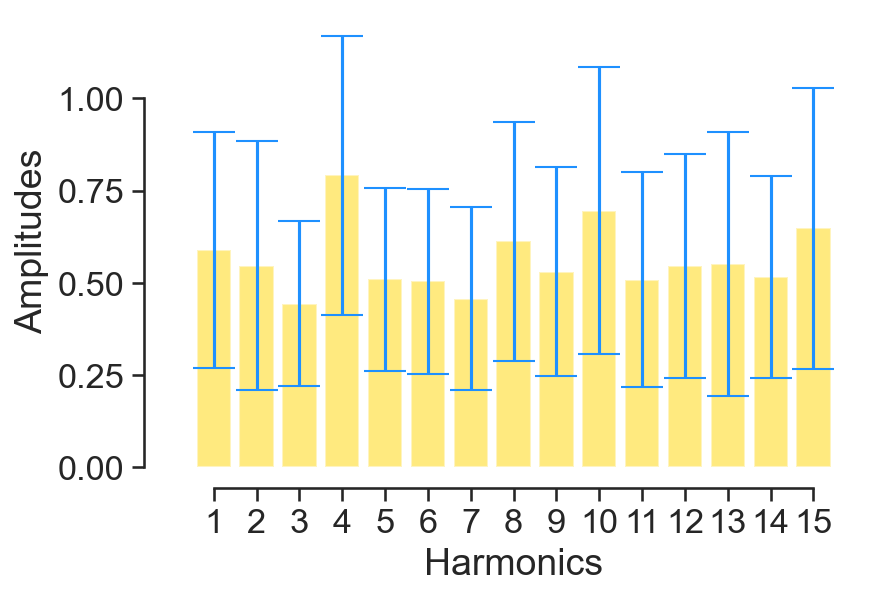}}
\subfigure[]{\includegraphics[]{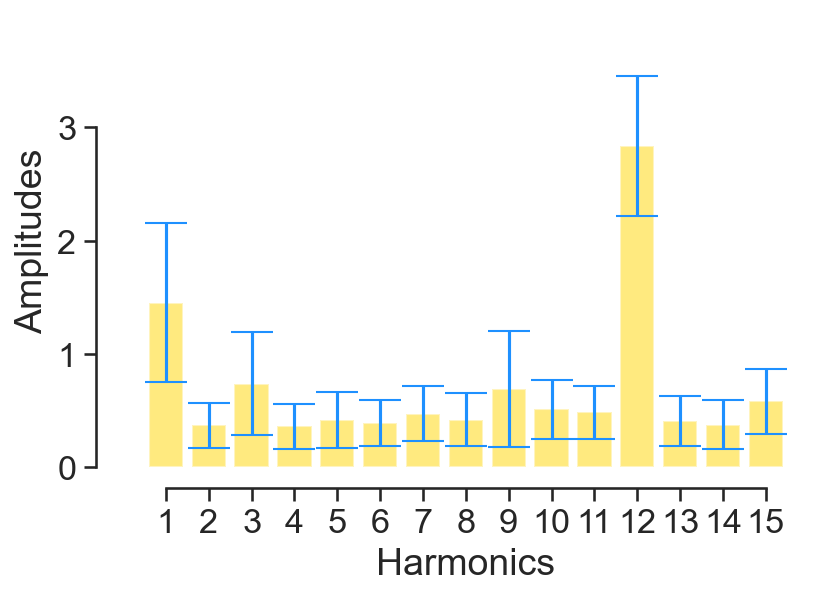}}
\end{subfigmatrix}
\caption{Comparison between the (a, c) single plane model and (b, d) the multi-plane transfer learning model at the mid-span location. Green circular markers are the true values from the rig; blue markers represent a subset of four rakes.}
\label{fig:multi_plane_rig_a}
\end{center}
\end{figure}

Circumferential plots of the single and the multi-plane yielded posterior distributions are contrasted in Figure~\ref{fig:multi_plane_rig_a} at the mid-span location. For the multi-plane result, only the result from the first rake arrangement is shown. While it is apparent that in both cases the true pattern (shown with black circular markers) is well-captured within two standard deviations, in (b) the uncertainty is significantly reduced partly owing to the improved prediction of the mean. This comparison is important to emphasise, as it demonstrates the accuracy of the model's predictions.

\subsection{Transfer learning with two adjacent planes from the same research engine}
Next, we consider two temperature measurement planes located axially adjacent to each other in a research aero-engine. The first plane comprises 7 rakes each fitted with 7 temperature probes. The second plane comprises 24 thermocouples all placed at mid-span. As there are no rotating components between these two measurement stations, and owing to the fact that the flow is predominantly axial, it is hypothesised that they should have very similar temperature behaviour.

\begin{figure}
\begin{center}
\begin{subfigmatrix}{2}
\subfigure[]{\includegraphics[]{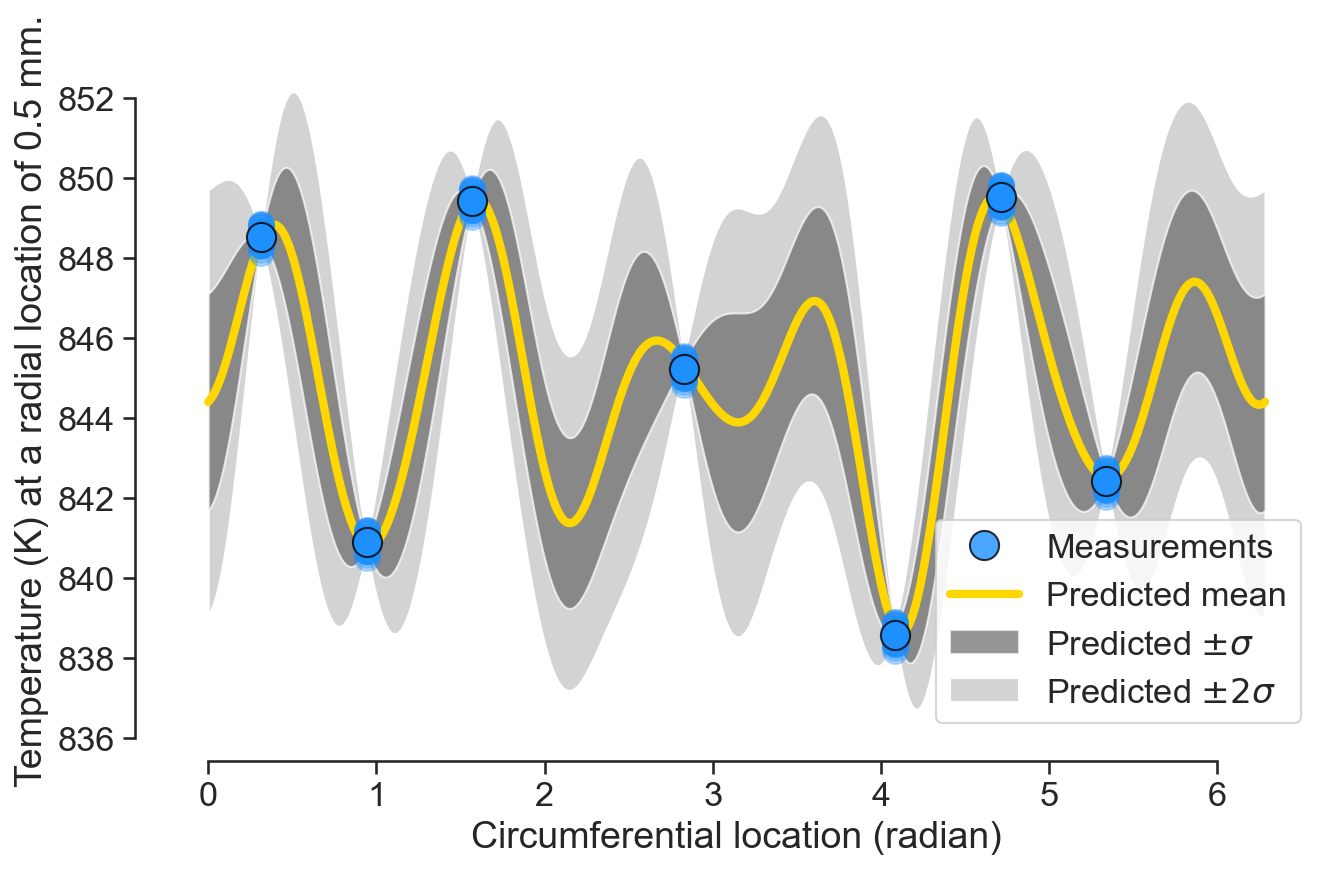}}
\subfigure[]{\includegraphics[]{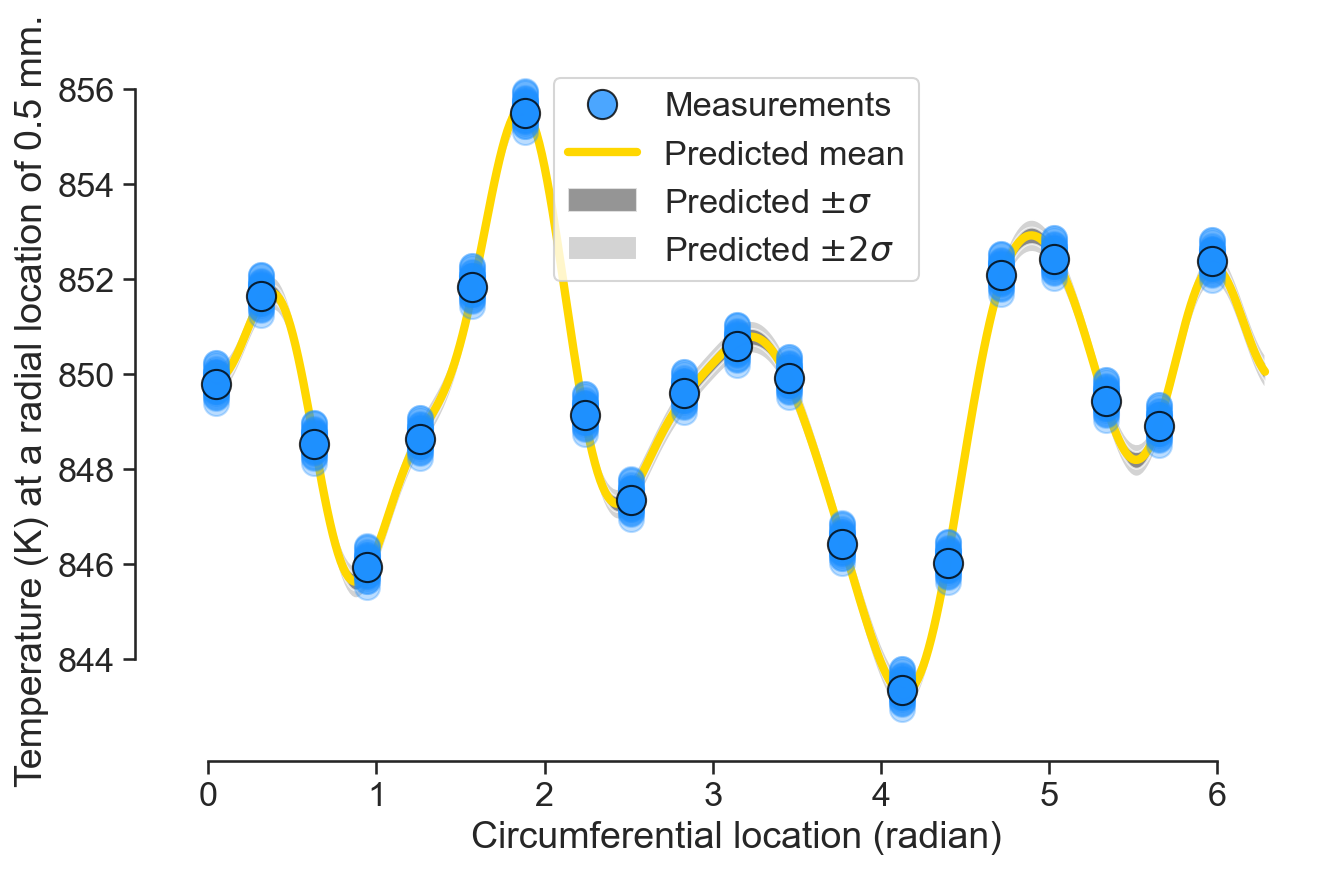}}
\subfigure[]{\includegraphics[]{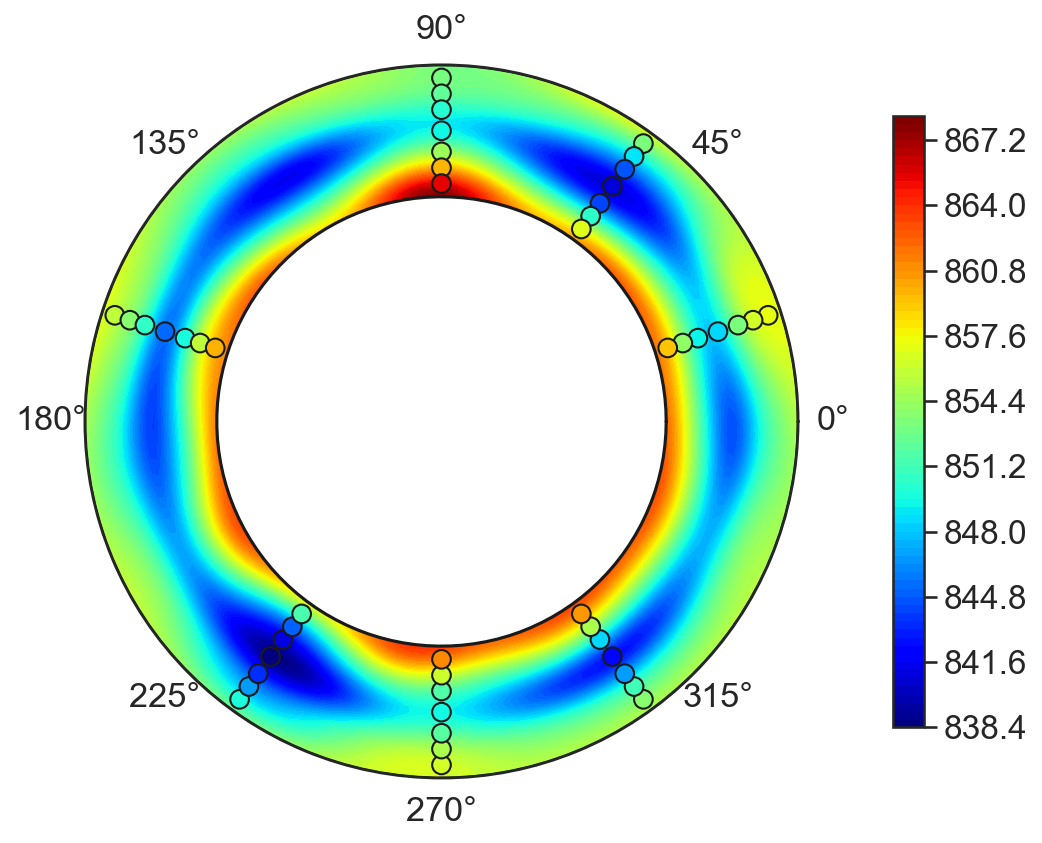}}
\subfigure[]{\includegraphics[]{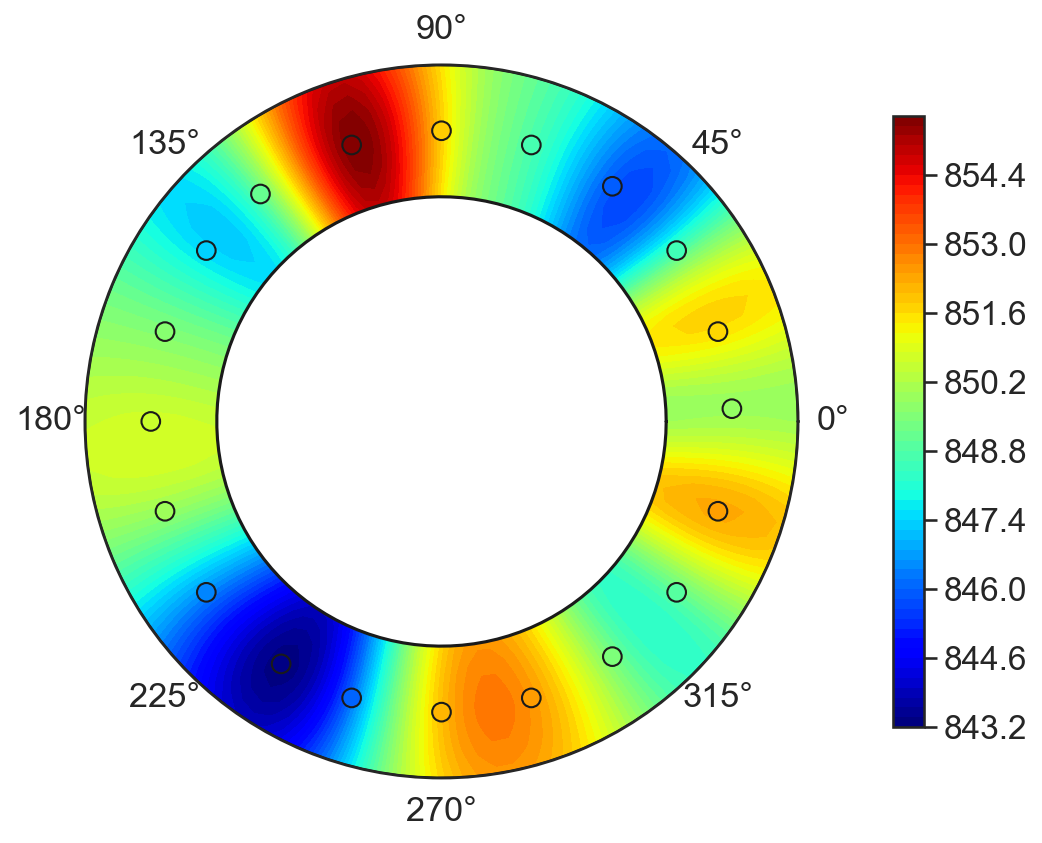}}
\subfigure[]{\includegraphics[]{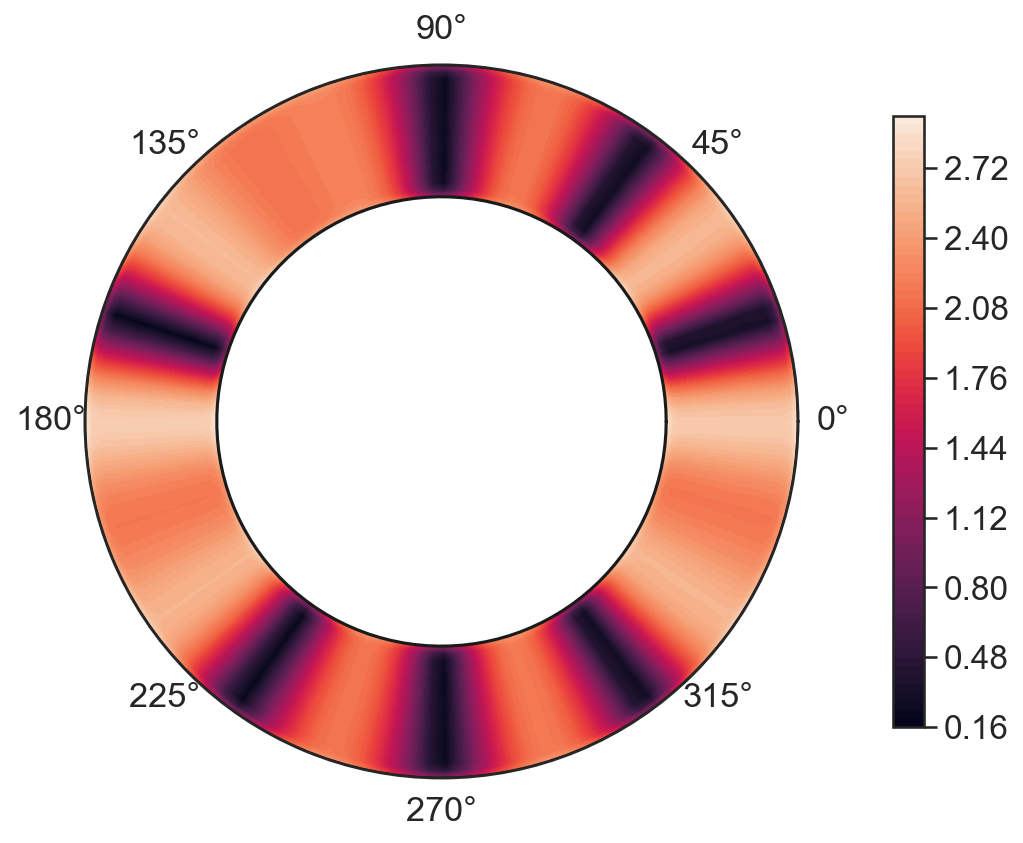}}
\subfigure[]{\includegraphics[]{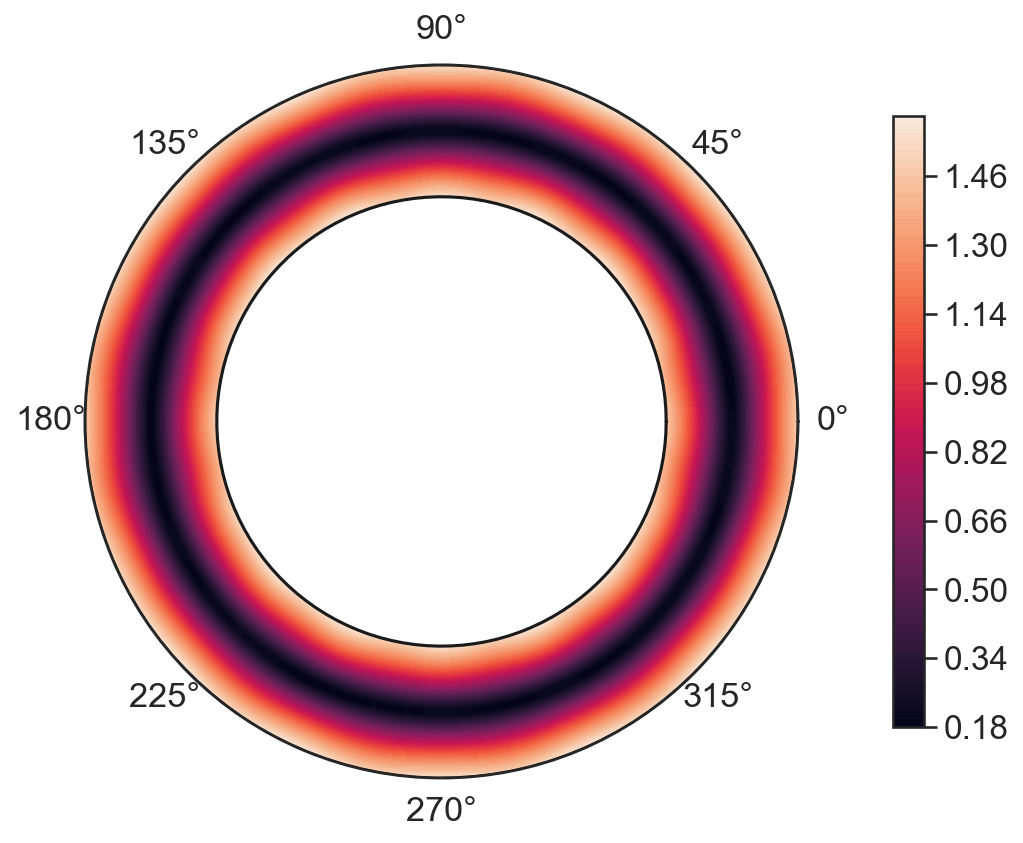}}
\end{subfigmatrix}
\caption{Single model results for the first plane in (a, c, e) and the second plane in (b, d, f); here each plane was run individually.}
\label{fig:multi_plane_same_engine_a}
\end{center}
\end{figure}

The results of evaluating each measurement plane in isolation are captured in Figure~\ref{fig:multi_plane_same_engine_a} with circumferential distributions at mid-span for each plane. For these results, the sparsity priors were used with wave numbers $\nu = \left(1, 2, \ldots, 9 \right)$. This choice was set by the fact that the inclusion of wave numbers above 9 in the first plane leads to aliasing as the minimum angular distance between probes is $36^{\circ}$. 

It is clear that owing to the number of measurements in the second plane, there is little uncertainty in the overall circumferential distribution. The same cannot be said for the upstream stator plane in (a). Thus, the goal here is to explore whether the transfer learning enabled multi-plane model can (i) reduce the circumferential uncertainty in the first plane, whilst (ii) reducing the radial uncertainty in the second plane.

As before, for the transfer learning model we set $\vs=(1,1)$. Note that this explicitly assumes that both planes have the same set of wave numbers, although their precise amplitudes and phases may moderately differ. 

\begin{figure}
\begin{center}
\begin{subfigmatrix}{2}
\subfigure[]{\includegraphics[]{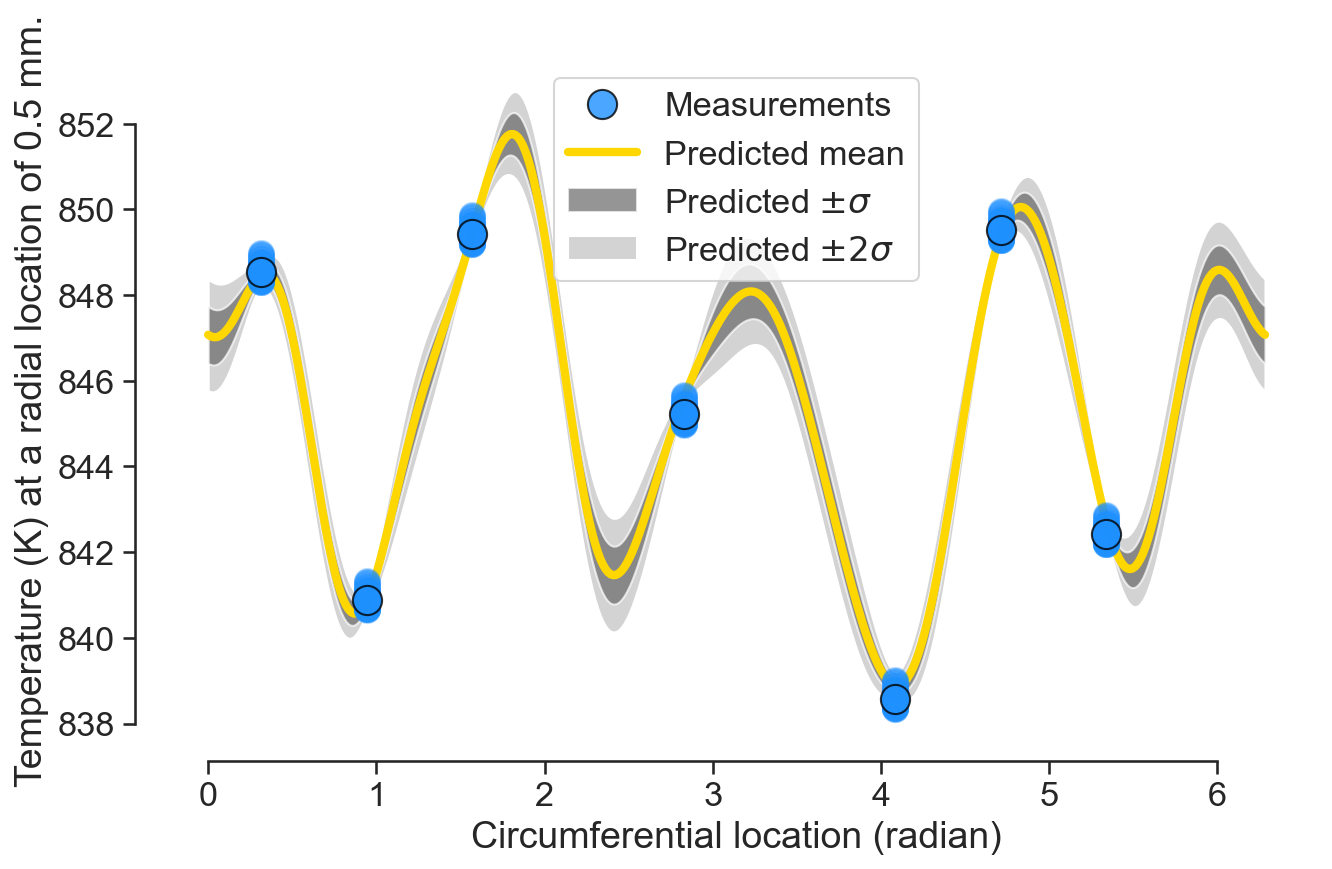}}
\subfigure[]{\includegraphics[]{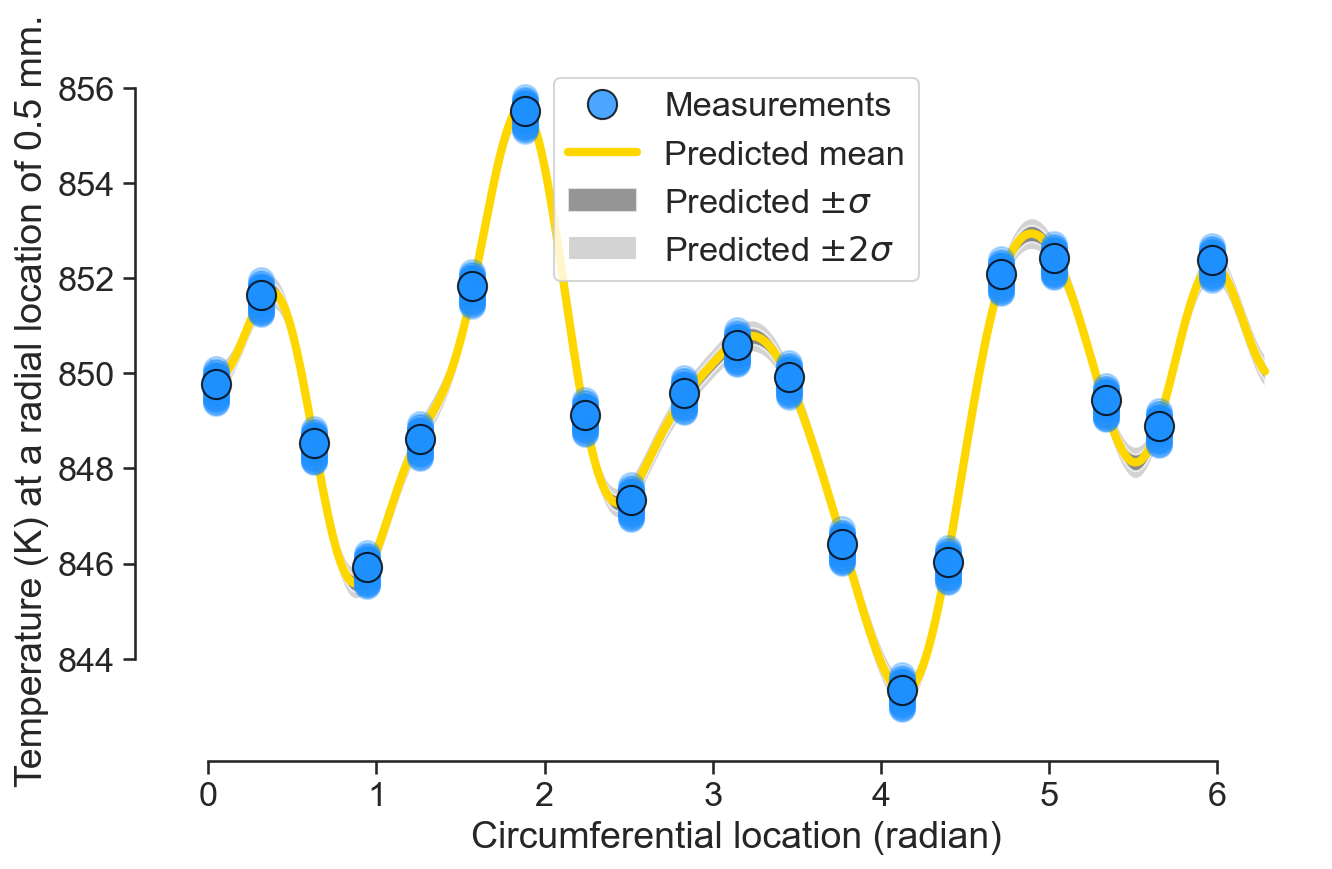}}
\subfigure[]{\includegraphics[]{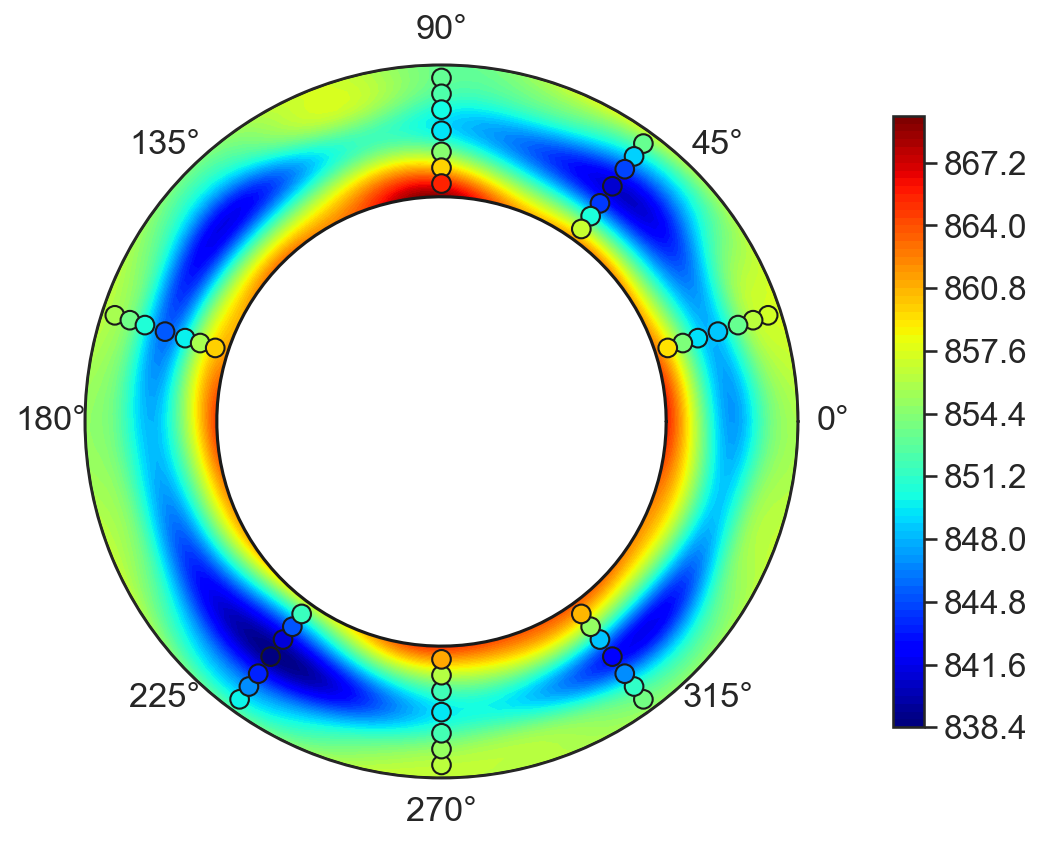}}
\subfigure[]{\includegraphics[]{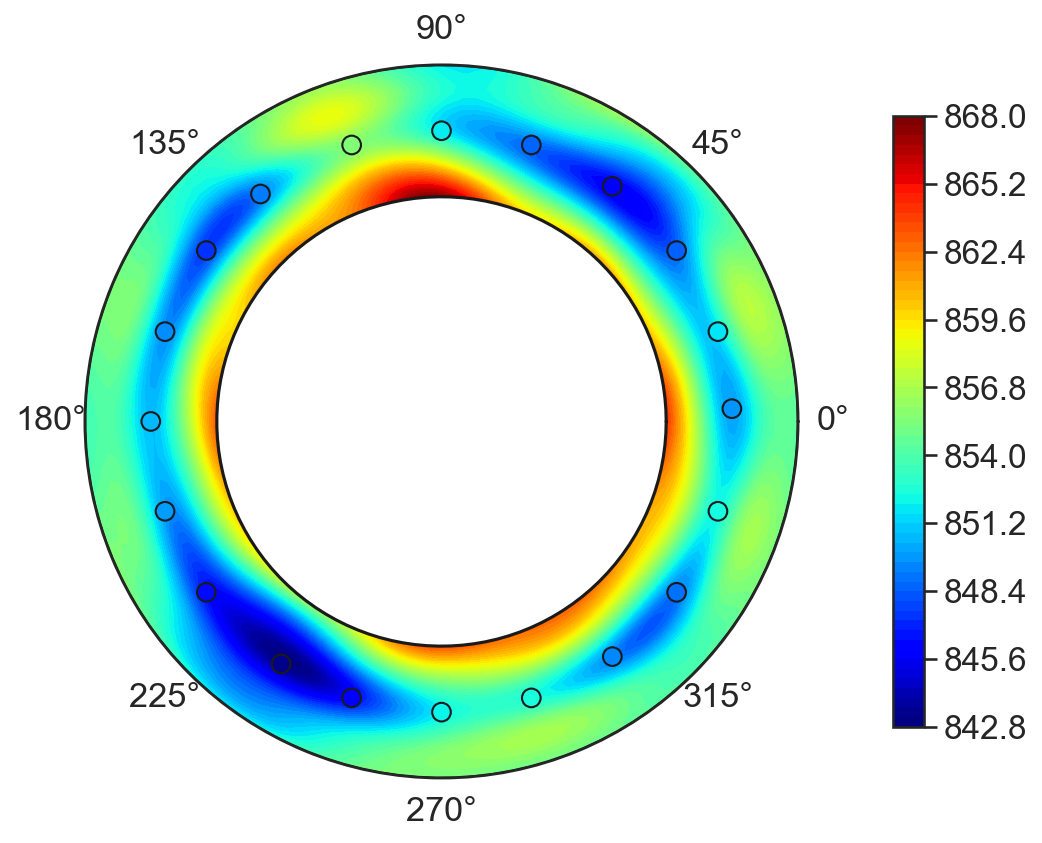}}
\subfigure[]{\includegraphics[]{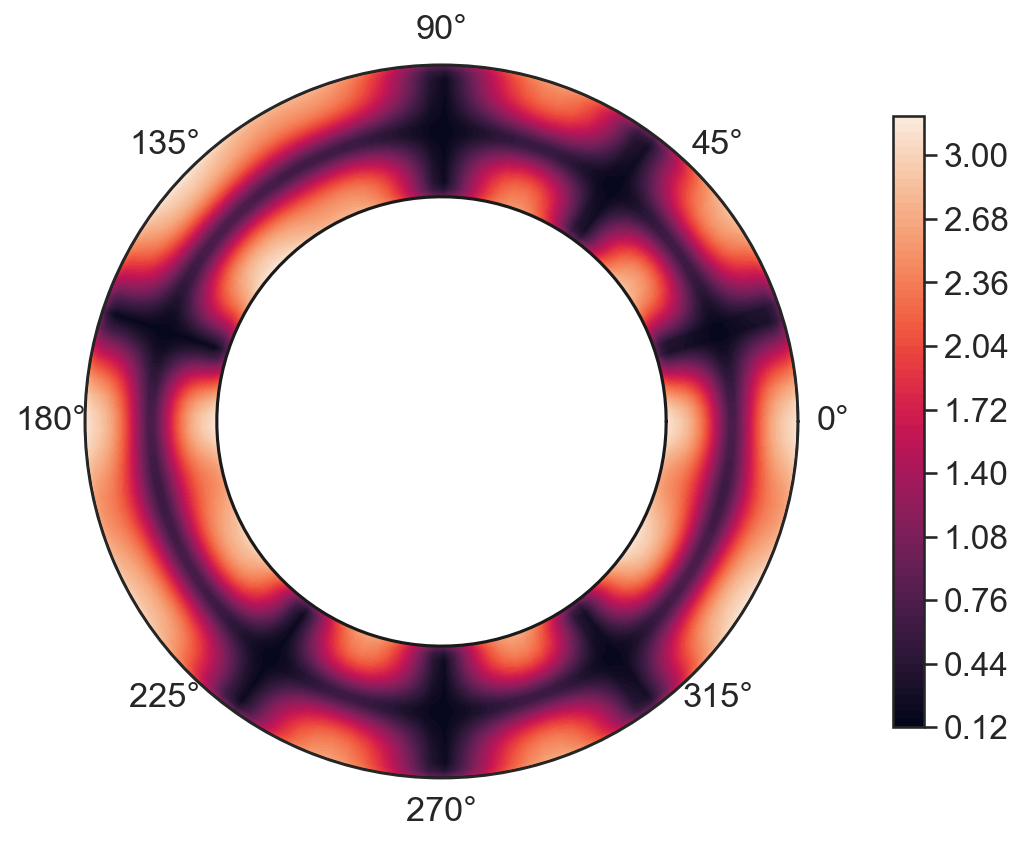}}
\subfigure[]{\includegraphics[]{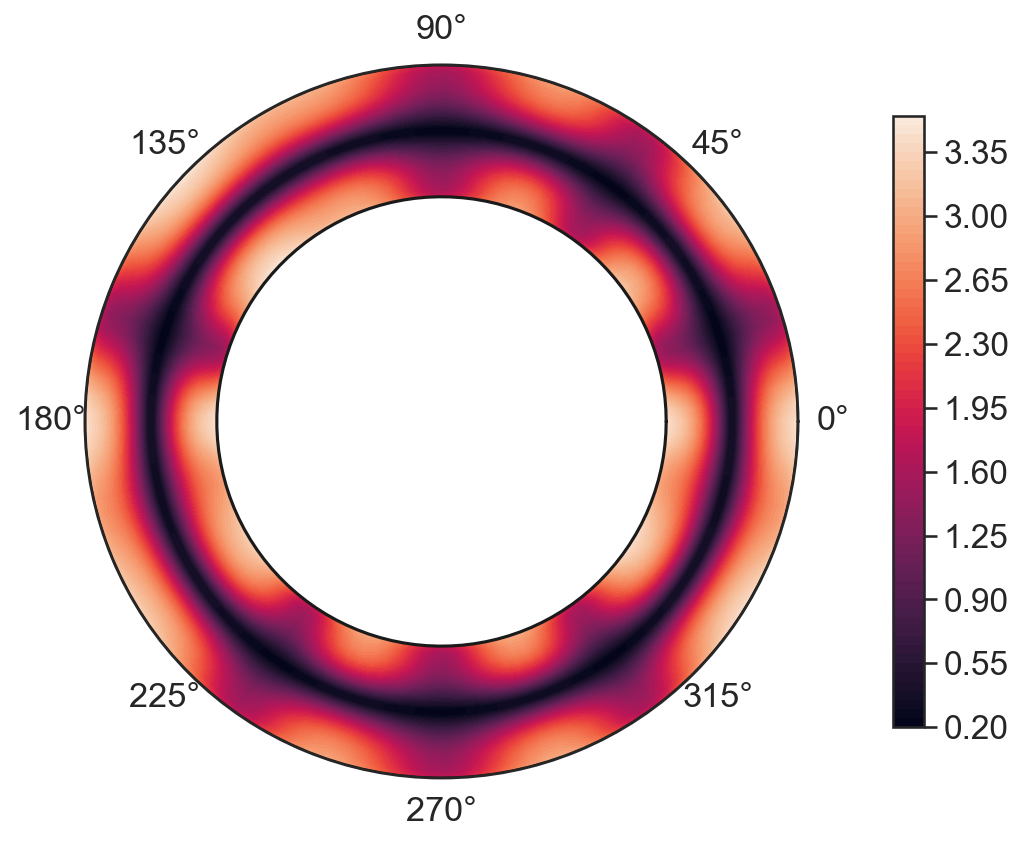}}
\end{subfigmatrix}
\caption{Multi-plane model results for the first plane in (a, c, e) and the second plane in (b, d, f).}
\label{fig:multi_plane_same_engine_b}
\end{center}
\end{figure}

Figure~\ref{fig:multi_plane_same_engine_b} shows the results of the proposed model. It is clear that there is a reduction in the uncertainties in the radial direction in the burner plane, corresponding to the rake locations in the stator plane. There is also a significant reduction in the circumferential direction at mid-span region in the stator plane. Additionally, note how the radial distribution of temperature in the second plane resembles that seen on the first plane.

\subsection{Transfer learning across a fleet}
One criticism of the work thus far is the reliance on $\vs$. Whilst in many cases, it is easy to establish whether two sets of measurements are similar, there may be equally many instances where such connections are difficult to draw. Ideally in such scenarios it will be useful if the model itself can shed some light on the relative similarity between measurement planes, by virtue of radial and circumferential characteristics. 

In this last example, we study the results of the multi-plane model on 8 planes. The data chosen for this study corresponds to the temperature measurements taken from the same measurement at approximately the same throttle setting for 8 different research engines. Planes E1 to E3 belong to the same family, and planes E4, E5 and E7 belong to another family. Plane E8 is similar to E1 to E3, but does have a different blade numbers. Additionally, planes E6 and E7 are more closely related to E4 and E5 than to planes E1 to E3. 

\begin{figure}
\begin{center}
\includegraphics[width=0.7\textwidth]{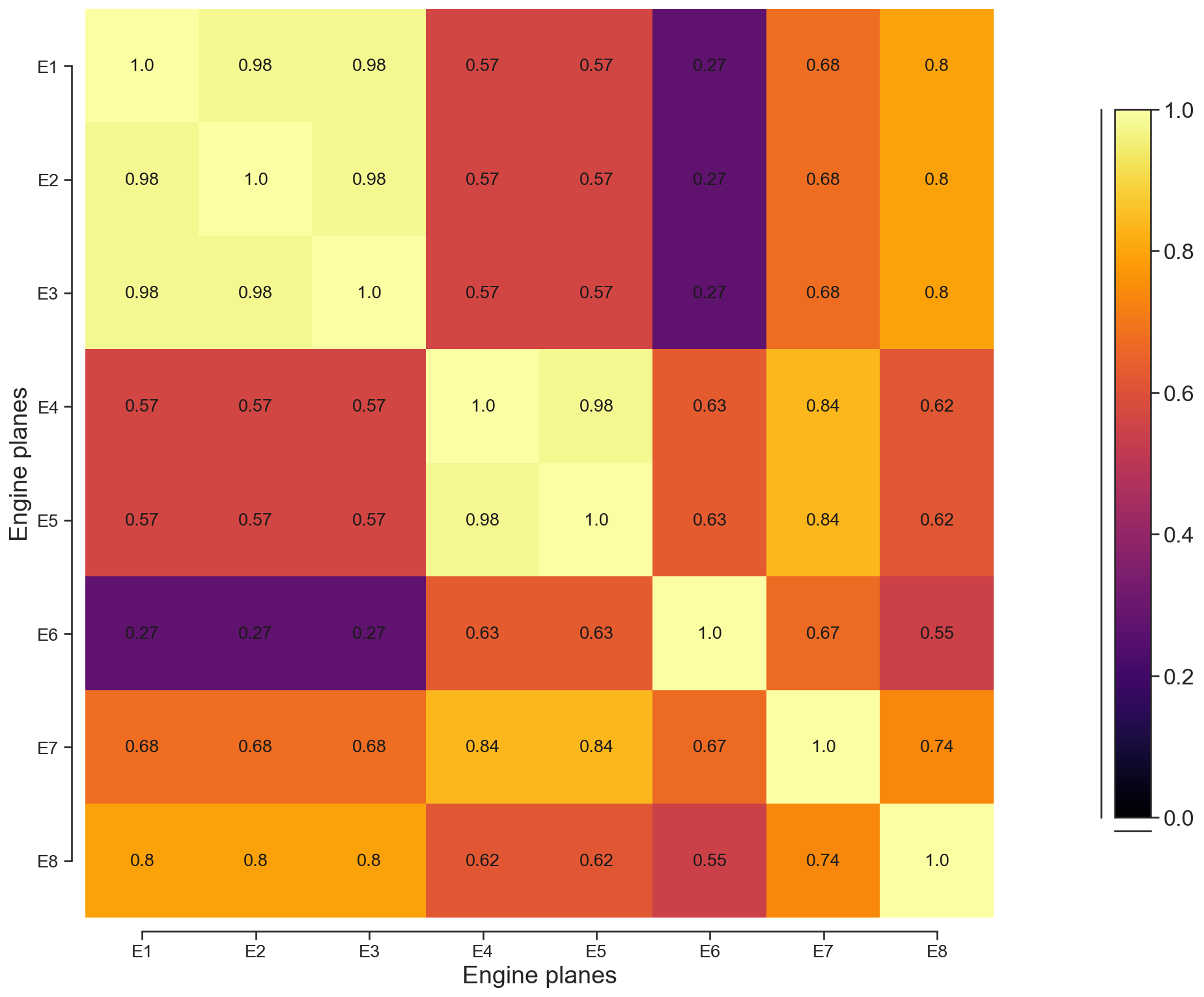}
\caption{A planar correlation plot developed by averaging posterior distributions of the parameters in $\mS$.}
\label{fig:last}
\end{center}
\end{figure}

Rather than encode all these relationships in $\vs$, we intentionally capture only the first few and set $\vs=(1,1,1,2,2,3,4,5)$. From the resulting posterior distributions of $\\xi_1, \ldots, \xi_{W}$ and their placement in $\mS$, we can construct the correlation matrix shown in Figure~\ref{fig:last} by taking the mean of all the relevant hyperparameters $\boldsymbol{\xi}$. To reiterate, these correlation values stem from the constants in \eqref{equ:kernel_planar}. 

\begin{figure}
\begin{center}
\includegraphics[scale=0.7]{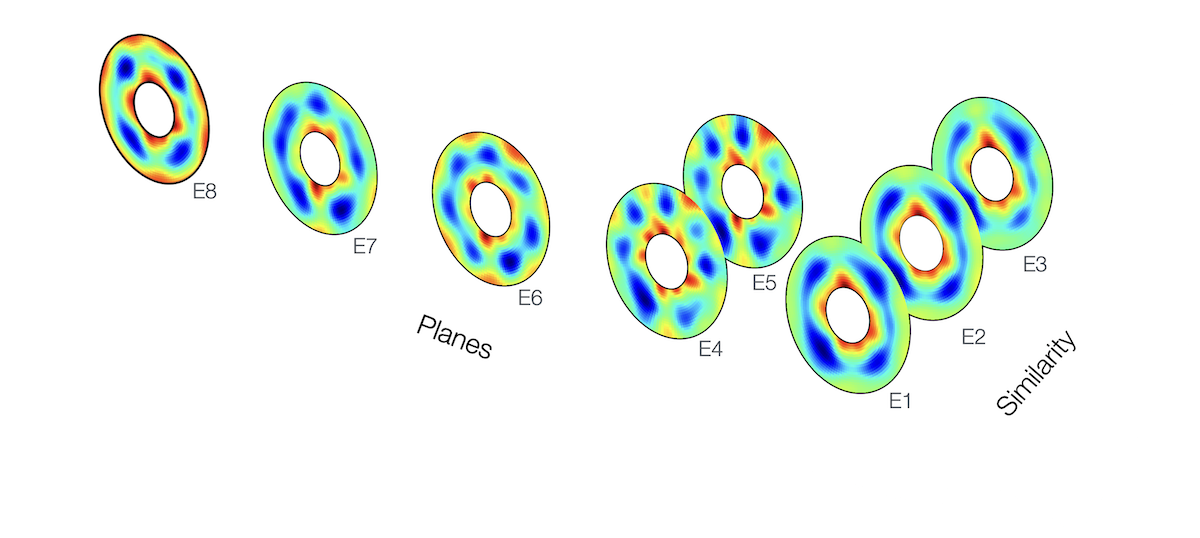}
\caption{Posterior spatial means of the different planes.}
\label{fig:final}
\end{center}
\end{figure}

From this correlation plot we observe that many of the relationships previously mentioned but not captured in $s$ are apparent. For instance, E6 is observed to be more closely related to E4 and E5 respectively, compared to E1 (and by extension E2 and E3) with a value of 0.27. The model also rated E8's similarity to E1 at 0.8, which seems reasonable given that both have a dominant mode four pattern. This value is higher than the correlation between E1, E2, and E3 and any of the other engines, which also aligns with our expectations. For completeness we include the posterior mean distributions in Figure~\ref{fig:final}.

\section*{Conclusions}
Understanding the spatial annular pattern born from engine measurements provides valuable aerothermal insight. This paper presents a transfer learning model suited for engine temperature and pressure measurements. It represents a step-up from prior averaging, uncertainty assessment and spatial extrapolation works. Central to our contribution is the ability to transfer information across planes with a planar kernel and a user-defined input on the similarity between the different measurement planes. Beyond the results presented in this paper, the proposed model has been extensively tested on measurement planes with 1-2 rakes of instrumentation when paired with planes with 6-7 rakes of instrumentation---with the goal of improving the spatial prediction even with 1-2 rakes. Across all cases, the multi-plane model yielded improved predictions. With respect to the proposed transfer learning model, future work could leverage Dirichlet prior models to learn $\mS$ from data directly as part of the inference process.

Given the utility of the proposed model to test and measurement, future engine design, and engine health monitoring programmes, we anticipate many forthcoming advances within this modelling paradigm, especially for prognostic and diagnostic efforts.

\paragraph{Funding Statement}
The work was part funded by the Fan and Nacelle Future Aerodynamic Research (FANFARE) project under grant number 113286, which receives UK national funding through the Aerospace Technology Institute (ATI) and Innovate UK, together with work funded by Rolls-Royce plc. The authors are grateful to Rolls-Royce plc for permission to publish this paper.

\paragraph{Data Availability Statement}
The data that support the findings of this study is available from Rolls-Royce. Restrictions apply to the availability of this data. 

\paragraph{Competing Interests  Statement}
All authors indicate no competing interests.

\bibliography{references}
\bibliographystyle{asmems4}

\end{document}